\documentclass[a4paper,11pt]{article}

\usepackage{jheppub} 

\usepackage[T1]{fontenc} 
\usepackage{amssymb}
\usepackage{tikz}
\usepackage{enumitem}
\usepackage{amsmath}
\usepackage{mathrsfs}
\usepackage{comment}
\usepackage{float}
\usepackage{xcolor}

\title{\boldmath The two-loop eight-point amplitude in ABJM theory}

\author[a,b,c]{Song He}
\author[d,e]{Yu-tin Huang}
\author[d]{Chia-Kai Kuo}
\author[a]{Zhenjie Li}

\affiliation[a]{CAS Key Laboratory of Theoretical Physics, Institute of Theoretical Physics, Chinese Academy of Sciences, Beijing 100190, China}
\affiliation[b]{
School of Fundamental Physics and Mathematical Sciences, Hangzhou Institute for Advanced Study, UCAS, Hangzhou 310024, China; ICTP{-}AP,
Beijing/Hangzhou, China}
\affiliation[c]{Peng Huanwu Center for Fundamental Theory, Hefei, Anhui 230026, P. R. China}
\affiliation[d]{Department of Physics and Center for Theoretical Physics, National Taiwan University, Taipei 10617, Taiwan}
\affiliation[e]{Physics Division, National Center for Theoretical Sciences, Taipei 10617, Taiwan}

\emailAdd{songhe@itp.ac.cn}
\emailAdd{yutinyt@gmail.com}
\emailAdd{chiakaikuo@gmail.com}
\emailAdd{lizhenjie@itp.ac.cn}

\abstract{In this paper, we present the two-loop correction to scattering amplitudes in three-dimensional $\mathcal{N}=6$ Chern-Simons matter theory. We use the eight-point case as our main example, but the method generalizes to all multiplicities. The integrand is completely fixed by dual conformal symmetry, maximal cuts, constraints from soft-collinear behavior, and from the vanishing of odd-multiplicity amplitudes. After performing integrations with Higgs regularizations, the integrated results demonstrate that the infrared divergence is again identical to that of ${\cal N}=4$ super Yang-Mills. After subtracting divergences, the finite part is dual conformal invariant, and respects various symmetries; it has uniform transcendentality weight two and exhibits a nice analytic structure.}

\begin{document} 
\maketitle
\flushbottom

\section{Introduction}

\label{sec:intro}
For the past decade, there has been profound progress in the understanding of the perturbative S-matrix of $\mathcal{N}=4$ super-Yang-Mills (sYM). Beginning with the all-loop result of the four- and five-point amplitude, known as the Bern-Dixon-Smirnov (BDS)~\cite{Bern:2005iz} ansatz, combined with its strong coupling string theory dual~\cite{Alday:2007hr}, represents the complete answer for $n\leq 5$-pt $\mathcal{N}=4$ super-Yang-Mills. This result can be understood from the hidden dual conformal symmetry in the planar limit~\cite{Drummond:2007aua} (see~\cite{Berkovits:2008ic} for its string theory origin), which has its origin in the duality between the amplitude and null polygonal Wilson Loops~\cite{Alday:2007hr,Alday:2007he,Alday:2009yn,Brandhuber:2007yx, Drummond:2007aua, Drummond:2007cf,
Drummond:2007bm, Drummond:2008aq,Bern:2008ap}. The symmetry fixes the result up to functions of invariant cross ratios which are absent for $n\leq5$~\cite{Drummond:2007au}. Starting at $n=6$, one finds deviation from BDS ansatz due to finite functions of conformal cross-ratios (remainder functions)~\cite{Alday:2007he}. However, by inclusion of the fermionic part of the dual superconformal symmetry, one obtains differential equations that iteratively determine the finite part of the amplitude
~\cite{Caron-Huot:2011dec}. This, combined with \textit{symbol} technology~\cite{Goncharov:2010jf}, initiated the bootstrap program, which culminated to the state-of-the-art $n=6$ MHV amplitude to seven-loops (NMHV to six loops) and $n=7$ to four loops~\cite{Dixon:2011pw, Dixon:2011nj, Dixon:2013eka, Dixon:2014voa, Dixon:2014iba, Drummond:2014ffa, Dixon:2015iva, Caron-Huot:2016owq, Dixon:2016nkn, Drummond:2018caf, Caron-Huot:2019vjl, Caron-Huot:2019bsq,Dixon:2020cnr} (see ~\cite{Caron-Huot:2020bkp} for a recent review). Starting from the seminal work of~\cite{CaronHuot:2011ky}, the frontier for higher multiplicities has been pushed to two-loop $n=8,9$ MHV~\cite{Golden:2021ggj}, (the symbol of) NMHV and N${}^2$MHV~\cite{Zhang:2019vnm,He:2020vob}, as well as that of three-loop $n=8$ MHV~\cite{Li:2021bwg}. 

Remarkably the dual superconformal symmetry that sits at the heart of the progress in $\mathcal{N}=4$ sYM, was found to also emerge in the planar limit of a three-dimensional theory~\cite{Bargheer:2010hn, Huang:2010qy, Gang:2010gy}.  This is the  $\mathcal{N}=6$ three-dimensional Chern-Simons matter theory~\cite{Hosomichi:2008jb, Aharony:2008ug} commonly referred to as ABJM. It was speculated that since the same type of symmetry is present, based on a different supergroup OSp(6|4) instead of SU(4|4) for $\mathcal{N}=4$ sYM, the scattering amplitudes of the two theory must share a similar structure. Indeed, on the one hand, the BDS ansatz of $\mathcal{N}=4$ sYM appears to completely capture the four-point amplitude of ABJM theory~\cite{Chen:2011vv}, on the other hand, topological features unique to the kinematics of massless scattering in three-dimensions reveal itself as novel non-analytic behavior of the amplitude at one-loop to all (even) multiplicity~\cite{Bianchi:2012cq, Bargheer:2012cp, Brandhuber:2012un, Brandhuber:2012wy}. These two features are fully fleshed out in the six-point two-loop amplitude~\cite{Caron-Huot:2012sos}, which is the current frontier. 

In this paper, we would like to extend this frontier to two-loop eight (and higher) points. There are multiple motivations. The first, of course, is the verification that the BDS ansatz continues to capture the infrared (IR) divergence of this theory. Second, the eight-point amplitude provides valuable data for perturbative calculations as well as a non-perturbative flux-tube program~\cite{Basso:2018tif}: while the pentagon OPE~\cite{Basso:2013vsa} has proved to be extremely successful for bootstrapping and non-perturbative computations of sYM amplitudes, explicit higher-point data is needed to push such a program for ABJM amplitudes. By continuing to the eight-point, our results will also provide us a window into the potential patterns of the symbol alphabet of the theory, which may control the analytic property to higher loops and provide the starting point for a bootstrap program. Finally, we would like to see how the non-analytic pieces (in terms of sign functions), are extrapolated from the six- to the eight-points. This may allow us to make a connection with anyon effects that was proposed for the Chern-Simons matter theory with fundamental matter~\cite{Jain:2014nza}. 

Our strategy for computing two-loop amplitudes can already be illustrated with a warm-up exercise at one-loop, which gives new integrands for $n\geq 8$. Already for one-loop, we begin with a set of the dual conformally invariant basis of the triangle and tensor box integrals,  and fix their coefficients using maximal cuts and constraints from soft-collinear behavior; the integration is trivial, which gives well-known sign functions. For two loops, the integral basis consists of kissing-triangle, double-triangle, box-triangle, and double-box topologies. It's highly non-trivial to find the correct set of numerators for these topologies.  Very nicely, as we show explicitly for eight-points, matching soft cuts, maximal cuts, as well as vanishing collinear-soft limits, and three-point cuts will be sufficient to fix the integrand completely. The integration is then performed using Higgs regularization: we find to our satisfaction that IR divergences are again given by eight-point BDS ansatz. Very nicely, we identify a subset of integral that directly gives the BDS ansatz, which we conjecture to generalize to all multiplicities. After subtracting divergences, the finite part of the amplitude is not only manifestly dual conformally invariant, and passes various stringent consistency checks such as little-group parity and reflection symmetry. It is given by only three kinds of uniform weight-two functions, which is the result of highly nontrivial cancellations among higher weight contributions (and pieces beyond multiple polylogs) from individual integrals. These functions have very simple symbols which satisfy physical-discontinuity conditions, and the alphabet consists of letters that are simple polynomials of cross-ratios as well as phases. The latter is unique to ABJM theory. They are dressed with non-analytic sign functions, and all these interesting structures nicely generalize those in the six-point amplitude~\cite{Caron-Huot:2012sos}.

The paper is organized as follows. In section~\ref{sec:preliminary}, we review the basics of scattering amplitudes in ABJM theory needed in this paper, including various symmetries they satisfy, tree amplitudes, and leading singularities for maximal cuts up to two loops. In section~\ref{sec:one-loop}, we compute one-loop integrands and amplitudes from integral basis (dressed with maximal cuts), and in particular, give explicit results for eight-point amplitudes. In section~\ref{sec:two-loop}, we move to the construction of two-loop eight-point integrands, using soft cuts, maximal cuts, vanishing collinear-soft limits, and vanishing three-point cuts; we show that the cancellation of unphysical cuts ensures that elliptic pieces, which appear in individual (double-box) integrals, all cancel in the final answer. In section~\ref{Section5}, we integrate all the integrals in the eight-point amplitude with Higgs regulators. The final amplitude satisfies all consistency checks, with IR divergences captured by BDS ansatz, and we comment on the analytical structure, including all the symbol letters, in section~\ref{Section6}. We end with conclusions and outlook and collect more results in the appendices. 

\section{Preliminaries} \label{sec:preliminary}

We begin with a lightning review of the relevant ingredients. We will be interested in the ordered amplitude of ABJM theory, denoted as $A_n(\bar{1}2\bar{3}\cdots n)$, where the external legs alternate between two on-shell super-multiplets, denoted as $(\Phi^I,\bar{\Psi}_I)$. Here $I=1,2,3$ are SU(3) indices, the linearly realized subgroup of SO(6) R-symmetry. The on-shell mutltiplets transform as bi-fundamental representation $(N, \bar{N})$ and $(\bar{N}, N)$ of $SU(N)\times SU(N)$ gauge group. Thus only an even number of legs can form a color singlet, and the amplitude is non-vanishing.

Since $(\Phi^I,\bar{\Psi}_I)$ is a bosonic and a fermionic multiplet, respectively, the super-amplitude is cyclic by two sites invariants up to a sign:
\begin{equation}
    A_n(\bar{1}2\bar{3}\cdots n)=(-)^{n/2-1}\,A_n(\bar{3}4\bar{5}\cdots 2)\,.
\end{equation}
Furthermore, under $Z_2$ little group transformation of individual on-shell variables $\Lambda_{i,a}=(\lambda_{i,\alpha}, \eta_{i,I})$ here $p_i^{\alpha \beta}=(-1)^i \lambda_i^{\alpha} \lambda_i^{\beta}$,  $\Lambda_{i,a}\rightarrow -\Lambda_{i,a}$, the amplitude will attain a minus sign for odd legs. Finally, due to the reflection invariance of the fundamental vertices, the amplitude enjoys the following reflection symmetry~\cite{Bargheer:2012cp},
\begin{equation}\label{ReflectionRule}
    A_n(\bar{1}2\bar{3}\cdots n)=(-)^{n(n-2)/8{+}\ell}\,A_n(\bar{1}n\cdots 4\bar{3}2)\,,
\end{equation}
where $\ell$ denotes the number of loops. 

Due to the dual-conformal covariance of the planar theory, it is often useful to express part of the amplitude in terms of dual variables $x_i$, defined through $p_i=x_{i{+}1}{-}x_{i}=x_{i,i+1}$. Dual conformal invariance is then manifest by embedding $x_i$ in embedding space, i.e., a projective plane in 5 dimensions $y_i=(x_i,1,x_i^2)$, and 
\begin{equation}
    (i\cdot j):=y_i\cdot y_{j}=(x_i-x_j)^2=x_{i,j}^2\,. 
\end{equation}

The $OSp(6|4)$ Yangian symmetry~\cite{Bargheer:2010hn} of planar ABJM theory suggests that the amplitude can be written in terms of Yangian invariants. While infrared divergences render part of the symmetry anomalous, it must be proportional to the tree amplitude and hence the sum of Yangian invariants. Such invariants are nicely captured by the residues of the integral over the orthogonal Grassmannian~\cite{Lee:2010du, Gang:2010gy}:
\begin{equation}\label{eq:grass}
\int \frac{d^{2k^2}C^{\alpha i}}{{\rm GL}(k)}\frac{1}{M_1M_2\cdots M_k}\delta^{\frac{k(k{+}1)}{2}}\left(CC^T\right)\delta^{2k|3k}(C^\alpha\cdot\Lambda)
\end{equation}
where $C^{\alpha i}$ are matrix elements of an $k\times 2k$ matrix, and $M_i$ are the $k\times k$ consecutive minors of $C$ beginning with column $i$, i.e. $M_i\equiv {\rm det}(i,i{+}1,i{+}2,\cdots,i{+}k{-}1)$. The integral is $k^2$-dimensional subject to $\frac{k(k{+}1)}{2}{+}2k{-}3$ delta functions constraints, where the ${-}3$ is due to momentum conservation. Thus the remaining dimension is $\frac{(k{-}3)(k{-}2)}{2}$, to be localized on the minors. 

For eight points ($k=4$) we have a one-dimensional integral to be localized by the vanishing of one of the minors. Note that since the orthogonal constraint is quadratic in $C$, the solution is split into a positive and negative branch. Furthermore, as the minors are quadratic functions in integration variables, the solutions come in pairs. Thus we will label the residues for $M_i=0$, or the leading singularities, as ${\rm LS}_{\pm,1,2}[i]$ where $\pm$ labels the branch and $1,2$ the solutions. The explicit form of ${\rm LS}_{\pm,1,2}[i]$ is given in appendix~\ref{LSApp}.

Importantly, the on-shell data (unitarity cuts) necessary to determine the multi-loop amplitude can be expressed in terms of linear combinations of these leading singularities. Beginning with the eight-point tree amplitude, which was identified as the sum over residues of minors $M_2$ and $M_4$ in~\cite{Gang:2010gy}, amounts to
\begin{eqnarray}\label{TreeLS}
A^{\mathrm{tree}}_8&=&\sum_{a=1,2}{\rm LS}_{+,a}[2]+{\rm LS}_{-,a}[2]+{\rm LS}_{+,a}[4]+{\rm LS}_{-,a}[4]\,\nonumber\\
&=&-\left(\sum_{a=1,2}{\rm LS}_{+,a}[1]+{\rm LS}_{-,a}[1]+{\rm LS}_{+,a}[3]+{\rm LS}_{-,a}[3]\right)\,.
\end{eqnarray} 
Note that the eight-point super-amplitude should pick up a minus sign under a cyclic shift by two. This is manifested from the denominator of the Grassmannian integral, where $M_1M_2M_3M_4\rightarrow M_3M_4M_5M_6$, and using that $M_5M_6=-M_1M_2$ due to orthogonal conditions~\cite{Lee:2010du}. This can also be seen from properties of the leading singularities listed in eq.(\ref{ShiftSym}). At eight points, the tree level amplitude is even under reflection symmetry eq.(\ref{ReflectionRule}), which also can be read off from the behavior of leading singularities under reflection in eq.(\ref{ReflectSym}).

Similarly, we can identify the one-loop maximal cut with linear combinations of leading singularities. At the eight-point, we will be interested in triangle cuts where the loop region satisfies $(a\cdot i)=(a\cdot i{+}2)=(a\cdot i{+}4)=0$ (for $i=1,2, \cdots, 8$). The cut is then given by the product of a 6-point and two 4-point  amplitudes
$$
\vcenter{\hbox{\scalebox{1}{
\begin{tikzpicture}[x=0.75pt,y=0.75pt,yscale=-1,xscale=1]

\draw [color={rgb, 255:red, 0; green, 0; blue, 0 }  ,draw opacity=1 ][line width=0.75]    (200.79,150.38) -- (183.04,132.62) ;
\draw [color={rgb, 255:red, 0; green, 0; blue, 0 }  ,draw opacity=1 ][line width=0.75]    (200.79,150.38) -- (218.25,131.75) ;
\draw [color={rgb, 255:red, 0; green, 0; blue, 0 }  ,draw opacity=1 ][line width=0.75]    (259.45,200.25) -- (234.16,200.38) ;
\draw [color={rgb, 255:red, 0; green, 0; blue, 0 }  ,draw opacity=1 ][line width=0.75]    (167.42,200.38) -- (145.38,205.92) ;
\draw  [color={rgb, 255:red, 0; green, 0; blue, 0 }  ,draw opacity=1 ][fill={rgb, 255:red, 0; green, 0; blue, 0 }  ,fill opacity=1 ] (236.3,189.06) .. controls (242.86,189.06) and (248.18,194.38) .. (248.18,200.94) .. controls (248.18,207.5) and (242.86,212.81) .. (236.3,212.81) .. controls (229.75,212.81) and (224.43,207.5) .. (224.43,200.94) .. controls (224.43,194.38) and (229.75,189.06) .. (236.3,189.06) -- cycle ;
\draw  [color={rgb, 255:red, 0; green, 0; blue, 0 }  ,draw opacity=1 ][fill={rgb, 255:red, 0; green, 0; blue, 0 }  ,fill opacity=1 ] (200.79,140.75) .. controls (206.11,140.75) and (210.42,145.06) .. (210.42,150.38) .. controls (210.42,155.69) and (206.11,160) .. (200.79,160) .. controls (195.48,160) and (191.17,155.69) .. (191.17,150.38) .. controls (191.17,145.06) and (195.48,140.75) .. (200.79,140.75) -- cycle ;
\draw [color={rgb, 255:red, 0; green, 0; blue, 0 }  ,draw opacity=1 ][line width=0.75]    (154.72,219.58) -- (167.42,200.38) ;
\draw [color={rgb, 255:red, 0; green, 0; blue, 0 }  ,draw opacity=1 ][line width=0.75]    (238.45,225.25) -- (234.16,200.38) ;
\draw [color={rgb, 255:red, 0; green, 0; blue, 0 }  ,draw opacity=1 ][line width=0.75]    (250.12,220.25) -- (234.16,200.38) ;
\draw [color={rgb, 255:red, 0; green, 0; blue, 0 }  ,draw opacity=1 ][line width=0.75]    (257.78,210.92) -- (234.16,200.38) ;
\draw   (200.79,150.38) -- (234.16,200.38) -- (167.42,200.38) -- cycle ;
\draw  [color={rgb, 255:red, 0; green, 0; blue, 0 }  ,draw opacity=1 ][fill={rgb, 255:red, 0; green, 0; blue, 0 }  ,fill opacity=1 ] (167.42,190.75) .. controls (172.74,190.75) and (177.05,195.06) .. (177.05,200.38) .. controls (177.05,205.69) and (172.74,210) .. (167.42,210) .. controls (162.11,210) and (157.8,205.69) .. (157.8,200.38) .. controls (157.8,195.06) and (162.11,190.75) .. (167.42,190.75) -- cycle ;
\draw [color={rgb, 255:red, 139; green, 87; blue, 42 }  ,draw opacity=1 ][line width=1.5]    (174.5,169.75) -- (193.5,180.75) ;
\draw [color={rgb, 255:red, 139; green, 87; blue, 42 }  ,draw opacity=1 ][line width=1.5]    (200.5,190.25) -- (200.5,210.75) ;
\draw [color={rgb, 255:red, 139; green, 87; blue, 42 }  ,draw opacity=1 ][line width=1.5]    (210.5,181.25) -- (228.5,169.25) ;

\draw (197.13,215.15) node [anchor=north west][inner sep=0.75pt]   [align=left] {$\displaystyle i$};
\draw (126.13,157.65) node [anchor=north west][inner sep=0.75pt]   [align=left] {$\displaystyle i+2$};
\draw (231.63,160.65) node [anchor=north west][inner sep=0.75pt]   [align=left] {$\displaystyle i+4$};
\draw (268,215.15) node [anchor=north west][inner sep=0.75pt]   [align=left] {$\displaystyle .$};

\end{tikzpicture}

}}}$$

As there are two solutions to the cut condition, we denote them as $\mathcal{C}_{i,i{+}2,i{+}4}^\pm$: 
\begin{equation}\label{1LoopLS1}
	\mathcal{C}_{i,i{+}2,i{+}4}^\pm=\int\prod_{I=1}^{3}d\eta_{\ell_{1}}^{I}d\eta_{l_{2}}^{I}d\eta_{l_{3}}^{I}\mathcal{A}_4^{\mathrm{tree}}\mathcal{A}_4^{\mathrm{tree}}\mathcal{A}_6^{\mathrm{tree}}\bigg|_{\ell_1=\ell^{\pm}_1}
\end{equation}
where the state sum is given by Grassmann-odd integrals, and the $\pm$ is defined through their relation with the leading singularities, 
\begin{equation}\label{1LoopLS2}
\frac{\mathcal{C}^{\pm}_{2,4,8}}{2 \sqrt{(2\cdot 4\cdot 8)}}=\pm \left({\rm LS}_{+,2(1)}[4]{+}{\rm LS}_{-,1(2)}[4]\right),\quad
\frac{\mathcal{C}^\pm_{4,6,8}}{2 \sqrt{(4\cdot 6\cdot 8)}}=\pm \left({\rm LS}_{+,1(2)}[4]{+}{\rm LS}_{-,1(2)}[4]\right), \quad
\end{equation}
where $\sqrt{(i\cdot j\cdot k)}\equiv \sqrt{(i\cdot j)(i\cdot k)(j\cdot k)}$ and $\mathcal{C}^+_{2,4,8}(\mathcal{C}^-_{2,4,8})$ is proportional to ${\rm LS}_{+,1}[4]{+}{\rm LS}_{-,1}[4]$ $({\rm LS}_{+,2}[4]{+}{\rm LS}_{-,2}[4])$ . That these two cuts are identified with the same leading singularities reflects the fact that they can be written in terms of the same on-shell diagram~\cite{Huang:2013owa} 
. The remaining cuts are similarly identified, and we list them for completeness. 
\begin{eqnarray}
&&\frac{\mathcal{C}^{\pm}_{2,6,8}}{2 \sqrt{(2\cdot 6\cdot 8)}}=\pm \left({\rm LS}_{+,2(1)}[2]{+}{\rm LS}_{-,1(2)}[2]\right),\quad
\frac{\mathcal{C}^\pm_{2,4,6}}{2 \sqrt{(2\cdot 4\cdot 6)}}=\pm \left({\rm LS}_{+,1(2)}[2]{+}{\rm LS}_{-,1(2)}[2]\right),\nonumber\\
&&\frac{\mathcal{C}^{\pm}_{1,3,7}}{2 \sqrt{(1\cdot 3\cdot 7)}}=\pm \left({\rm LS}_{+,1(2)}[3]{+}{\rm LS}_{-,2(1)}[3]\right),\quad
\frac{\mathcal{C}^\pm_{3,5,7}}{2 \sqrt{(3\cdot 5\cdot 7)}}=\pm \left({\rm LS}_{+,1(2)}[3]{+}{\rm LS}_{-,1(2)}[3]\right), \nonumber\\
&&\frac{\mathcal{C}^{\pm}_{1,5, 7}}{2 \sqrt{(1\cdot 5\cdot 7)}}=\pm \left({\rm LS}_{+,1(2)}[1]{+}{\rm LS}_{-,2(1)}[1]\right),\quad
\frac{\mathcal{C}^\pm_{1,3,5}}{2 \sqrt{(1\cdot 3\cdot 5)}}=\pm \left({\rm LS}_{+,1(2)}[1]{+}{\rm LS}_{-,1(2)}[1]\right)\,. \nonumber
\end{eqnarray}
We reminder readers that under the cyclic by two sites $\Lambda_i\rightarrow\Lambda_{i-2}$, the one-loop maximal cuts transform as:
\begin{eqnarray}\label{CutShifts}
    &&\frac{\mathcal{C}^{\pm}_{2,4,8}}{ \sqrt{(2\cdot 4\cdot 8)}}\rightarrow-\frac{\mathcal{C}^{\pm}_{2,6,8}}{ \sqrt{(2\cdot 6\cdot 8)}}\rightarrow\frac{\mathcal{C}^{\pm}_{4,6,8}}{ \sqrt{(4\cdot 6\cdot 8)}}\rightarrow-\frac{\mathcal{C}^{\pm}_{2,4,6}}{\sqrt{(2\cdot 4\cdot 6)}},\nonumber\\
    &&\frac{\mathcal{C}^{\pm}_{1,3,7}}{\sqrt{(1\cdot 3\cdot 7)}}\rightarrow-\frac{\mathcal{C}^{\pm}_{1,5,7}}{\sqrt{(1\cdot 5\cdot 7)}}\rightarrow\frac{\mathcal{C}^{\pm}_{3,5,7}}{\sqrt{(3\cdot 5\cdot 7)}}\rightarrow-\frac{\mathcal{C}^{\pm}_{1,3,5}}{\sqrt{(1\cdot 3\cdot 5)}};
\end{eqnarray}
by reflection $\{\Lambda_1,\Lambda_2,\ldots, \Lambda_8 \}\rightarrow\{\Lambda_1,\Lambda_8,\ldots, \Lambda_2 \}$, they have
\begin{eqnarray}\label{OneLoopCutFlip}
    &&\frac{\mathcal{C}^{\pm}_{2,4,8}}{\sqrt{(2\cdot 4\cdot 8)}}\leftrightarrow \frac{\mathcal{C}^{\mp}_{1,3,7}}{\sqrt{(1\cdot 3\cdot 7)}}, \quad \frac{\mathcal{C}^{\pm}_{4,6,8}}{\sqrt{(4\cdot 6\cdot 8)}}\leftrightarrow \frac{\mathcal{C}^{\mp}_{3,5,7}}{\sqrt{(3\cdot 5\cdot 7)}},\nonumber\\
    &&\frac{\mathcal{C}^{\pm}_{2,6,8}}{\sqrt{(2\cdot 6\cdot 8)}}\leftrightarrow \frac{\mathcal{C}^{\mp}_{1,3,5}}{\sqrt{(1\cdot 3\cdot 5)}}, \quad \frac{\mathcal{C}^{\pm}_{2,4,6}}{\sqrt{(2\cdot 4\cdot 6)}}\leftrightarrow \frac{\mathcal{C}^{\mp}_{1,5,7}}{\sqrt{(1\cdot 5\cdot 7)}};
\end{eqnarray}
and for parity $\Lambda_i\rightarrow -\Lambda_i$:
\begin{equation}\label{eq:parity}
    \begin{aligned}
    &\mathcal{C}^{+}_{2,4,8}\leftrightarrow(-)^{F_i{+}1}\mathcal{C}^{-}_{2,4,8}\, \text{for $i=8,1,2,3$}, \quad \mathcal{C}^{+}_{4,6,8}\leftrightarrow(-)^{F_i{+}1}\mathcal{C}^{-}_{4,6,8}\, \text{for $i=4,5,6,7$},\\ 
        &\mathcal{C}^{+}_{2,6,8}\leftrightarrow(-)^{F_i{+}1}\mathcal{C}^{-}_{2,6,8}\, \text{for $i=1,6,7,8$},
         \quad \mathcal{C}^{+}_{2,4,6}\leftrightarrow(-)^{F_i{+}1}\mathcal{C}^{-}_{2,4,6}\, \text{for $i=2,3,4,5$},\\
        &\mathcal{C}^{+}_{1,3,5}\leftrightarrow(-)^{F_i{+}1}\mathcal{C}^{-}_{1,3,5}\, \text{for $i=1,2,3,4$}, \quad \mathcal{C}^{+}_{1,5,7}\leftrightarrow(-)^{F_i{+}1}\mathcal{C}^{-}_{1,5,7}\, \text{for $i=5,6,7,8$},\\ 
        &\mathcal{C}^{+}_{1,3,7}\leftrightarrow(-)^{F_i{+}1}\mathcal{C}^{-}_{1,3,7}\, \text{for $i=1,2,7,8$},
         \quad \mathcal{C}^{+}_{3,5,7}\leftrightarrow(-)^{F_i{+}1}\mathcal{C}^{-}_{3,5,7}\, \text{for $i=3,4,5,6$},
    \end{aligned}
\end{equation}
where $F_i$ is the fermion number of leg $i$, and one-loop cuts remain unchanged under the parity transformation in other cases. We see that for the legs displayed, the cuts pick up additional $Z_2$ weights. These excess weights should be canceled against the functions that dress them. For the remaining legs the $Z_2$ weight is canonical: 
\begin{equation}
    \mathcal{C}_{i,j,k}^\pm\rightarrow (-)^{F_i} \mathcal{C}_{i,j,k}^\pm\,. 
\end{equation}

Finally, we consider the two-loop maximal cuts. At eight points, there are two types of topology for maximal cuts, kissing triangle and box-triangle, which we denote as 
\begin{equation}
\mathcal{C}^i_{\includegraphics[scale=0.12]{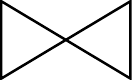} }:\quad  \vcenter{\hbox{\scalebox{0.8}{
\begin{tikzpicture}[x=0.75pt,y=0.75pt,yscale=-1,xscale=1]

\draw  [line width=1.2]  (201,299.5) -- (251,299.5) -- (251,349.5) -- (201,349.5) -- cycle ;
\draw  [line width=1.2]  (251,299.5) -- (301,299.5) -- (301,349.5) -- (251,349.5) -- cycle ;
\draw [color={rgb, 255:red, 0; green, 0; blue, 0 }  ,draw opacity=1 ][line width=1.2]    (301,299.5) -- (301,286.54) -- (301,279.25) ;
\draw [color={rgb, 255:red, 0; green, 0; blue, 0 }  ,draw opacity=1 ][line width=1.2]    (301,299.5) -- (313.96,299.5) -- (321.25,299.5) ;
\draw [color={rgb, 255:red, 0; green, 0; blue, 0 }  ,draw opacity=1 ][line width=1.2]    (301,349.5) -- (313.96,349.5) -- (321.25,349.5) ;
\draw [color={rgb, 255:red, 0; green, 0; blue, 0 }  ,draw opacity=1 ][line width=1.2]    (301,369.75) -- (301,356.79) -- (301,349.5) ;
\draw [color={rgb, 255:red, 0; green, 0; blue, 0 }  ,draw opacity=1 ][line width=1.2]    (180.75,349.5) -- (193.71,349.5) -- (201,349.5) ;
\draw [color={rgb, 255:red, 0; green, 0; blue, 0 }  ,draw opacity=1 ][line width=1.2]    (201,369.75) -- (201,356.79) -- (201,349.5) ;
\draw [color={rgb, 255:red, 0; green, 0; blue, 0 }  ,draw opacity=1 ][line width=1.2]    (201,299.5) -- (201,286.54) -- (201,279.25) ;
\draw [color={rgb, 255:red, 0; green, 0; blue, 0 }  ,draw opacity=1 ][line width=1.2]    (180.75,299.5) -- (193.71,299.5) -- (201,299.5) ;
\draw  [color={rgb, 255:red, 155; green, 155; blue, 155 }  ,draw opacity=1 ][fill={rgb, 255:red, 155; green, 155; blue, 155 }  ,fill opacity=1 ] (192.75,299.5) .. controls (192.75,294.94) and (196.44,291.25) .. (201,291.25) .. controls (205.56,291.25) and (209.25,294.94) .. (209.25,299.5) .. controls (209.25,304.06) and (205.56,307.75) .. (201,307.75) .. controls (196.44,307.75) and (192.75,304.06) .. (192.75,299.5) -- cycle ;
\draw  [color={rgb, 255:red, 155; green, 155; blue, 155 }  ,draw opacity=1 ][fill={rgb, 255:red, 155; green, 155; blue, 155 }  ,fill opacity=1 ] (192.75,349.5) .. controls (192.75,344.94) and (196.44,341.25) .. (201,341.25) .. controls (205.56,341.25) and (209.25,344.94) .. (209.25,349.5) .. controls (209.25,354.06) and (205.56,357.75) .. (201,357.75) .. controls (196.44,357.75) and (192.75,354.06) .. (192.75,349.5) -- cycle ;
\draw  [color={rgb, 255:red, 155; green, 155; blue, 155 }  ,draw opacity=1 ][fill={rgb, 255:red, 155; green, 155; blue, 155 }  ,fill opacity=1 ] (292.75,299.5) .. controls (292.75,294.94) and (296.44,291.25) .. (301,291.25) .. controls (305.56,291.25) and (309.25,294.94) .. (309.25,299.5) .. controls (309.25,304.06) and (305.56,307.75) .. (301,307.75) .. controls (296.44,307.75) and (292.75,304.06) .. (292.75,299.5) -- cycle ;
\draw  [color={rgb, 255:red, 155; green, 155; blue, 155 }  ,draw opacity=1 ][fill={rgb, 255:red, 155; green, 155; blue, 155 }  ,fill opacity=1 ] (292.75,349.5) .. controls (292.75,344.94) and (296.44,341.25) .. (301,341.25) .. controls (305.56,341.25) and (309.25,344.94) .. (309.25,349.5) .. controls (309.25,354.06) and (305.56,357.75) .. (301,357.75) .. controls (296.44,357.75) and (292.75,354.06) .. (292.75,349.5) -- cycle ;
\draw  [color={rgb, 255:red, 155; green, 155; blue, 155 }  ,draw opacity=1 ][fill={rgb, 255:red, 155; green, 155; blue, 155 }  ,fill opacity=1 ] (250.42,295.17) .. controls (257.46,295.17) and (263.17,308.37) .. (263.17,324.67) .. controls (263.17,340.96) and (257.46,354.17) .. (250.42,354.17) .. controls (243.38,354.17) and (237.67,340.96) .. (237.67,324.67) .. controls (237.67,308.37) and (243.38,295.17) .. (250.42,295.17) -- cycle ;

\draw (246.7,360.9) node [anchor=north west][inner sep=0.75pt]   [align=left] {$\displaystyle i$};
\draw (166,317.5) node [anchor=north west][inner sep=0.75pt]   [align=left] {$\displaystyle i+2$};
\draw (235.7,270.4) node [anchor=north west][inner sep=0.75pt]   [align=left] {$\displaystyle i+4$};

\end{tikzpicture}

}
}},\quad \mathcal{C}^i_{\includegraphics[scale=0.13]{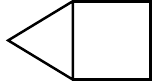} }:\quad \vcenter{\hbox{\scalebox{0.8}{
\begin{tikzpicture}[x=0.75pt,y=0.75pt,yscale=-1,xscale=1]

\draw  [line width=1.2]  (251,299.5) -- (301,299.5) -- (301,349.5) -- (251,349.5) -- cycle ;
\draw [color={rgb, 255:red, 0; green, 0; blue, 0 }  ,draw opacity=1 ][line width=1.2]    (209.5,324.33) -- (191.75,342.08) ;
\draw [color={rgb, 255:red, 0; green, 0; blue, 0 }  ,draw opacity=1 ][line width=1.2]    (209.5,324.33) -- (205.38,320.21) -- (191.75,306.58) ;
\draw [color={rgb, 255:red, 0; green, 0; blue, 0 }  ,draw opacity=1 ][line width=1.2]    (301,299.5) -- (301,286.54) -- (301,279.25) ;
\draw [color={rgb, 255:red, 0; green, 0; blue, 0 }  ,draw opacity=1 ][line width=1.2]    (301,299.5) -- (313.96,299.5) -- (321.25,299.5) ;
\draw [color={rgb, 255:red, 0; green, 0; blue, 0 }  ,draw opacity=1 ][line width=1.2]    (301,349.5) -- (313.96,349.5) -- (321.25,349.5) ;
\draw [color={rgb, 255:red, 0; green, 0; blue, 0 }  ,draw opacity=1 ][line width=1.2]    (301,369.75) -- (301,356.79) -- (301,349.5) ;
\draw [color={rgb, 255:red, 0; green, 0; blue, 0 }  ,draw opacity=1 ][line width=1.2]    (251,279.25) -- (251,292.21) -- (251,299.5) ;
\draw [color={rgb, 255:red, 0; green, 0; blue, 0 }  ,draw opacity=1 ][line width=1.2]    (251,349.5) -- (251,362.46) -- (251,369.75) ;
\draw  [line width=1.2]  (209.5,324.33) -- (251,299.5) -- (251,349.17) -- cycle ;
\draw  [color={rgb, 255:red, 155; green, 155; blue, 155 }  ,draw opacity=1 ][fill={rgb, 255:red, 155; green, 155; blue, 155 }  ,fill opacity=1 ] (201.25,324.33) .. controls (201.25,319.78) and (204.94,316.08) .. (209.5,316.08) .. controls (214.06,316.08) and (217.75,319.78) .. (217.75,324.33) .. controls (217.75,328.89) and (214.06,332.58) .. (209.5,332.58) .. controls (204.94,332.58) and (201.25,328.89) .. (201.25,324.33) -- cycle ;
\draw  [color={rgb, 255:red, 155; green, 155; blue, 155 }  ,draw opacity=1 ][fill={rgb, 255:red, 155; green, 155; blue, 155 }  ,fill opacity=1 ] (242.75,300.46) .. controls (242.75,295.91) and (246.44,292.21) .. (251,292.21) .. controls (255.56,292.21) and (259.25,295.91) .. (259.25,300.46) .. controls (259.25,305.02) and (255.56,308.71) .. (251,308.71) .. controls (246.44,308.71) and (242.75,305.02) .. (242.75,300.46) -- cycle ;
\draw  [color={rgb, 255:red, 155; green, 155; blue, 155 }  ,draw opacity=1 ][fill={rgb, 255:red, 155; green, 155; blue, 155 }  ,fill opacity=1 ] (292.75,299.5) .. controls (292.75,294.94) and (296.44,291.25) .. (301,291.25) .. controls (305.56,291.25) and (309.25,294.94) .. (309.25,299.5) .. controls (309.25,304.06) and (305.56,307.75) .. (301,307.75) .. controls (296.44,307.75) and (292.75,304.06) .. (292.75,299.5) -- cycle ;
\draw  [color={rgb, 255:red, 155; green, 155; blue, 155 }  ,draw opacity=1 ][fill={rgb, 255:red, 155; green, 155; blue, 155 }  ,fill opacity=1 ] (242.75,349.5) .. controls (242.75,344.94) and (246.44,341.25) .. (251,341.25) .. controls (255.56,341.25) and (259.25,344.94) .. (259.25,349.5) .. controls (259.25,354.06) and (255.56,357.75) .. (251,357.75) .. controls (246.44,357.75) and (242.75,354.06) .. (242.75,349.5) -- cycle ;
\draw  [color={rgb, 255:red, 155; green, 155; blue, 155 }  ,draw opacity=1 ][fill={rgb, 255:red, 155; green, 155; blue, 155 }  ,fill opacity=1 ] (292.75,349.5) .. controls (292.75,344.94) and (296.44,341.25) .. (301,341.25) .. controls (305.56,341.25) and (309.25,344.94) .. (309.25,349.5) .. controls (309.25,354.06) and (305.56,357.75) .. (301,357.75) .. controls (296.44,357.75) and (292.75,354.06) .. (292.75,349.5) -- cycle ;

\draw (175.5,316.5) node [anchor=north west][inner sep=0.75pt]   [align=left] {$\displaystyle i$};

\end{tikzpicture}
}
}}\,.
\end{equation}
Each of them is given by a product of five 4-point tree amplitudes, summed over the internal states. For example, with a specific choice of $i$ they are given by:
  \begin{equation}
 	\begin{aligned}
 		\mathcal{C}^8_{\includegraphics[scale=0.12]{Cut1Sym.pdf} }[a_\pm, b_\pm]=\int \prod_{i=1}^6 \prod_{I=1}^3  d\eta_{\ell_i}^I  &\mathcal{A}_4(\bar{1},\ell_2,-\bar{\ell_1},8) \mathcal{A}_4(\bar{2},3,\bar{\ell_3},-\ell_2)  \mathcal{A}_4(\bar{\ell_1},-\ell_3,\bar{\ell_5},-\ell_4)\\
 		&\times \mathcal{A}_4(-\bar{\ell_5},4,\bar{5},\ell_6)  \mathcal{A}_4(-\bar{\ell_6},6,\bar{7},\ell_4) \\
 \mathcal{C}^2_{\includegraphics[scale=0.13]{Cut2Sym.pdf} }[a_\pm, b_\pm]=\int \prod_{i=1}^6 \prod_{I=1}^3  d\eta_{\ell'_i}^I  &\mathcal{A}_4(\bar{1},2,\bar{\ell^\prime_2},-\ell^\prime_1) \mathcal{A}_4(\bar{3},-\ell^\prime_4,\bar{\ell^\prime_3},-\ell^\prime_2)  \mathcal{A}_4(\bar{\ell^\prime_1},-\ell^\prime_3,\bar{\ell^\prime_5},8)\\
 		&\times \mathcal{A}_4(\bar{\ell^\prime_4},4,\bar{\ell^\prime_5},-\ell^\prime_6)  \mathcal{A}_4(\bar{\ell^\prime_6},6,\bar{7},-\ell^\prime_5)\,. 
 	\end{aligned}
\end{equation}
In the above the argument $a_{\pm}$, $b_{\pm}$ simply labels the two solutions for each loop region on the maximal cut, so there are four solutions for a given two-loop maximal cut. Again these cuts can be mapped to the leading singularities. These two topologies are actually ``equivalent'' in a sense they are given by  the same leading singularities. In the language of on-shell diagrams, this equivalence is a reflection of ``triangle-move'':
\begin{figure}[H]   
\begin{center}

$\vcenter{\hbox{\scalebox{0.7}{
\begin{tikzpicture}[x=0.75pt,y=0.75pt,yscale=-1,xscale=1]

\draw  [line width=1.5]  (250.25,209) -- (309.65,279.8) -- (190.85,279.8) -- cycle ;
\draw [line width=1.5]    (162.6,109.95) -- (190.85,138.2) ;
\draw [line width=1.5]    (163.27,165.78) -- (190.85,138.2) ;
\draw [line width=1.5]    (337.9,166.45) -- (309.65,138.2) ;
\draw [line width=1.5]    (338.25,110.07) -- (310.67,137.65) ;
\draw [line width=1.5]    (162.6,251.55) -- (190.85,279.8) ;
\draw [line width=1.5]    (163.27,307.38) -- (190.85,279.8) ;
\draw [line width=1.5]    (339.05,308.17) -- (310.8,279.92) ;
\draw [line width=1.5]    (338.38,252.33) -- (310.8,279.92) ;

\draw  [line width=1.5]  (518.75,198.41) -- (633.19,198.41) -- (633.19,286.5) -- (518.75,286.5) -- cycle ;
\draw [line width=1.5]    (661.44,314.75) -- (633.19,286.5) ;
\draw [line width=1.5]    (660.78,258.92) -- (633.19,286.5) ;
\draw [line width=1.5]    (490.5,258.25) -- (518.75,286.5) ;
\draw [line width=1.5]    (491.17,314.08) -- (518.75,286.5) ;
\draw [line width=1.5]    (490.5,170.16) -- (518.75,198.41) ;
\draw [line width=1.5]    (661.08,170.82) -- (633.5,198.41) ;
\draw [line width=1.5]    (604.53,99.36) -- (576.28,127.61) ;
\draw [line width=1.5]    (548.7,100.02) -- (576.28,127.61) ;
\draw  [color={rgb, 255:red, 144; green, 19; blue, 254 }  ,draw opacity=1 ][line width=1.5]  (250.25,209) -- (190.85,138.2) -- (309.65,138.2) -- cycle ;
\draw  [color={rgb, 255:red, 144; green, 19; blue, 254 }  ,draw opacity=1 ][line width=1.5]  (576.28,127.61) -- (633.5,198.41) -- (519.06,198.41) -- cycle ;

\draw (144.53,97) node [anchor=north west][inner sep=0.75pt]  [font=\large] [align=left] {$\displaystyle \overline{1}$};
\draw (346.6,158.4) node [anchor=north west][inner sep=0.75pt]  [font=\large] [align=left] {$\displaystyle \overline{3}$};
\draw (344.8,296.73) node [anchor=north west][inner sep=0.75pt]  [font=\large] [align=left] {$\displaystyle \overline{5}$};
\draw (145.07,239.53) node [anchor=north west][inner sep=0.75pt]  [font=\large] [align=left] {$\displaystyle \overline{7}$};
\draw (144.93,157.53) node [anchor=north west][inner sep=0.75pt]  [font=\large] [align=left] {$\displaystyle 8$};
\draw (346,97.67) node [anchor=north west][inner sep=0.75pt]  [font=\large] [align=left] {$\displaystyle 2$};
\draw (346.27,241.13) node [anchor=north west][inner sep=0.75pt]  [font=\large] [align=left] {$\displaystyle 4$};
\draw (145.6,299.13) node [anchor=north west][inner sep=0.75pt]  [font=\large] [align=left] {$\displaystyle 6$};
\draw (390,201.67) node [anchor=north west][inner sep=0.75pt]  [font=\huge] [align=left] {$\displaystyle =$};
\draw (673,302.33) node [anchor=north west][inner sep=0.75pt]  [font=\large] [align=left] {$\displaystyle \overline{5}$};
\draw (673.17,249.33) node [anchor=north west][inner sep=0.75pt]  [font=\large] [align=left] {$\displaystyle 4$};
\draw (466.67,245.83) node [anchor=north west][inner sep=0.75pt]  [font=\large] [align=left] {$\displaystyle \overline{7}$};
\draw (470,303.33) node [anchor=north west][inner sep=0.75pt]  [font=\large] [align=left] {$\displaystyle 6$};
\draw (473.83,155.83) node [anchor=north west][inner sep=0.75pt]  [font=\large] [align=left] {$\displaystyle 8$};
\draw (670,152.5) node [anchor=north west][inner sep=0.75pt]  [font=\large] [align=left] {$\displaystyle \overline{3}$};
\draw (533.83,86) node [anchor=north west][inner sep=0.75pt]  [font=\large] [align=left] {$\displaystyle \overline{1}$};
\draw (615,87.17) node [anchor=north west][inner sep=0.75pt]  [font=\large] [align=left] {$\displaystyle 2$};
\draw (243.33,109.83) node [anchor=north west][inner sep=0.75pt]  [font=\large] [align=left] {$\displaystyle \ell _{2}$};
\draw (197.13,176.23) node [anchor=north west][inner sep=0.75pt]  [font=\large] [align=left] {$\displaystyle \ell _{1}$};
\draw (289.43,175.53) node [anchor=north west][inner sep=0.75pt]  [font=\large] [align=left] {$\displaystyle \ell _{3}$};
\draw (192.43,225.63) node [anchor=north west][inner sep=0.75pt]  [font=\large] [align=left] {$\displaystyle \ell _{4}$};
\draw (287.23,225.43) node [anchor=north west][inner sep=0.75pt]  [font=\large] [align=left] {$\displaystyle \ell _{5}$};
\draw (241.33,287.23) node [anchor=north west][inner sep=0.75pt]  [font=\large] [align=left] {$\displaystyle \ell _{6}$};
\draw (524.33,136.83) node [anchor=north west][inner sep=0.75pt]  [font=\large] [align=left] {$\displaystyle \ell _{1}^{\prime }$};
\draw (618.83,137.33) node [anchor=north west][inner sep=0.75pt]  [font=\large] [align=left] {$\displaystyle \ell _{2}^{\prime }$};
\draw (566.83,203.33) node [anchor=north west][inner sep=0.75pt]  [font=\large] [align=left] {$\displaystyle \ell _{3}^{\prime }$};
\draw (645.83,221.33) node [anchor=north west][inner sep=0.75pt]  [font=\large] [align=left] {$\displaystyle \ell _{4}^{\prime }$};
\draw (491.83,222.33) node [anchor=north west][inner sep=0.75pt]  [font=\large] [align=left] {$\displaystyle \ell _{6}^{\prime }$};
\draw (567.33,290.83) node [anchor=north west][inner sep=0.75pt]  [font=\large] [align=left] {$\displaystyle \ell _{5}^{\prime }$};
\draw (700,290) node [anchor=north west][inner sep=0.75pt]   [align=left] {$\displaystyle .$};

\end{tikzpicture}

}}}$

\end{center}
\end{figure}

\noindent In particular, we have:
\begin{equation}
 	\begin{aligned}
 		&{\rm LS}_{\pm,1}[4]=\left.\frac{\mathcal{C}^4_{\includegraphics[scale=0.12]{Cut1Sym.pdf}} [a_+,b_\pm]}{16\, \text{det}(\ell_1,\ell_2,\ell_3)\text{det}(\ell_4,\ell_5,\ell_6)}\right|_{a,b=a_+,b_\pm}=\pm\frac{\mathcal{C}^4_{\includegraphics[scale=0.12]{Cut1Sym.pdf} } [a_+,b_\pm]}{4\sqrt{(8\cdot2\cdot4)}\sqrt{(4\cdot6\cdot8)}}\\
 		&{\rm LS}_{\pm,2}[4]=\left.\frac{\mathcal{C}^4_{\includegraphics[scale=0.12]{Cut1Sym.pdf}} [a_-,b_\mp]}{16\, \text{det}(\ell_1,\ell_2,\ell_3)\text{det}(\ell_4,\ell_5,\ell_6)}\right|_{a,b=a_-,b_\mp}=\pm\frac{\mathcal{C}^4_{\includegraphics[scale=0.12]{Cut1Sym.pdf} }[a_-,b_\mp]}{4\sqrt{(8\cdot2\cdot4)}\sqrt{(4\cdot6\cdot8)}}.
		 	\end{aligned}
\end{equation}
 Note that the Jacobian flips sign between the two solutions, $4\left.\text{det}(\ell_1,\ell_2,\ell_3)\right|_{a=a_\pm}=\pm  2\sqrt{(8\cdot 2\cdot 4)}$ and $4 \left.\text{det}(\ell_4,\ell_5,\ell_6)\right|_{b=b_\pm}{=}\pm 2\sqrt{(4\cdot 6\cdot 8)}$. 
 Other leading singularities and cuts correspond to each other in the following manner:
\begin{equation}
 	\begin{aligned}
 		&{\rm LS}_{\pm,1}[1]=\pm\frac{\mathcal{C}^1_{\includegraphics[scale=0.12]{Cut1Sym.pdf} } [a_+,b_\pm]}{4\sqrt{(1\cdot3\cdot5)}\sqrt{(1\cdot5\cdot7)}}, \quad {\rm LS}_{\pm,2}[1]=\pm\frac{\mathcal{C}^1_{\includegraphics[scale=0.12]{Cut1Sym.pdf} }[a_-,b_\mp]}{4\sqrt{(1\cdot3\cdot5)}\sqrt{(1\cdot5\cdot7)}};\\
 		&{\rm LS}_{\pm,1}[2]=\pm\frac{\mathcal{C}^2_{\includegraphics[scale=0.12]{Cut1Sym.pdf} } [a_+,b_\pm]}{4\sqrt{(2\cdot4\cdot6)}\sqrt{(2\cdot6\cdot8)}}, \quad {\rm LS}_{\pm,2}[2]=\pm\frac{\mathcal{C}^2_{\includegraphics[scale=0.12]{Cut1Sym.pdf} }[a_-,b_\mp]}{4\sqrt{(2\cdot4\cdot6)}\sqrt{(2\cdot6\cdot8)}};\\
 		&{\rm LS}_{\pm,1}[3]=\pm\frac{\mathcal{C}^3_{\includegraphics[scale=0.12]{Cut1Sym.pdf} } [a_+,b_\pm]}{4\sqrt{(1\cdot3\cdot7)}\sqrt{(3\cdot5\cdot7)}}, \quad {\rm LS}_{\pm,2}[2]=\pm\frac{\mathcal{C}^3_{\includegraphics[scale=0.12]{Cut1Sym.pdf} }[a_-,b_\mp]}{4\sqrt{(1\cdot3\cdot7)}\sqrt{(3\cdot5\cdot7)}}.
		 	\end{aligned}
\end{equation}
Similarly, the box-triangle cut can also be identified with the same set of leading singularities:
  \begin{equation}
 	\begin{aligned}
 		&{\rm LS}_{\pm,1}[4]=\left.\frac{\mathcal{C}^2_{\includegraphics[scale=0.13]{Cut2Sym.pdf}}[a_\pm,b_+]}{16\, \text{det}(\ell^\prime_1,\ell^\prime_2,\ell^\prime_3)\text{det}(\ell^\prime_4,\ell^\prime_5,\ell^\prime_6)}\right|_{a,b=a_\pm,b_+}=\pm\frac{\mathcal{C}^2_{\includegraphics[scale=0.13]{Cut2Sym.pdf}}[a_\pm,b_+]}{4 \sqrt{(b_+\cdot1\cdot3)}\sqrt{(4\cdot6\cdot8)}}\\
 		&{\rm LS}_{\pm,2}[4]=\left.\frac{\mathcal{C}^2_{\includegraphics[scale=0.13]{Cut2Sym.pdf}}[a_\mp,b_-]}{16\, \text{det}(\ell^\prime_1,\ell^\prime_2,\ell^\prime_3)\text{det}(\ell^\prime_4,\ell^\prime_5,\ell^\prime_6)}\right|_{a,b=a_\mp,b_-}=\pm\frac{\mathcal{C}^2_{\includegraphics[scale=0.13]{Cut2Sym.pdf}}[a_\mp,b_-]}{4\sqrt{(b_-\cdot1\cdot3)}\sqrt{(4\cdot6\cdot8)}}\,.
 	\end{aligned}
 \end{equation}
 Notice that here the Jacobian factor depends on the cut-solution for one of the loop regions, i.e. $b$. This hints at a more involved integrand to reproduce the maximal cut as we will see.

Before closing, since all the on-shell data are given by leading singularities, the two-loop maximal cut can be linearly related to the one-loop maximal cut. This relation will be useful when one combines the constraint of soft-cuts, which reduces the two-loop  to one-loop integrand. Explicitly they are given as: 
\begin{equation}\label{OneMatchTwoLoop}
  \begin{aligned}
	&\frac{\mathcal{C}^+_{2,4,8}}{2\sqrt{(2\cdot4\cdot8)}}{=}\frac{\mathcal{C}^8_{\includegraphics[scale=0.12]{Cut1Sym.pdf}}[a_-,b_-]{-}\mathcal{C}^8_{\includegraphics[scale=0.12]{Cut1Sym.pdf}}[a_-,b_+]}{4\sqrt{(8\cdot2\cdot4)}\sqrt{(4\cdot6\cdot8)}}{=}\frac{\mathcal{C}^2_{\includegraphics[scale=0.13]{Cut2Sym.pdf}}[a_-,b_-]}{4\sqrt{(b_-\cdot1\cdot3)}\sqrt{(4\cdot6\cdot8)}} {-} \frac{\mathcal{C}^2_{\includegraphics[scale=0.13]{Cut2Sym.pdf}}[a_-,b_+]}{4\sqrt{(b_+\cdot1\cdot3)}\sqrt{(4\cdot6\cdot8)}},\\
	-&\frac{\mathcal{C}^-_{2,4,8}}{2\sqrt{(2\cdot4\cdot8)}}{=}\frac{\mathcal{C}^8_{\includegraphics[scale=0.12]{Cut1Sym.pdf}}[a_+,b_+]{-}\mathcal{C}^2_{\includegraphics[scale=0.12]{Cut1Sym.pdf}}[a_+,b_-]}{4\sqrt{(8\cdot2\cdot4)}\sqrt{(4\cdot6\cdot8)}}{=}\frac{\mathcal{C}^2_{\includegraphics[scale=0.13]{Cut2Sym.pdf}}[a_+,b_+]}{4\sqrt{(b_+\cdot1\cdot3)}\sqrt{(4\cdot6\cdot8)}} {-} \frac{\mathcal{C}^2_{\includegraphics[scale=0.13]{Cut2Sym.pdf}}[a_+,b_-]}{4\sqrt{(b_-\cdot1\cdot3)}\sqrt{(4\cdot6\cdot8)}},\\
 	&\frac{\mathcal{C}^+_{4,6,8}}{2\sqrt{(4\cdot6\cdot8)}}=\frac{\mathcal{C}^8_{\includegraphics[scale=0.12]{Cut1Sym.pdf}}[a_+,b_+]-\mathcal{C}^2_{\includegraphics[scale=0.12]{Cut1Sym.pdf}}[a_-,b_+]}{4\sqrt{(8\cdot2\cdot4)}\sqrt{(4\cdot6\cdot8)}}=\frac{\mathcal{C}^2_{\includegraphics[scale=0.13]{Cut2Sym.pdf}}[a_+,b_+]-\mathcal{C}^2_{\includegraphics[scale=0.13]{Cut2Sym.pdf}}[a_-,b_+]}{4\sqrt{(b_+\cdot1\cdot3)}\sqrt{(4\cdot6\cdot8)}}, \\
	-&\frac{\mathcal{C}^-_{4,6,8}}{2\sqrt{(4\cdot6\cdot8)}}=\frac{-\mathcal{C}^8_{\includegraphics[scale=0.12]{Cut1Sym.pdf}}[a_+,b_-]+\mathcal{C}^8_{\includegraphics[scale=0.12]{Cut1Sym.pdf}}[a_-,b_-]}{4\sqrt{(8\cdot2\cdot4)}\sqrt{(4\cdot6\cdot8)}}=\frac{-\mathcal{C}^2_{\includegraphics[scale=0.13]{Cut2Sym.pdf}}[a_+,b_-]+\mathcal{C}^2_{\includegraphics[scale=0.13]{Cut2Sym.pdf}}[a_-,b_-]}{4\sqrt{(b_-\cdot1\cdot3)}\sqrt{(4\cdot6\cdot8)}}\,.
  \end{aligned}
\end{equation}

\section{The one-loop eight-point integrand and amplitude}\label{sec:one-loop}
Let us begin by constructing the one-loop integrand. Note that since the one-loop amplitude is finite, in principle, no regularization is needed. However, since it will appear in soft limits of two-loop amplitudes, we will derive integral representations which are manifestly dual conformal invariant and thus amenable for Higgs branch regularization, {\it i.e.} it can be expanded in $m^2$ the Higgs branch vacuum expectation values (vev). 

\subsection{Maximal Cuts}
There are two types of on-shell data that constrain the one-loop integrand: 1. the maximal cut where three propagators connecting three massive corners are put on-shell, and 2. when the momenta between consecutive massless corners are soft. On the maximal cuts, the one-loop amplitude must satisfy:
\begin{equation}\label{CutCond}
\left.A^{1-\text{loop}}_8\right|_{\substack{\textrm{cut}\\i,\,i{+}2,\,i{+}4}}=\frac{1}{\sqrt{(i\cdot i{+}2 \cdot i{+}4)}}\mathcal{C}_{i,i{+}2,i{+}4}^\pm
\end{equation}
where $1/\sqrt{(i\cdot i{+}2 \cdot i{+}4)}$ is the Jacobian for putting propagators on shell. For latter convenience, we define shorthand notations for the sum and difference of one-loop maximal cuts, weighted by the Jacobian:
\begin{equation}\label{BDCut}
    \mathcal{B}_{i,j,k}\equiv \frac{\mathcal{C}^+_{i,j,k}+\mathcal{C}^-_{i,j,k}}{2\sqrt{(i\cdot j\cdot k)}},\quad  \mathcal{D}_{i,j,k}\equiv \frac{\mathcal{C}^+_{i,j,k}-\mathcal{C}^-_{i,j,k}}{2\sqrt{(i\cdot j\cdot k)}}\,.
\end{equation}

These maximal cuts are intimately related to terms in the tree-level BCFW recursion~\cite{Brandhuber:2012un}. In particular, combining eq.(\ref{TreeLS}) with eq.(\ref{1LoopLS1}) and eq,(\ref{1LoopLS2}),the tree-level amplitudes are given by sum over the  \textit{difference} of maximal cut on the two cut solutions:  
\begin{eqnarray}\label{TreeRel}
		A_{8}^{\mathrm{tree}}	&=&\mathcal{D}_{2,4,8}+\mathcal{D}_{2,6,8}=-\mathcal{D}_{1,3,5}-\mathcal{D}_{1,3,7}\,\nonumber\\
		&=&\mathcal{D}_{4,6,8}+\mathcal{D}_{2,4,6}=-\mathcal{D}_{1,5,7}-\mathcal{D}_{3,5,7}\,.
\end{eqnarray}
Note that the above representation also exhibits shift by one-site symmetry, where the extra minus sign reflects the fact that exchanging the gauge groups corresponds to exchanging $k\leftrightarrow -k$.

Since there are no odd multiplicity amplitudes in ABJM, we cannot have massless corners. However in the case of consecutive massless corners, one can have three propagators becoming on-shell, $(a\cdot i{-}1)=(a\cdot i)=(a\cdot i{+}1)=0$, when the exchanged momenta between two massless legs become soft. In terms of the dual region, this corresponds to the limit $y_a\rightarrow y_i$ and  the amplitude reduces to that with one loop lower:
\begin{equation}\label{SoftCond}
	\left.A_{n}^{\ell-\text{loop}}\right|_{\substack{\textrm{cut}\\i{-}1,\,i,\,i{+}1}}=(-1)^{i}A_{n}^{(\ell-1)-\text{loop}}.
\end{equation}
We will refer to such cuts as ``soft cuts''.

\subsection{The integral basis and one-loop amplitudes}

We will determine the full integrand of the one-loop ABJM amplitude by requiring that all maximal cut matches, i.e. eq.(\ref{CutCond}) and eq.(\ref{SoftCond}) holds.  To achieve this, we define a pair of ``chiral-box'' integrals for any triplet $(i,j,k)$,
\begin{eqnarray}\label{ChiralBox}
	I^{\pm}\left(i,j,k\right)&=&\frac{1}{2}\left(\int_{a}\frac{\sqrt{\left(i\cdot j\cdot k\right)}}{\left(a\cdot i\right)\left(a\cdot j\right)\left(a\cdot k\right)}\pm\frac{-\epsilon\left(a,i,j,k,X\right)}{\sqrt{2}\left(a\cdot i\right)\left(a\cdot j\right)\left(a\cdot k\right)\left(a\cdot X\right)}\right)\nonumber\\
	&\equiv& \frac{1}{2}\left(I_{tri}\left(i,j,k\right)\pm I_{box}\left(i,j,k,X\right)\right),
\end{eqnarray}
and the triangle integral is absent if any pair of the labels are adjacent; for example,  $I^{\pm}\left(i{-}1,i,i{+}1\right)=\pm I_{box}\left(i{-}1, i, i{+}1\right)$. Note that in the formula we have introduced a reference point, $X$, which should cancel out in the physical one-loop integrand.\\

This combination has the desired property that it evaluates to $1$ or $0$ on the maximal cuts. For adjacent soft-cuts, $\left(a\cdot i{-}1\right){=}\left(a\cdot i\right){=}\left(a\cdot i{+}1\right)=0$, $I^{\pm}\left(i{-}1,j, i{+}1\right)$ which only involves the box integral, evaluates to   
\begin{equation}\mathrm{Cut}_{i{-}1,i,i{+}1}\ I_{box}\left(i{-}1,i,i{+}1,X\right)
=\int_a \delta((a\cdot i{-}1))\delta((a\cdot i)) \delta((a\cdot i{+}1))  \frac{\epsilon\left(a,i{-}1,i,i{+}1,X\right)}{\sqrt{2}\left(a\cdot X\right)}=1\,.
\end{equation}
For massive maximal cut the triangle integrals give $\mathrm{Cut}_{i,i{+}2,i{+}4}^{\pm}I_{tri}\left(i, i{+}2, i{+}4\right)=1$. For the box, due to the numerator we have 
\begin{equation}\label{CutSign}
			\left.\epsilon\left(a,i,i{+}2,i{+}4,X\right)\right|_{a \rightarrow a^\pm}=\mp\sqrt{2}\sqrt{(i\cdot i{+}2\cdot i{+}4)}\left(a^{\pm}\cdot X\right),
\end{equation}
where $a^\pm$ represents the two cut solutions. On the cut the combination integral give:
\begin{equation}
		\mathrm{Cut}_{i,i{+}2,i{+}4}^{\pm}\ I^{\pm}\left(i,i{+}2,i{+}4\right)=1, \quad \mathrm{Cut}_{i,i{+}2,i{+}4}^{\pm} I^{\mp}\left(i,i{+}2,i{+}4\right)=0\,.
\end{equation}

We are now ready to write the general one-loop amplitude. By manifesting all massive and soft maximal cut, we write:
\begin{eqnarray}\label{OneLoopX}
		A_n^{\text{1-loop}}=-A_n^{\text{tree}}\sum_{i=1}^n (-)^i I_{box}\left(i{-}1,i,i{+}1,X\right)+\sum_{\rm{massive}}\frac{\mathcal{C}^{\pm}_{i,j,k}}{\sqrt{\left(i\cdot j\cdot k\right)}}I^{\pm}\left(i,j,k\right).
\end{eqnarray}
If one chooses one of the external regions as the reference point, $X=y_i$, some of the box integrals will be absent due to the tensor numerator. However, identities amongst the maximal cut and the tree amplitude will ensure that all cuts are faithfully reproduced. This is an alternative way of saying that the representation is $X$ independent. For example, if we choose $X=y_1$, then the box integrals for the soft cut $(n, 1, 2)$ will be absent from the part proportional to $A_n^{\text{tree}}$. They however will be present in the massive cut part 
\begin{eqnarray}
&&\sum_{i}\frac{\mathcal{C}^{\pm}_{i,n,2}}{\sqrt{\left(i\cdot n\cdot 2\right)}}I^{\pm}\left(i,n,2\right)\bigg|_{X=y_1}\\
&=&\sum_{i}\frac{\mathcal{C}^{\pm}_{i,n,2}}{2\sqrt{\left(i\cdot n\cdot 2\right)}}\left(\int_{a}\frac{\sqrt{\left(i\cdot n\cdot 2\right)}}{\left(a\cdot i\right)\left(a\cdot n\right)\left(a\cdot 2\right)}\pm\frac{-\epsilon\left(a,i,n,2,1\right)}{\sqrt{2}\left(a\cdot i\right)\left(a\cdot n\right)\left(a\cdot 2\right)\left(a\cdot 1\right)}\right)\nonumber
\end{eqnarray}
where the $i$ only sums over massive maximal cuts. This simply reproduces the ($n,1$)-shift for the BCFW representation of the tree amplitude. Similarly, choosing  $X=y_i$ utilizes the BCFW representation for the ($n,i$)-shift.

As an example, consider $8$-points there are now 4$+$4 massive maximal cuts, $(2,4,6)$ and 3 cyclic by two as well as $(1,3,5)$ and its counterpart. Similarly, there are 8 massless maximal cuts. By choosing the reference point $X=y_1$, we find the following combination that satisfies all cuts:
\begin{equation}\label{OneLoopAnsw}
	\begin{aligned}
		A_{8}^{\mathrm{1-loop}}	&=A_{8}^{\mathrm{tree}}\left[I_{box}\left(1,2,3,4\right){-}I_{box}\left(1,3,4,5\right){+}I_{box}\left(1,4,5,6\right){-}I_{box}\left(1,5,6,7\right){+}I_{box}\left(1,6,7,8\right)\right]\\
	&{+}\sum_{\pm}\frac{\mathcal{C}^{\pm}_{3,5,7}}{\sqrt{\left(3\cdot5\cdot7\right)}}I^\pm\left(3,5,7\right){+}\mathcal{B}_{1,3,5}I_{tri}\left(1,3,5\right){+}\mathcal{B}_{1,3,7}I_{tri}\left(1,3,7\right){+}\mathcal{B}_{1,5,7}I_{tri}\left(1,5,7\right)\\
	&{+}\left[\sum_{\pm}\frac{\mathcal{C}^{\pm}_{2,4,6}}{\sqrt{\left(2\cdot4\cdot6\right)}}I^\pm\left(2,4,6\right){+}(i\rightarrow i{+}2){+}(i\rightarrow i{+}4){+}(i\rightarrow i{+}6)\right]\,.
	\end{aligned}
\end{equation}
The above manifests all even maximal cuts, i.e.  $(2,4,6)$ and its orbits, as well as $(3,5,7)$. For the massive cuts that involve $1$, $(1,3,5)$ $(5,7,1)$ $(7,1,3)$, they are reproduced when combined with the contribution from the part proportional to the tree amplitude. For example, the following terms contribute to cut $(1,3,5)$:
\begin{eqnarray}
-A_{8}^{\mathrm{tree}}I_{box}\left(1,3,4,5\right)+\mathcal{D}_{3,5,7}I_{box}\left(1,3,5,7\right)+\mathcal{B}_{1,3,5}I_{tri}\left(1,3,5\right)\,.\nonumber
\end{eqnarray}
Using the relation between the tree amplitude and maximal cut in eq.~\eqref{BDCut}, as well as eq.~\eqref{TreeRel}, we see that indeed it evaluates to $\mathcal{C}^{\pm}_{1,3,5}$ on both cut solutions respectively. Similarly, the soft-cut constraint eq.(\ref{SoftCond}), is manifested by the box integrals in the first line of eq.(\ref{OneLoopAnsw}) except for the cut $(8,1,2)$. This is reproduced by 
\begin{eqnarray}
\mathcal{D}_{2,6,8}I_{box}\left(1,2,6,8\right)+\mathcal{D}_{2,4,8}I_{box}\left(1,2,4,8\right)\,,
\end{eqnarray}
which reproduces the soft limit again using eq.(\ref{TreeRel}). 

For completeness, we give the integrated result. Since the tensor box integrals can be written as a total derivative, it vanishes, and one only needs to retain the massive triangles. The massive scale one-loop triangle integral is 
\begin{equation}
I_{tri}(i,j,k)=\int_{a}\frac{\sqrt{\left(i\cdot j\cdot k\right)}}{\left(a\cdot i\right)\left(a\cdot j\right)\left(a\cdot k\right)},
\end{equation}
whose result is well-known
\begin{equation}
I_{tri}(i,j,k)=-\frac{i \pi}{4}\frac{\sqrt{(i\cdot j)(j\cdot k)(k\cdot i)}}{\sqrt{-(i\cdot j)-i\epsilon}\sqrt{-(j\cdot k)-i\epsilon}\sqrt{-(k\cdot i)-i\epsilon}},
\end{equation}
and it can be further simplified to
\begin{equation}
I_{tri}(i,j,k)=\frac{\pi}{4} {\rm sgn}_c[(i\cdot j)]{\rm sgn}_c[(j\cdot k)]{\rm sgn}_c[(k\cdot i)]
\end{equation}
by defining the sign function~\cite{Brandhuber:2012un}:
\begin{equation}\label{SignDef}
{\rm sgn}_c[(i \cdot  j)]\equiv -\frac{\sqrt{-(i\cdot j)}}{\sqrt{-(i\cdot j)-i\epsilon}}=\pm1,
\end{equation}
and we can further canonically calculate the square root on the numerator by introducing the spinor
\[
(i\cdot j)=p_{i,i+1,\cdots,j-1}^2=-\frac{1}{4}\langle  \mu\, \nu \rangle^2\quad \text{for $p_{i,i+1,\cdots,j-1}^{\alpha \beta}=|\mu\rangle^{(\alpha} |\nu\rangle^{\beta)}$},
\]
so
\begin{equation}
{\rm sgn}_c[(i \cdot  j)]\equiv -\frac{\langle  \mu\, \nu \rangle}{\sqrt{-(i\cdot j)-i\epsilon}}=\pm1.
\end{equation}


The one-loop eight-point amplitude now is
\begin{equation}\label{OneLoopResult}
\boxed{	\begin{aligned}
		A_{8}^{\mathrm{1-loop}}	&=\frac{N}{k}\frac{\pi}{4}\left[\mathcal{B}_{246}\;{\rm sgn}_c[(2\cdot 4)]{\rm sgn}_c[(4\cdot 6)]{\rm sgn}_c[(6\cdot 2)]+{\rm cyclic}\;\;\right]\,.
	\end{aligned} }
\end{equation}
Thus the one-loop amplitude is given by the combinations of maximal cut, each summed over the two solutions, weighted by the sign function associated with the cut.

Before closing, let us verify that the result eq.(\ref{OneLoopResult}) satisfied cyclic shift and reflection symmetries. Since from eq.(\ref{CutShifts}), the one-loop maximal cuts attain a minus sign under cyclic shift by two sites, the above one-loop amplitude inherits this property. Under reflection, the cuts interchange according to eq.(\ref{OneLoopCutFlip}) and as a sum of two cut solutions, the sign of $\mathcal{B}_{i,j,k}$ is not affected by exchanging of $\pm$, while the sign function attains a minus sign. To see this note that the sign functions come in the combination\footnote{Here, we use $\sqrt{{-}p_I^2}$ implicitly represent the spinor bracket $\langle\mu \nu\rangle$ with $p_I^{\alpha\beta}=|\mu \rangle^{(\alpha}|\nu\rangle^{\beta)}$. }:
\begin{equation}
    {\rm sgn}_c[(i\cdot i{+}2)]{\rm sgn}_c[(i{+}2\cdot i{+}4)]{\rm sgn}_c[(i\cdot i{+}4)]{=}\frac{\langle i\,i{+}1\rangle}{\sqrt{{-}p_{i, i{+}1}^2{-}i\epsilon}}\frac{\langle i{+}2\,i{+}3\rangle}{\sqrt{{-}p_{i{+}2,i{+}3}^2{-}i\epsilon}}\frac{\sqrt{{-}p_{i,i{+1},i{+}2,i{+}3}^2}}{\sqrt{{-}p_{i,i{+1},i{+}2,i{+}3}^2{-}i\epsilon}}\,,
\end{equation}
where $p_I:=\sum_{i\in I}p_i$. Under reflection, again say  $\{\Lambda_1,\Lambda_2,\ldots, \Lambda_8 \}\rightarrow\{\Lambda_1,\Lambda_8,\ldots, \Lambda_2 \}$, we see 
\begin{equation}
   \frac{\langle 12\rangle}{\sqrt{{-}p_{1,2}^2{-}i\epsilon}}\frac{\langle 34\rangle}{\sqrt{{-}p_{3,4}^2{-}i\epsilon}}\frac{\sqrt{{-}p_{1,2,3,4}^2}}{\sqrt{{-}p_{1,2,3,4}^2{-}i\epsilon}} \rightarrow \frac{\langle 18\rangle}{\sqrt{{-}p_{8,1}^2{-}i\epsilon}}\frac{\langle 76\rangle}{\sqrt{{-}p_{6,7}^2{-}i\epsilon}}\frac{\sqrt{{-}p_{6,7,8,1}^2}}{\sqrt{{-}p_{6,7,8,1}^2{-}i\epsilon}}.
\end{equation}
The last square root in the numerator can be removed if we note that any vector in three dimensions can be written in a bi-spinor form as $K_{\alpha\beta}=\mu_{(\alpha}\nu_{\beta)}$, such that $K^2=-\langle \mu,\nu\rangle^2$. We see that the three spinor brackets are mapped to reversed ordering,\footnote{To see this sign for $\sqrt{{-}p_{i,i{+}1,i{+}2,i{+}3}^2}$, one can simply take double collinear limit where $p_i\parallel p_{i{+}1}$ and $p_{i{+}2}\parallel p_{i{+}3}$ and apply the reflection. }  and thus giving the requisite minus sign satisfying eq.(\ref{ReflectionRule}). 



\section{The two-loop eight-point integrand}\label{sec:two-loop}

The four and six-point two-loop integrands were constructed via imposing various physical constraints on a set of dual conformal integrands~\cite{Chen:2011vv, Caron-Huot:2012sos}. These include the matching of (1) soft cuts, (2) maximal cuts, (3) vanishing collinear-soft limits, and (4) vanishing three-point cuts. The first two were already discussed in our construction of the one-loop integrand: at two loops, the only new ingredient is that there are two topologically distinct maximal cuts,
\begin{equation}
 \vcenter{\hbox{\scalebox{0.8}{
\begin{tikzpicture}[x=0.75pt,y=0.75pt,yscale=-1,xscale=1]

\draw  [line width=1.2]  (201,299.5) -- (251,299.5) -- (251,349.5) -- (201,349.5) -- cycle ;
\draw  [line width=1.2]  (251,299.5) -- (301,299.5) -- (301,349.5) -- (251,349.5) -- cycle ;
\draw [color={rgb, 255:red, 0; green, 0; blue, 0 }  ,draw opacity=1 ][line width=1.2]    (301,299.5) -- (301,286.54) -- (301,279.25) ;
\draw [color={rgb, 255:red, 0; green, 0; blue, 0 }  ,draw opacity=1 ][line width=1.2]    (301,299.5) -- (313.96,299.5) -- (321.25,299.5) ;
\draw [color={rgb, 255:red, 0; green, 0; blue, 0 }  ,draw opacity=1 ][line width=1.2]    (301,349.5) -- (313.96,349.5) -- (321.25,349.5) ;
\draw [color={rgb, 255:red, 0; green, 0; blue, 0 }  ,draw opacity=1 ][line width=1.2]    (301,369.75) -- (301,356.79) -- (301,349.5) ;
\draw [color={rgb, 255:red, 0; green, 0; blue, 0 }  ,draw opacity=1 ][line width=1.2]    (180.75,349.5) -- (193.71,349.5) -- (201,349.5) ;
\draw [color={rgb, 255:red, 0; green, 0; blue, 0 }  ,draw opacity=1 ][line width=1.2]    (201,369.75) -- (201,356.79) -- (201,349.5) ;
\draw [color={rgb, 255:red, 0; green, 0; blue, 0 }  ,draw opacity=1 ][line width=1.2]    (201,299.5) -- (201,286.54) -- (201,279.25) ;
\draw [color={rgb, 255:red, 0; green, 0; blue, 0 }  ,draw opacity=1 ][line width=1.2]    (180.75,299.5) -- (193.71,299.5) -- (201,299.5) ;
\draw  [color={rgb, 255:red, 155; green, 155; blue, 155 }  ,draw opacity=1 ][fill={rgb, 255:red, 155; green, 155; blue, 155 }  ,fill opacity=1 ] (192.75,299.5) .. controls (192.75,294.94) and (196.44,291.25) .. (201,291.25) .. controls (205.56,291.25) and (209.25,294.94) .. (209.25,299.5) .. controls (209.25,304.06) and (205.56,307.75) .. (201,307.75) .. controls (196.44,307.75) and (192.75,304.06) .. (192.75,299.5) -- cycle ;
\draw  [color={rgb, 255:red, 155; green, 155; blue, 155 }  ,draw opacity=1 ][fill={rgb, 255:red, 155; green, 155; blue, 155 }  ,fill opacity=1 ] (192.75,349.5) .. controls (192.75,344.94) and (196.44,341.25) .. (201,341.25) .. controls (205.56,341.25) and (209.25,344.94) .. (209.25,349.5) .. controls (209.25,354.06) and (205.56,357.75) .. (201,357.75) .. controls (196.44,357.75) and (192.75,354.06) .. (192.75,349.5) -- cycle ;
\draw  [color={rgb, 255:red, 155; green, 155; blue, 155 }  ,draw opacity=1 ][fill={rgb, 255:red, 155; green, 155; blue, 155 }  ,fill opacity=1 ] (292.75,299.5) .. controls (292.75,294.94) and (296.44,291.25) .. (301,291.25) .. controls (305.56,291.25) and (309.25,294.94) .. (309.25,299.5) .. controls (309.25,304.06) and (305.56,307.75) .. (301,307.75) .. controls (296.44,307.75) and (292.75,304.06) .. (292.75,299.5) -- cycle ;
\draw  [color={rgb, 255:red, 155; green, 155; blue, 155 }  ,draw opacity=1 ][fill={rgb, 255:red, 155; green, 155; blue, 155 }  ,fill opacity=1 ] (292.75,349.5) .. controls (292.75,344.94) and (296.44,341.25) .. (301,341.25) .. controls (305.56,341.25) and (309.25,344.94) .. (309.25,349.5) .. controls (309.25,354.06) and (305.56,357.75) .. (301,357.75) .. controls (296.44,357.75) and (292.75,354.06) .. (292.75,349.5) -- cycle ;
\draw  [color={rgb, 255:red, 155; green, 155; blue, 155 }  ,draw opacity=1 ][fill={rgb, 255:red, 155; green, 155; blue, 155 }  ,fill opacity=1 ] (250.42,295.17) .. controls (257.46,295.17) and (263.17,308.37) .. (263.17,324.67) .. controls (263.17,340.96) and (257.46,354.17) .. (250.42,354.17) .. controls (243.38,354.17) and (237.67,340.96) .. (237.67,324.67) .. controls (237.67,308.37) and (243.38,295.17) .. (250.42,295.17) -- cycle ;

\draw (246.7,360.9) node [anchor=north west][inner sep=0.75pt]   [align=left] {$\displaystyle i$};
\draw (166,317.5) node [anchor=north west][inner sep=0.75pt]   [align=left] {$\displaystyle i+2$};
\draw (235.7,270.4) node [anchor=north west][inner sep=0.75pt]   [align=left] {$\displaystyle i+4$};

\end{tikzpicture}

}
}},\quad\quad 
\vcenter{\hbox{\scalebox{0.8}{
\begin{tikzpicture}[x=0.75pt,y=0.75pt,yscale=-1,xscale=1]

\draw  [line width=1.2]  (251,299.5) -- (301,299.5) -- (301,349.5) -- (251,349.5) -- cycle ;
\draw [color={rgb, 255:red, 0; green, 0; blue, 0 }  ,draw opacity=1 ][line width=1.2]    (209.5,324.33) -- (191.75,342.08) ;
\draw [color={rgb, 255:red, 0; green, 0; blue, 0 }  ,draw opacity=1 ][line width=1.2]    (209.5,324.33) -- (205.38,320.21) -- (191.75,306.58) ;
\draw [color={rgb, 255:red, 0; green, 0; blue, 0 }  ,draw opacity=1 ][line width=1.2]    (301,299.5) -- (301,286.54) -- (301,279.25) ;
\draw [color={rgb, 255:red, 0; green, 0; blue, 0 }  ,draw opacity=1 ][line width=1.2]    (301,299.5) -- (313.96,299.5) -- (321.25,299.5) ;
\draw [color={rgb, 255:red, 0; green, 0; blue, 0 }  ,draw opacity=1 ][line width=1.2]    (301,349.5) -- (313.96,349.5) -- (321.25,349.5) ;
\draw [color={rgb, 255:red, 0; green, 0; blue, 0 }  ,draw opacity=1 ][line width=1.2]    (301,369.75) -- (301,356.79) -- (301,349.5) ;
\draw [color={rgb, 255:red, 0; green, 0; blue, 0 }  ,draw opacity=1 ][line width=1.2]    (251,279.25) -- (251,292.21) -- (251,299.5) ;
\draw [color={rgb, 255:red, 0; green, 0; blue, 0 }  ,draw opacity=1 ][line width=1.2]    (251,349.5) -- (251,362.46) -- (251,369.75) ;
\draw  [line width=1.2]  (209.5,324.33) -- (251,299.5) -- (251,349.17) -- cycle ;
\draw  [color={rgb, 255:red, 155; green, 155; blue, 155 }  ,draw opacity=1 ][fill={rgb, 255:red, 155; green, 155; blue, 155 }  ,fill opacity=1 ] (201.25,324.33) .. controls (201.25,319.78) and (204.94,316.08) .. (209.5,316.08) .. controls (214.06,316.08) and (217.75,319.78) .. (217.75,324.33) .. controls (217.75,328.89) and (214.06,332.58) .. (209.5,332.58) .. controls (204.94,332.58) and (201.25,328.89) .. (201.25,324.33) -- cycle ;
\draw  [color={rgb, 255:red, 155; green, 155; blue, 155 }  ,draw opacity=1 ][fill={rgb, 255:red, 155; green, 155; blue, 155 }  ,fill opacity=1 ] (242.75,300.46) .. controls (242.75,295.91) and (246.44,292.21) .. (251,292.21) .. controls (255.56,292.21) and (259.25,295.91) .. (259.25,300.46) .. controls (259.25,305.02) and (255.56,308.71) .. (251,308.71) .. controls (246.44,308.71) and (242.75,305.02) .. (242.75,300.46) -- cycle ;
\draw  [color={rgb, 255:red, 155; green, 155; blue, 155 }  ,draw opacity=1 ][fill={rgb, 255:red, 155; green, 155; blue, 155 }  ,fill opacity=1 ] (292.75,299.5) .. controls (292.75,294.94) and (296.44,291.25) .. (301,291.25) .. controls (305.56,291.25) and (309.25,294.94) .. (309.25,299.5) .. controls (309.25,304.06) and (305.56,307.75) .. (301,307.75) .. controls (296.44,307.75) and (292.75,304.06) .. (292.75,299.5) -- cycle ;
\draw  [color={rgb, 255:red, 155; green, 155; blue, 155 }  ,draw opacity=1 ][fill={rgb, 255:red, 155; green, 155; blue, 155 }  ,fill opacity=1 ] (242.75,349.5) .. controls (242.75,344.94) and (246.44,341.25) .. (251,341.25) .. controls (255.56,341.25) and (259.25,344.94) .. (259.25,349.5) .. controls (259.25,354.06) and (255.56,357.75) .. (251,357.75) .. controls (246.44,357.75) and (242.75,354.06) .. (242.75,349.5) -- cycle ;
\draw  [color={rgb, 255:red, 155; green, 155; blue, 155 }  ,draw opacity=1 ][fill={rgb, 255:red, 155; green, 155; blue, 155 }  ,fill opacity=1 ] (292.75,349.5) .. controls (292.75,344.94) and (296.44,341.25) .. (301,341.25) .. controls (305.56,341.25) and (309.25,344.94) .. (309.25,349.5) .. controls (309.25,354.06) and (305.56,357.75) .. (301,357.75) .. controls (296.44,357.75) and (292.75,354.06) .. (292.75,349.5) -- cycle ;

\draw (175.5,316.5) node [anchor=north west][inner sep=0.75pt]   [align=left] {$\displaystyle i$};

\end{tikzpicture}
}
}}\,.
\end{equation}
Each maximal cut consists of four cut solutions. Collinear-soft limits occur when two-loop momenta are collinear to a massless external leg, and lead to non-factorizable singularities, which should not occur at two-loops since the leading IR divergence occurs at this order (see ~\cite{Caron-Huot:2012sos} and reference therein for a more detailed discussion). Finally, at two loops, one can have internal trivalent vertices which must vanish on the cut due to vanishing three-point amplitudes. We will demonstrate that the eight-point integrand can be fully determined using such on-shell data. As a further consistency check,  we will show that the resulting integrand vanishes on any multi-particle cut that involves odd-multiplicity amplitudes.

We begin by first introducing the topologies of two-loop integrals that participate in  the constraints discussed above. First, for soft cuts, we have:

\begin{eqnarray}\label{SoftCuts}
&&I^{db}_{A, i,j}:\;\vcenter{\hbox{\scalebox{0.7}{


}
}}
\end{eqnarray}
where we have used superscripts $db$ and $bt$ to represent double-box and box-triangle topology, respectively. Note that $I^{db}_{A, i,j}$ correspond to two cases for fixed $i$, with $j=i{+}4$ and $i{-}2$. The same is true for other topologies with an extra label $j$; for $ I^{db}_{A, i,j}$, $ I^{db}_{B, i, j}$ and $ I^{db}_{C, i}$ topologies participate in two distinct soft cuts.

The topologies entering the two types of maximal cuts are:
\begin{eqnarray}
&&I^{db}_{G, i}:\;\vcenter{\hbox{\scalebox{0.7}{
%

}
}}.
\end{eqnarray}

At this point, all double-boxes and  box-triangles are fully determined. What remains is the double triangles which can be fixed by the remaining two constraints. 

The collinear-soft singularities occur for the following topologies:
\begin{eqnarray}
&&I^{db}_{A, i,j}:\;\vcenter{\hbox{\scalebox{0.7}{
%

}
}}\,.
\end{eqnarray}
Once again, $I^{db}_{F, i,j}$, $I^{bt}_{B, i}$ and $I^{bt}_{C, i}$ contains two separate soft-collinear singularities. Finally, the topology that is involved in the three-point amplitude sub-cuts is:
\begin{eqnarray}
&&I^{db}_{B, i, j}:\;\vcenter{\hbox{\scalebox{0.7}{
%

}
}}\,.
\end{eqnarray}

The on-shell data controlling the soft cuts are the one-loop maximal cuts. Since the double-box integrals $I^{db}_{F, i}$ are subject to both soft-cut and two-loop maximal-cut constraints, this tells us that the tree, one and two-loop on-shell data are not independent. Indeed they can all be expressed as linear combinations of  leading singularities as previously discussed.



\subsection{Soft constraints}\label{sec:SoftCut}

Let us begin with the soft constraints which restrict the numerators for the topologies listed in eq.(\ref{SoftCuts}). Without loss of generality, we consider the soft channel $1,2,3$, where the integrands share the common propagators  $1/(a\cdot 1)(a\cdot 2)(a\cdot 3)$. Following the one-loop discussion, it is natural to introduce the Levi-Civita numerators $\epsilon(a,1,2,3,\,^\mu)$ such that on the cut:
\begin{equation}
	 \int _a \delta((a\cdot1)) \delta((a\cdot2)) \delta((a\cdot3)) \;\; \epsilon(a,1,2,3,\,^\mu)=\sqrt{2} y_2^\mu\,.
\end{equation} 
The task is to find suitable vectors to contract with $\epsilon(a,1,2,3,\,^\mu)$, such that one reproduces the one-loop box and triangles in the soft limit. Starting with double box topologies, the one-loop box integrals can be obtained by contracting with the vector that is the box numerator with $y_2$ removed, while the triangles are reproduced if one contracts with $y_b$.

Beginning with topologies $I^{db}_{A, i,j}$, $I^{db}_{B, i,j}$, and $I^{db}_{C, i}$, since these contain two soft cuts, one for loop $a$ one for $b$, their soft limits should only lead to boxes. This is because there are no one-loop triangles with massless corners. For example, for  $I^{db}_{A, 1,5}$ we introduce the double epsilon numerator $n^A_{1,5}=\epsilon(a,1,2,3,\,^\mu)\epsilon(b,4,5,6,\,_\mu)/2$, we find
\begin{equation}
	 \int _a \delta((a\cdot1)) \delta((a\cdot2)) \delta((a\cdot3))  \; I^{db}_{A, 1,5} [n^{db}_{A,1,5}]\rightarrow I_{box}\left(4,5,6,2\right)=\int_{b}\frac{\epsilon\left(b,4,5,6,2\right)}{\sqrt{2}\left(b\cdot 4\right)\left(b\cdot 5\right)\left(b\cdot 6\right)\left(b\cdot 2\right)}.
\end{equation}
Similarly, on the soft cut for $b$ we find the one-loop box $I_{box}\left(1,2,3,5\right)$. Readers can verify by themselves the similar construction of the numerators for $I^{db}_{B, i,j}$, and $I^{db}_{C, i}$ can produce other one-loop box in~\eqref{OneLoopAnsw}. Here we simply give their results:
\begin{equation}\label{NumeratorD}
	\begin{aligned}
	    &n^{db}_{A,i,j}=\frac{1}{2}\epsilon\left(a,i,i{+}1,i{+}2,\,^\mu\right)\epsilon\left(b, j{-}1,j,j{+}1,\,_\mu\right),\\
	    &n^{db}_{B,i,j}=\frac{1}{2}\epsilon\left(a,i,i{+}1,i{+}2,\,^\mu\right)\epsilon\left(b, j{-}1,j,j{+}1,\,_\mu\right),\\
		&n^{db}_{C,i}=\frac{1}{2}\epsilon\left(a,i,i{+}1,i{+}2,\,^\mu\right)\epsilon\left(b, i{+}4,i{+}5,i{+}6,\,_\mu\right)\,.
	\end{aligned}
\end{equation}

The remaining double boxes in eq.(\ref{SoftCuts}) will produce both one-loop box and triangles. This suggest that each topology contains two distinct numerators. Let us use $I^{db}_{D, i,j}$ as an illustration. For $I^{db}_{D, 1,5}$, introducing the following numerators
\begin{eqnarray}\label{dbnum1}
n^{db}_{D,1,5,b}\equiv\frac{1}{2}\epsilon(a,1,2,3,\,^\mu)\epsilon(b,3,5,1,\,_\mu),\ n^{db}_{D,1,5,t}\equiv\frac{1}{\sqrt{2}}\epsilon(a,1,2,3,\,b)\sqrt{(3\cdot5\cdot 1)}\,
\end{eqnarray}
where we use the subscript $b, t$ to indicate their fate under soft cuts, i.e., they reduce to one-loop box and triangles respectively. Indeed one can verify that: 
\begin{eqnarray}
	 \int _a \delta((a\cdot1)) \delta((a\cdot2)) \delta((a\cdot3))\; I^{db}_{D, 1,5}[n^{db}_{D,1,5,b}]\rightarrow\; I_{box}\left(3,5,1,2\right),\nonumber\\
	 \int _a \delta((a\cdot1)) \delta((a\cdot2)) \delta((a\cdot3))\; I^{db}_{D, 1,5}[ n^{db}_{D,1,5,t}] \rightarrow\; I_{tri}\left(1,3,5\right).
\end{eqnarray}
Similar definitions of numerators apply for topologies $I^{db}_{E, i,j}$ and $I^{db}_{F, i}$.

Finally, for the box-triangles $I^{bt}_{A, i,j}$ and $I^{bt}_{B, i}$, we expect their soft cuts to reproduce the one-loop triangles. However, this is insufficient to fix the form of their numerators uniquely. We will postpone the determination of $n^{ bt}_{A,i,j}$ and $n^{ bt}_{B,i}$ till the discussion of vanishing collinear soft where they can be fixed. For now, we will simply assume that their soft-cuts lead to the correct one-loop triangle.

Since each diagram under the soft cut is matched to the one-loop box and triangle integrals in a one-to-one fashion, the coefficient in front of each integrand is uniquely consequently determined by the one-loop amplitude. Summing all the soft-channels $i$, $i{+}1$, $i{+}2$, we write down the ``soft'' part of the two-loop integrand:
\begin{equation}\label{SoftAnsw}
\boxed{	\begin{aligned}
		&A_{8,soft}^{\text{$2$-loop}}=A_{8}^{\mathrm{tree}}\times \left[-I^{db}_{A,1,5}+I^{db}_{B,1,4}+I^{db}_{C,1}+\mathrm{cyclic}\right]\\
	&+\left[-\mathcal{D}_{1,3,5}I^{db}_{D,1,5}[n_{b}]-\mathcal{D}_{1,3,7}I^{db}_{D,1,7}[n_{b}]-\mathcal{D}_{3,5,7}I^{db}_{E, 1,5}[n_{b}]-\mathcal{D}_{1,5,7}I^{db}_{E, 1,7}[n_{b}]-\mathcal{D}_{4,6,8}I^{db}_{F,1}[n_{b}]\right.\\
	&\quad\;+\mathcal{B}_{1,3,5}I^{db}_{D,1,5}[n_{t}]+\mathcal{B}_{1,3,7}I^{db}_{D,1,7}[n_{t}]+\mathcal{B}_{3,5,7}I^{db}_{E, 1,5}[n_{t}]+\mathcal{B}_{1,5,7}I^{db}_{E, 1,7}[n_{t}]+\mathcal{B}_{4,6,8}I^{db}_{F,1}[n_{t}]\\
	&\left.\quad\;{+}\mathcal{B}_{2,4,6}I^{bt}_{A,1,5}+\mathcal{B}_{2,6,8}I^{bt}_{A,1,7}+\mathcal{B}_{2,4,8}I^{bt}_{B,1}+(-)\mathrm{cyclic}\right]\,\\
	\end{aligned} }
\end{equation}
where $(-)\mathrm{cyclic}$ indicates that one sum over cyclic permutations with alternating signs. From here on we only display numerators of integrals that are not unique. For example, there are two numerators for $I^{db}_{D,1,5}$ indicated by $I^{db}_{D,1,5}[n_{b,t}]$, whereas the numerator for $I^{db}_{A,1,5}$ is uniquely given in eq.(\ref{NumeratorD}), and we suppress its display.


\subsection{Maximal cuts}\label{sec:maximal_cut}

We now move on to the matching of maximal cuts, where for each cut there are four solutions labeled with $\{a_{\pm}, b_{\pm}\}$. There are two topologies entering each cut, so we aim to construct numerators such that one can have combinations that evaluate to 1 on one of the solutions, and zero for the remaining. As it turns out, from the point of view of individual topologies, it will be easier to identify numerators that evaluate to $\pm1$ on the cut. As long as the sign pattern is distinct for each numerator, their linear combination can lead to the desired result. We discuss two cuts separately.

\paragraph{Kissing-triangle maximal-cut $\mathcal{C}^i_{\includegraphics[scale=0.12]{Cut1Sym.pdf} }$}
We first consider the maximal cut corresponding to the kissing triangle:
$$\mathcal{C}^i_{\includegraphics[scale=0.12]{Cut1Sym.pdf} }:\; \vcenter{\hbox{\scalebox{0.8}{
\begin{tikzpicture}[x=0.75pt,y=0.75pt,yscale=-1,xscale=1]

\draw  [line width=1.2]  (201,299.5) -- (251,299.5) -- (251,349.5) -- (201,349.5) -- cycle ;
\draw  [line width=1.2]  (251,299.5) -- (301,299.5) -- (301,349.5) -- (251,349.5) -- cycle ;
\draw [color={rgb, 255:red, 0; green, 0; blue, 0 }  ,draw opacity=1 ][line width=1.2]    (301,299.5) -- (301,286.54) -- (301,279.25) ;
\draw [color={rgb, 255:red, 0; green, 0; blue, 0 }  ,draw opacity=1 ][line width=1.2]    (301,299.5) -- (313.96,299.5) -- (321.25,299.5) ;
\draw [color={rgb, 255:red, 0; green, 0; blue, 0 }  ,draw opacity=1 ][line width=1.2]    (301,349.5) -- (313.96,349.5) -- (321.25,349.5) ;
\draw [color={rgb, 255:red, 0; green, 0; blue, 0 }  ,draw opacity=1 ][line width=1.2]    (301,369.75) -- (301,356.79) -- (301,349.5) ;
\draw [color={rgb, 255:red, 0; green, 0; blue, 0 }  ,draw opacity=1 ][line width=1.2]    (180.75,349.5) -- (193.71,349.5) -- (201,349.5) ;
\draw [color={rgb, 255:red, 0; green, 0; blue, 0 }  ,draw opacity=1 ][line width=1.2]    (201,369.75) -- (201,356.79) -- (201,349.5) ;
\draw [color={rgb, 255:red, 0; green, 0; blue, 0 }  ,draw opacity=1 ][line width=1.2]    (201,299.5) -- (201,286.54) -- (201,279.25) ;
\draw [color={rgb, 255:red, 0; green, 0; blue, 0 }  ,draw opacity=1 ][line width=1.2]    (180.75,299.5) -- (193.71,299.5) -- (201,299.5) ;
\draw  [color={rgb, 255:red, 155; green, 155; blue, 155 }  ,draw opacity=1 ][fill={rgb, 255:red, 155; green, 155; blue, 155 }  ,fill opacity=1 ] (192.75,299.5) .. controls (192.75,294.94) and (196.44,291.25) .. (201,291.25) .. controls (205.56,291.25) and (209.25,294.94) .. (209.25,299.5) .. controls (209.25,304.06) and (205.56,307.75) .. (201,307.75) .. controls (196.44,307.75) and (192.75,304.06) .. (192.75,299.5) -- cycle ;
\draw  [color={rgb, 255:red, 155; green, 155; blue, 155 }  ,draw opacity=1 ][fill={rgb, 255:red, 155; green, 155; blue, 155 }  ,fill opacity=1 ] (192.75,349.5) .. controls (192.75,344.94) and (196.44,341.25) .. (201,341.25) .. controls (205.56,341.25) and (209.25,344.94) .. (209.25,349.5) .. controls (209.25,354.06) and (205.56,357.75) .. (201,357.75) .. controls (196.44,357.75) and (192.75,354.06) .. (192.75,349.5) -- cycle ;
\draw  [color={rgb, 255:red, 155; green, 155; blue, 155 }  ,draw opacity=1 ][fill={rgb, 255:red, 155; green, 155; blue, 155 }  ,fill opacity=1 ] (292.75,299.5) .. controls (292.75,294.94) and (296.44,291.25) .. (301,291.25) .. controls (305.56,291.25) and (309.25,294.94) .. (309.25,299.5) .. controls (309.25,304.06) and (305.56,307.75) .. (301,307.75) .. controls (296.44,307.75) and (292.75,304.06) .. (292.75,299.5) -- cycle ;
\draw  [color={rgb, 255:red, 155; green, 155; blue, 155 }  ,draw opacity=1 ][fill={rgb, 255:red, 155; green, 155; blue, 155 }  ,fill opacity=1 ] (292.75,349.5) .. controls (292.75,344.94) and (296.44,341.25) .. (301,341.25) .. controls (305.56,341.25) and (309.25,344.94) .. (309.25,349.5) .. controls (309.25,354.06) and (305.56,357.75) .. (301,357.75) .. controls (296.44,357.75) and (292.75,354.06) .. (292.75,349.5) -- cycle ;
\draw  [color={rgb, 255:red, 155; green, 155; blue, 155 }  ,draw opacity=1 ][fill={rgb, 255:red, 155; green, 155; blue, 155 }  ,fill opacity=1 ] (250.42,295.17) .. controls (257.46,295.17) and (263.17,308.37) .. (263.17,324.67) .. controls (263.17,340.96) and (257.46,354.17) .. (250.42,354.17) .. controls (243.38,354.17) and (237.67,340.96) .. (237.67,324.67) .. controls (237.67,308.37) and (243.38,295.17) .. (250.42,295.17) -- cycle ;

\draw (246.7,360.9) node [anchor=north west][inner sep=0.75pt]   [align=left] {$\displaystyle i$};
\draw (166,317.5) node [anchor=north west][inner sep=0.75pt]   [align=left] {$\displaystyle i+2$};
\draw (235.7,270.4) node [anchor=north west][inner sep=0.75pt]   [align=left] {$\displaystyle i+4$};

\end{tikzpicture}

}
}}$$
The cut involves two topologies, a double box $I^{db}_{G, i}$ and a kissing triangle $I^{kt}_{G, i}$.  Without loss of generality we take $i=1$ where the cut conditions are $(a\cdot1)=(a\cdot3)=(a\cdot5)=0$ and $(b\cdot5)=(b\cdot7)=(b\cdot1)=0$. This generates a Jacobian factors  $1/\sqrt{(1\cdot 3\cdot 5)}$ from $a$ and  $1/\sqrt{(5\cdot 7\cdot 1)}$ from $b$.

If we choose the numerator for the kissing triangle integral to be the inverse of the Jacobian factors, then the integral evaluates to one on all four solutions on the cut: 
\begin{equation}
I^{kt}_{G,i}:=I^{kt}_{G,i}[n^{kt}_i],\quad  n^{kt}_i\equiv \sqrt{(i{+}4\cdot i{+}2\cdot i)}\sqrt{(i{+}4\cdot i{-}2\cdot i)}\,.
\end{equation} 
For the double box integral, we then consider the numerators involves in the five-component Levi-Civita tensors, which evaluate to inverse Jacobian factors with $\pm1$ depending on the cut solution:  
\begin{equation}
	\begin{aligned}
		\left.\frac{\epsilon(a,1,3,5,\,^\mu)}{\sqrt{2}}\right|_{a=a_\pm}= \pm \sqrt{(5\cdot 3\cdot 1)} a_{\pm}^\mu\, ,\\
		\left.\frac{\epsilon(b,5,7,1,\,^\mu)}{\sqrt{2}}\right|_{b=b_\pm}= \pm \sqrt{(5\cdot 7\cdot 1)} b_{\pm}^\mu\, .
	\end{aligned}
\end{equation} 
One then only needs to ensure that they are contracted with appropriate vectors such that when both $a,b$ are on the cut, they evaluate to $(a\cdot b)$ to cancel the remaining uncut propagator. This leads to three potential numerators: 
\begin{equation}
	\begin{aligned}
		n^{db}_{G,1,a}:\quad &\frac{\epsilon(a,1,3,5,\,^\mu)\epsilon(b,5,7,1,\,_\mu)}{2}\\
		n^{db}_{G,1,b}:\quad &\frac{\sqrt{(1\cdot 3\cdot 5)} \epsilon(b,5,7,1,a)}{\sqrt{2}}\\
		n^{db}_{G,1,c}:\quad &\frac{\epsilon(a,1,3,5,b)\sqrt{(5\cdot 7\cdot 1)}}{\sqrt{2}}\,.
	\end{aligned}
\end{equation}
We summarize the residues of the integral under four solutions of  the  maximal cut below

\begin{equation}\label{eq:unity}
    \begin{aligned}
    		I^{db}_{G,1}[n_a]:& \left\{\begin{array}{c}
\left(a^{+},b^{+}\right){=}1\\
\left(a^{+},b^{-}\right){=-}1\\
\left(a^{-},b^{+}\right){=-}1\\
\left(a^{-},b^{-}\right){=}1
\end{array}\right.,\!\quad\quad I^{db}_{G,1}[n_b]:\left\{ \begin{array}{c}
\left(a^{+},b^{+}\right)=1\\
\left(a^{+},b^{-}\right)=-1\\
\left(a^{-},b^{+}\right)=1\\
\left(a^{-},b^{-}\right)=-1
\end{array}\right.,\\
I^{db}_{G,1}[n_c]:&\left\{ \begin{array}{c}
\left(a^{+},b^{+}\right)=1\\
\left(a^{+},b^{-}\right)=1\\
\left(a^{-},b^{+}\right)=-1\\
\left(a^{-},b^{-}\right)=-1
\end{array}\right.,\!\quad\quad \quad I^{kt}_{G,1}:\left\{ \begin{array}{c}
\left(a^{+},b^{+}\right)=1\\
\left(a^{+},b^{-}\right)=1\\
\left(a^{-},b^{+}\right)=1\\
\left(a^{-},b^{-}\right)=1
\end{array}\right. \,.
    \end{aligned}
\end{equation}
We see that the sign for the first integral depends on both $\{a_\pm, b_\pm\}$, the second only on $b_\pm$, the third on $a_\pm$ and the last is independent. As the sign pattern for each integral is distinct, requiring that the linear combination of integrals  $\alpha_1 I^{db}_{G,1}[n_a]+\beta_1 I^{db}_{G,1}[n_b]+\gamma_1 I^{db}_{G,1}[n_c]+\delta_1  I^{kt}_{G,1}$ produce a correct two-loop maximal cut, completely determines the coefficient to be: 
\begin{equation}
	\begin{aligned}
		&\alpha_1=\frac{1}{4}{\times}\frac{{}\mathcal{C}^1_{\includegraphics[scale=0.12]{Cut1Sym.pdf} }[a_+,b_+]{+}\mathcal{C}^1_{\includegraphics[scale=0.12]{Cut1Sym.pdf} }[a_+,b_-]{-}\mathcal{C}^1_{\includegraphics[scale=0.12]{Cut1Sym.pdf} }[a_-,b_+]{-}\mathcal{C}^1_{\includegraphics[scale=0.12]{Cut1Sym.pdf} }[a_-,b_-]}{ \sqrt{(1 \cdot 3\cdot 5)}\sqrt{(1\cdot 5\cdot 7)}}={}\mathcal{D}_{1,3,5}={}\mathcal{D}_{1,5,7}\\
		&\beta_1=\frac{1}{4}{\times}\frac{{-}\mathcal{C}^1_{\includegraphics[scale=0.12]{Cut1Sym.pdf} }[a_+,b_+]{+}\mathcal{C}^1_{\includegraphics[scale=0.12]{Cut1Sym.pdf} }[a_+,b_-]{+}\mathcal{C}^1_{\includegraphics[scale=0.12]{Cut1Sym.pdf} }[a_-,b_+]{-}\mathcal{C}^1_{\includegraphics[scale=0.12]{Cut1Sym.pdf} }[a_-,b_-]}{ \sqrt{(1 \cdot 3\cdot 5)}\sqrt{(1\cdot 5\cdot 7)}}={-}\mathcal{B}_{1,3,5}\\
		&\gamma_1=\frac{1}{4}{\times} \frac{{-}\mathcal{C}^1_{\includegraphics[scale=0.12]{Cut1Sym.pdf} }[a_+,b_+]{+}\mathcal{C}^1_{\includegraphics[scale=0.12]{Cut1Sym.pdf} }[a_+,b_-]{-}\mathcal{C}^1_{\includegraphics[scale=0.12]{Cut1Sym.pdf} }[a_-,b_+]{+}\mathcal{C}^1_{\includegraphics[scale=0.12]{Cut1Sym.pdf} }[a_-,b_-]}{ \sqrt{(1 \cdot 3\cdot 5)}\sqrt{(1\cdot 5\cdot 7)}}={-}\mathcal{B}_{1,5,7}\\
		&\delta_1=\frac{1}{4}{\times} \frac{{}\mathcal{C}^1_{\includegraphics[scale=0.12]{Cut1Sym.pdf} }[a_+,b_+]{+}\mathcal{C}^1_{\includegraphics[scale=0.12]{Cut1Sym.pdf} }[a_+,b_-]{+}\mathcal{C}^1_{\includegraphics[scale=0.12]{Cut1Sym.pdf} }[a_-,b_+]{+}\mathcal{C}^1_{\includegraphics[scale=0.12]{Cut1Sym.pdf} }[a_-,b_-]}{ \sqrt{(1 \cdot 3\cdot 5)}\sqrt{(1\cdot 5\cdot 7)}}\equiv {}\bar{\mathcal{D}}_{1,3,5}={}\bar{\mathcal{D}}_{1,5,7}.
	\end{aligned}
\end{equation}
The general solution of $\alpha_i I^{db}_{G,i}[n_a]+\beta_i I^{db}_{G,i}[n_b]+\gamma_i I^{db}_{G,i}[n_c]+\delta_i  I^{kt}_{G,i}$ for $i=1,2,3,4$ which produce correct two-loop maximal cut can be summarized as follows: 
\begin{equation}
	\begin{aligned}
		&\alpha_i=(-)^{i+1} \mathcal{D}_{i,i+2,i+4}=(-)^{i+1} \mathcal{D}_{i+4,i-2,i},\\
		&\beta_i=(-)^{i}\mathcal{B}_{i,i+2,i+4},\\
		&\gamma_i=(-)^{i} \mathcal{B}_{i+4,i-2,i},\\
		&\delta_i= \bar{\mathcal{D}}_{i,i+2,i+4}=\bar{\mathcal{D}}_{i+4,i-2,i}.
	\end{aligned}
\end{equation}

\paragraph{Box-triangle maximal-cut $\mathcal{C}^{i+1}_{\includegraphics[scale=0.13]{Cut2Sym.pdf} }$:}
We now move on to the box-triangle:
$$\mathcal{C}^{i+1}_{\includegraphics[scale=0.13]{Cut2Sym.pdf} }:\;\vcenter{\hbox{\scalebox{0.8}{
\begin{tikzpicture}[x=0.75pt,y=0.75pt,yscale=-1,xscale=1]

\draw  [line width=1.2]  (251,299.5) -- (301,299.5) -- (301,349.5) -- (251,349.5) -- cycle ;
\draw [color={rgb, 255:red, 0; green, 0; blue, 0 }  ,draw opacity=1 ][line width=1.2]    (209.5,324.33) -- (191.75,342.08) ;
\draw [color={rgb, 255:red, 0; green, 0; blue, 0 }  ,draw opacity=1 ][line width=1.2]    (209.5,324.33) -- (205.38,320.21) -- (191.75,306.58) ;
\draw [color={rgb, 255:red, 0; green, 0; blue, 0 }  ,draw opacity=1 ][line width=1.2]    (301,299.5) -- (301,286.54) -- (301,279.25) ;
\draw [color={rgb, 255:red, 0; green, 0; blue, 0 }  ,draw opacity=1 ][line width=1.2]    (301,299.5) -- (313.96,299.5) -- (321.25,299.5) ;
\draw [color={rgb, 255:red, 0; green, 0; blue, 0 }  ,draw opacity=1 ][line width=1.2]    (301,349.5) -- (313.96,349.5) -- (321.25,349.5) ;
\draw [color={rgb, 255:red, 0; green, 0; blue, 0 }  ,draw opacity=1 ][line width=1.2]    (301,369.75) -- (301,356.79) -- (301,349.5) ;
\draw [color={rgb, 255:red, 0; green, 0; blue, 0 }  ,draw opacity=1 ][line width=1.2]    (251,279.25) -- (251,292.21) -- (251,299.5) ;
\draw [color={rgb, 255:red, 0; green, 0; blue, 0 }  ,draw opacity=1 ][line width=1.2]    (251,349.5) -- (251,362.46) -- (251,369.75) ;
\draw  [line width=1.2]  (209.5,324.33) -- (251,299.5) -- (251,349.17) -- cycle ;
\draw  [color={rgb, 255:red, 155; green, 155; blue, 155 }  ,draw opacity=1 ][fill={rgb, 255:red, 155; green, 155; blue, 155 }  ,fill opacity=1 ] (201.25,324.33) .. controls (201.25,319.78) and (204.94,316.08) .. (209.5,316.08) .. controls (214.06,316.08) and (217.75,319.78) .. (217.75,324.33) .. controls (217.75,328.89) and (214.06,332.58) .. (209.5,332.58) .. controls (204.94,332.58) and (201.25,328.89) .. (201.25,324.33) -- cycle ;
\draw  [color={rgb, 255:red, 155; green, 155; blue, 155 }  ,draw opacity=1 ][fill={rgb, 255:red, 155; green, 155; blue, 155 }  ,fill opacity=1 ] (242.75,300.46) .. controls (242.75,295.91) and (246.44,292.21) .. (251,292.21) .. controls (255.56,292.21) and (259.25,295.91) .. (259.25,300.46) .. controls (259.25,305.02) and (255.56,308.71) .. (251,308.71) .. controls (246.44,308.71) and (242.75,305.02) .. (242.75,300.46) -- cycle ;
\draw  [color={rgb, 255:red, 155; green, 155; blue, 155 }  ,draw opacity=1 ][fill={rgb, 255:red, 155; green, 155; blue, 155 }  ,fill opacity=1 ] (292.75,299.5) .. controls (292.75,294.94) and (296.44,291.25) .. (301,291.25) .. controls (305.56,291.25) and (309.25,294.94) .. (309.25,299.5) .. controls (309.25,304.06) and (305.56,307.75) .. (301,307.75) .. controls (296.44,307.75) and (292.75,304.06) .. (292.75,299.5) -- cycle ;
\draw  [color={rgb, 255:red, 155; green, 155; blue, 155 }  ,draw opacity=1 ][fill={rgb, 255:red, 155; green, 155; blue, 155 }  ,fill opacity=1 ] (242.75,349.5) .. controls (242.75,344.94) and (246.44,341.25) .. (251,341.25) .. controls (255.56,341.25) and (259.25,344.94) .. (259.25,349.5) .. controls (259.25,354.06) and (255.56,357.75) .. (251,357.75) .. controls (246.44,357.75) and (242.75,354.06) .. (242.75,349.5) -- cycle ;
\draw  [color={rgb, 255:red, 155; green, 155; blue, 155 }  ,draw opacity=1 ][fill={rgb, 255:red, 155; green, 155; blue, 155 }  ,fill opacity=1 ] (292.75,349.5) .. controls (292.75,344.94) and (296.44,341.25) .. (301,341.25) .. controls (305.56,341.25) and (309.25,344.94) .. (309.25,349.5) .. controls (309.25,354.06) and (305.56,357.75) .. (301,357.75) .. controls (296.44,357.75) and (292.75,354.06) .. (292.75,349.5) -- cycle ;

\draw (162,316.5) node [anchor=north west][inner sep=0.75pt]   [align=left] {$\displaystyle i+1$};

\draw (340,345) node [anchor=north west][inner sep=0.75pt]   [align=left] {$\displaystyle .$};

\end{tikzpicture}
}
}}
$$

\noindent
In this case, we have a double box $I^{db}_{A,i}$ and a box-triangle $I^{bt}_{C,i}$ contributing. Since the numerator for $I^{db}_{A,i}$ has already been fixed by the soft-cut consideration, the task is to determine the numerator for the box-triangle integral. Let us consider $\mathcal{C}^2_{\includegraphics[scale=0.13]{Cut2Sym.pdf} }$, other cuts are related by cyclic permutations. An important distinction with the kissing-triangle $\mathcal{C}^i_{\includegraphics[scale=0.12]{Cut1Sym.pdf} }$, is that here the Jacobian factor from solving $(b\cdot 4)=(b\cdot 6)=(b\cdot 8)=0$ and $(a\cdot 1)=(a\cdot 3)=(a\cdot b_{\pm})=0$ is $1/(\sqrt{(4\cdot 6 \cdot 8)}\sqrt{(b_{\pm}\cdot 1\cdot 3)})$, i.e. it depends on the solution $b_\pm$. This suggests that the numerators will be more involved in reproducing the maximal cuts.

First the numerators for the double box integral of type $I^{db}_{F,1}$ was given in eq.(\ref{dbnum1}),
\begin{equation}
n^{db}_{F,1,b}=\epsilon(a,1,2,3,\,^\mu)\epsilon(b,4,6,8,_\mu)/2,\quad n^{db}_{F,1,t}=\epsilon(a,1,2,3,b)\sqrt{(4\cdot6)(6\cdot8)(8\cdot4)}/\sqrt{2}\,.
\end{equation}
The first numerator evaluates to $\pm1$ depending on $a_\pm$ and $b_\pm$, while the sign on the cut for $I^{db}_{F,1}[n_{t}]$ only depends on $a_\pm$. For the box-triangle $I^{bt}_{C,1}$ we will also introduce two sets of numerators $(n^{bt}_{C,i,\alpha}, n^{bt}_{C,i,\beta})$, so that the former evaluates to $+1$ on all four cut solutions and the other $\pm1$ depending on $b_\pm$. For $n^{bt}_{C,i,\alpha}$ the numerator can be further constrained by the cancellation of soft-collinear divergence. Here we simply give the result, leaving the details to appendix \ref{BoxTriaApp}: 
\begin{equation}\label{Boxtria1}
    \begin{split}
        n^{bt}_{C,i,\alpha}=&\frac{2\langle i\,i+1\rangle\langle i+3\,i+4\rangle\langle i+5\,i+6\rangle\langle i+2|p_{i,i+1}|i-1\rangle}{\epsilon(i,i+2,i+3,i+5,i-1)}\\
        &\times\left(\alpha_i (b\cdot i){+}\alpha_{i+2}(b\cdot i{+}2){+}\alpha_{i+3}(b\cdot i{+}3){+}\alpha_{i+5}(b\cdot i{+}5){+}\alpha_{i-1}(b\cdot i{-}1)\right)
    \end{split}
\end{equation}
where 
\begin{equation}
    \begin{split}
        &\alpha_i=(i+2\cdot i+5)(i+3\cdot i-1)-(i+2\cdot i-1)(i+3\cdot i+5),\\
        &\alpha_{i+2}=-(i\cdot i+3)(i+5\cdot i-1)+(i\cdot i+5)(i+3 \cdot i-1),\\
        &\alpha_{i+3}=(i\cdot i+2)(i+5\cdot i-1)-(i\cdot i+5)(i+2\cdot i-1),\\
        &\alpha_{i+5}=-(i\cdot i+2)(i+3\cdot i-1)+(i\cdot i+3)(i+2\cdot i-1),\\
        &\alpha_{i-1}=(i\cdot i+2)(i+3\cdot i+5)-(i\cdot i+3)(i+2\cdot i+5).
    \end{split}
\end{equation}
For $n^{bt}_{C,i,\beta}$ one has:
\begin{eqnarray}\label{Boxtria2}
n^{bt}_{C,i,\beta}&=&-\frac{1}{\sqrt{2}(i{+}1\cdot i{+}3)(i{+}1\cdot i{-}1)}\frac{\epsilon(i,i{+}1,i{+}2,i{+}3,i{-}1)}{\epsilon(i,i{+}2,i{+}3,i{-}3,i{-}1)}\sqrt{(i{+}3\cdot i{+}1 \cdot i{-}1)}\nonumber\\
&&\times\left[(i\cdot i{-}3)\epsilon(b,i{+}2,i{+}3,i{-}3,i{-}1){+}(i{-}1\cdot i{-}3)\epsilon(b,i,i{+}2,i{+}3,i{-}3)\right]\,,
\end{eqnarray}
one can numerically verify that with this numerator the maximal cut has a relative minus when evaluated on $b_\pm$. Note when generating kinematic data for the cut solutions, one should ensure that the resulting internal momenta can be identified with solutions to $\delta^{+}(\ell^2)$, i.e. $\ell^0>0$. We will discuss this in detail in appendix~\ref{AppKin}. 

We summarize the sign on the maximal cut as follows:
\begin{equation}\label{eq:unity2}
    \begin{aligned}
    		 I^{db}_{F,i}[n_b]:& \left\{\begin{array}{c}
\left(a^{+},b^{+}\right){=}1\\
\left(a^{+},b^{-}\right){=-}1\\
\left(a^{-},b^{+}\right){=-}1\\
\left(a^{-},b^{-}\right){=}1
\end{array}\right.,\!\quad\quad I^{bt}_{C,i}[n_{\beta}]:\left\{ \begin{array}{c}
\left(a^{+},b^{+}\right)=1\\
\left(a^{+},b^{-}\right)=-1\\
\left(a^{-},b^{+}\right)=1\\
\left(a^{-},b^{-}\right)=-1
\end{array}\right.,\\
I^{db}_{F,i}[n_t]:&\left\{ \begin{array}{c}
\left(a^{+},b^{+}\right)=1\\
\left(a^{+},b^{-}\right)=1\\
\left(a^{-},b^{+}\right)=-1\\
\left(a^{-},b^{-}\right)=-1
\end{array}\right.,\!\quad\quad I^{bt}_{C,i}[n_{\alpha}]\left\{ \begin{array}{c}
\left(a^{+},b^{+}\right)=1\\
\left(a^{+},b^{-}\right)=1\\
\left(a^{-},b^{+}\right)=1\\
\left(a^{-},b^{-}\right)=1
\end{array}\right. \,.
    \end{aligned}
\end{equation}

We can now solve for coefficients for the linear combination $\alpha_1 I^{db}_{F,1}[n_b]+\beta_1 I^{bt}_{C,1}[n_\beta]+\gamma_1 I^{db}_{F,1}[n_t]+\delta_1 I^{bt}_{C,1}[n_\alpha]$, the coefficient can be solved by the maximal cut:
\begin{equation}\label{eq:coefficient-maximal-cut}
	\begin{aligned}
		&\alpha_1=\frac{\mathcal{C}^2_{\includegraphics[scale=0.13]{Cut2Sym.pdf} }[a_+,b_+]-\mathcal{C}^2_{\includegraphics[scale=0.13]{Cut2Sym.pdf} }[a_-,b_+]}{4 \sqrt{(b_+ \cdot 1 \cdot 3)}\sqrt{(4\cdot 6\cdot 8)}}+\frac{\mathcal{C}^2_{\includegraphics[scale=0.13]{Cut2Sym.pdf} }[a_-,b_-]-\mathcal{C}^2_{\includegraphics[scale=0.13]{Cut2Sym.pdf} }[a_+,b_-]}{4 \sqrt{(b_- \cdot 1 \cdot 3)}\sqrt{(4\cdot 6\cdot 8)}}=-\mathcal{D}_{2,4,8}=-\mathcal{D}_{4,6,8}\\
		&\beta_1=-\frac{\mathcal{C}^2_{\includegraphics[scale=0.13]{Cut2Sym.pdf} }[a_+,b_+]+\mathcal{C}^2_{\includegraphics[scale=0.13]{Cut2Sym.pdf} }[a_-,b_+]]}{4 \sqrt{(b_+ \cdot 1 \cdot 3)}\sqrt{(4\cdot 6\cdot 8)}}+\frac{\mathcal{C}^2_{\includegraphics[scale=0.13]{Cut2Sym.pdf} }[a_+,b_-]+\mathcal{C}^2_{\includegraphics[scale=0.13]{Cut2Sym.pdf} }[a_-,b_-]}{4 \sqrt{(b_- \cdot 1 \cdot 3)}\sqrt{(4\cdot 6\cdot 8)}}=\mathcal{B}_{2,4,8}\\
		&\gamma_1=\frac{\mathcal{C}^2_{\includegraphics[scale=0.13]{Cut2Sym.pdf} }[a_+,b_+]-\mathcal{C}^2_{\includegraphics[scale=0.13]{Cut2Sym.pdf} }[a_-,b_+]}{4 \sqrt{(b_+ \cdot 1 \cdot 3)}\sqrt{(4\cdot 6\cdot 8)}}+\frac{\mathcal{C}^2_{\includegraphics[scale=0.13]{Cut2Sym.pdf} }[a_+,b_-]-\mathcal{C}^2_{\includegraphics[scale=0.13]{Cut2Sym.pdf} }[a_-,b_-]}{4 \sqrt{(b_- \cdot 1 \cdot 3)}\sqrt{(4\cdot 6\cdot 8)}}=\mathcal{B}_{4,6,8}\\
		&\delta_1=\frac{\mathcal{C}^2_{\includegraphics[scale=0.13]{Cut2Sym.pdf} }[a_+,b_+]+\mathcal{C}^2_{\includegraphics[scale=0.13]{Cut2Sym.pdf} }[a_-,b_+]}{4 \sqrt{(b_+ \cdot 1 \cdot 3)}\sqrt{(4\cdot 6\cdot 8)}}+\frac{\mathcal{C}^2_{\includegraphics[scale=0.13]{Cut2Sym.pdf} }[a_+,b_-]+\mathcal{C}^2_{\includegraphics[scale=0.13]{Cut2Sym.pdf} }[a_-,b_-]}{4 \sqrt{(b_- \cdot 1 \cdot 3)}\sqrt{(4\cdot 6\cdot 8)}}\equiv\bar{\mathcal{D}}_{2,4,8}=\bar{\mathcal{D}}_{4,6,8}\,.
	\end{aligned}
\end{equation}
We can obtain $i=2,3,\ldots,8$ in a similar way. The general solution of $\alpha_i I^{db}_{F,i}[n_b]+\beta_i I^{bt}_{C,i}[n_\beta]+\gamma_i I^{db}_{F,i}[n_t]+\delta_i I^{bt}_{C,i}[n_\alpha]$ for $i=1,2,\ldots,8$ which produces correct two-loop maximal cut $\mathcal{C}^{i+1}_{\includegraphics[scale=0.13]{Cut2Sym.pdf} }$ can be summarized as follows: 
\begin{equation}
	\begin{aligned}
		&\alpha_i=(-)^{i} \mathcal{D}_{i-1,i+1,i+3}=(-)^{i} \mathcal{D}_{i-3,i-1,i+3},\\
		&\beta_i=(-)^{i+1}\mathcal{B}_{i-1,i+1,i+3},\\
		&\gamma_i=(-)^{i+1} \mathcal{B}_{i-3,i-1,i+3},\\
		&\delta_i= \bar{\mathcal{D}}_{i-1,i+1,i+3}=\bar{\mathcal{D}}_{i-3,i-1,i+3}.
	\end{aligned}
\end{equation}

Thus we see that in addition to $A_{8,soft}^{\text{$2$-loop}}$, the matching to maximal cut requires us to introduce the following set of integrals: 
\begin{equation}\label{2LoopMaxCut}
\boxed{	\begin{aligned}
		A_{8,max-cut}^{\text{$2$-loop}}=&-\frac{1}{2} \mathcal{D}_{2,4,8} I^{db}_{G,1}[n_a]{+}\frac{1}{2} \mathcal{B}_{2,4,8} I^{db}_{G,1}[n_b]{+}\frac{1}{2} \mathcal{B}_{4,6,8} I^{db}_{G,1}[n_c]{+}\mathcal{B}_{2,4,8} I^{bt}_{C,1}[n_\beta]{+}(-)\mathrm{cyclic}\\
		&+\frac{1}{2} \bar{\mathcal{D}}_{2,4,8} I^{kt}_{G,1}+\bar{\mathcal{D}}_{2,4,8} I^{bt}_{C,1}[n_\alpha]+\mathrm{cyclic}\,.
	\end{aligned} }
\end{equation}
Note that the double box integral $ I^{db}_{F,1}$ was already included in $A_{8,soft}^{\text{$2$-loop}}$. One can check that its coefficient is fixed by maximal cuts,  $-\mathcal{D}_{4,6,8}$ for $I^{db}_{F,1}[n_b]$ and $\mathcal{B}_{4,6,8}$ for $I^{db}_{F,1}[n_t]$, is identical to that fixed by soft-cuts listed in  eq.(\ref{SoftAnsw}).  


\subsection{Collinear-soft constraints}\label{sec: collinear}
The implementation of soft-cut and maximal-cut constraints completely fixes all the double-box numerators and the numerator of one of the box-triangles, $I^{bt}_{C,i}$. The remaining box-triangles and double-triangle will be fixed by the requirement of vanishing collinear-soft limits. This corresponds to when the two-loop momentum becomes collinear with a massless external leg. In terms of region variables, the limit can be parameterized as:
\begin{equation} 
	y_a\rightarrow y_i+\tau_a y_{i+1}, \quad y_b\rightarrow y_i+\tau_b y_{i+1}
\end{equation}
where $p_i=y_{i{+}1}{-}y_i$ is the massless external momenta.  

As discussed previously, collinear-soft occurs for topologies $I^{db}_{A,i,j}$, $I^{db}_{F,i}$, $I^{bt}_{A,i,j}$, $I^{bt}_{B,i}$, $I^{bt}_{C,i}$, $I^{dt}_{i,i{+}2;i{+}3,i{-}3}$ and $I^{dt}_{i,i{+}2;i{+}3,i{+}1}$. We introduce the numerator for double-triangles as:
\begin{equation}\label{eq:numerator_bt}
I^{dt}_{i_1,i_2;i_3,i_4}:=I^{dt}_{i_1,i_2;i_3,i_4}[n^{dt}],\quad\quad n^{dt}=(i_1\cdot i_2)(i_3\cdot i_4)\,.
\end{equation}
As we will see, the collinear-soft limits of double boxes are canceled against double triangles, while triangle-boxes cancel among themselves. 
\paragraph{Double-box and double-triangles:}

Let us begin with the double box integral $I^{db}_{A,i,j}$. Choosing $j=i{+}4$, in the collinear limit where $y_a, y_b$ is sent to $(y_{i{+}2}, y_{i{+}3})$, the divergence becomes proportional to:
\begin{equation}
I^{db}_{A,i,i{+}4}\; \rightarrow \; -\frac{(i{+}3\cdot i{+}5)(i\cdot i{+}2)}{(a\cdot i)(b\cdot i{+}5)}\,.
\end{equation}
This can be directly cancelled by the double triangle $I^{dt}_{i,i{+}2;i{+}3,i{+}5}$. Thus the following combination is finite in the collinear-soft limit: 
\begin{equation}\label{DoubleTriangle1}
I^{db}_{A,i,i{+}4}+I^{dt}_{i,i{+}2;i{+}3,i{+}5}\,.
\end{equation}

Next, consider the double-box integral $I^{db}_{F,i}$, which has two potential collinear-soft limits,  $y_a, y_b$ is sent to $(y_{i}, y_{i{-}1})$ and $(y_{i{+}2}, y_{i{+}3})$. When dressed with the numerator $n^{db}_{F,i,t}$, the $\epsilon(a, i,i{+}1,i{+}2,b)$ vanishes in the limit, so we only need to consider $I^{db}_{F,i}[n_{b}]$. Similar to the previous case, the following combination with double triangles:
\begin{equation}\label{DoubleTriangle2}
I^{db}_{F,i}[n_{b}]+ I^{dt}_{i,i{+}2;i{+}3,i{-}1}-I^{dt}_{i,i{+}2;i{+}3,i{+}5}-I^{dt}_{i,i{+}2;i{+}5,i{-}1}\,,
\end{equation}
is finite in both collinear limits. 

\paragraph{Box triangles:} For the box-triangles, $I^{bt}_{A,i,j}$ and $I^{bt}_{B,i}$ were partially constrained by soft-cut conditions in subsection~\ref{sec:SoftCut}, it should reduce to one-loop triangle under soft-cut. This suggests that their numerator take the form 
\begin{equation}
\frac{\epsilon(a, i, i{+}1, i{+}2, X)}{\sqrt{2}(i{+}1\cdot X)}\sqrt{(I\cdot J\cdot K)}
\end{equation}
where $X$ is to be determined and $\sqrt{(I\cdot J\cdot K)}$ is the inverse Jacobian for the one-loop triangle. For $I^{bt}_{A,i,j}$, since there is only one-collinear soft regime, it can be rendered finite by choosing $X$. For $j=i{-}2$ we chose $X=i{-}1$, while for $j=i{+}4$ we chose $X=i{+}3$. 

For $I^{bt}_{B,i}$ there are two potential divergent regions, and choosing $X$ can only remove one of them. The remaining is to be canceled by $I^{bt}_{C,i}$. There are two numerators for $I^{bt}_{C,i}$, $n^{bt}_{C,i,\alpha}$ and $n^{bt}_{C,i, \beta}$ denoted in eq.(\ref{Boxtria1}) and eq.(\ref{Boxtria2}). Only the latter has non-trivial collinear-soft contributions for $y_a, y_b$ is sent to $(y_{i}, y_{i{-}1})$  
 \begin{equation}
 n^{bt}_{C,i, \beta}\rightarrow -\frac{(b\cdot i{-}3)\epsilon(i,i{+}1,i{+}2,i{+}3,i{-}1)\sqrt{(i{+}3\cdot i{-}3\cdot i{-}1)}}{\sqrt{2}(i{+}1\cdot i{+}1)(i{+}1\cdot i{-}1)}\,.
 \end{equation}
This cancels against $I^{bt}_{B,i}$ with a numerator:
\begin{equation}
I^{bt}_{B,i}:=I^{bt}_{B,i}[n^{bt}_{B,i}],\quad n^{bt}_{B,i}=\frac{\epsilon(a, i, i{+}1, i{+}2, i{+}3)}{\sqrt{2}(i{+}1\cdot i{+}3)}\sqrt{(i{+}1\cdot i{-}1\cdot i{-}3)}\,.
\end{equation}
Thus all the box-triangle coefficients are completely determined.


\subsection{Vanishing unphysical cuts}
So far using soft, maximal-cut, and collinear-soft constraints we have determined the double-box, box-triangle and a large class of double-triangle numerators. The remaining can be fixed by requiring the absence of unphysical cuts, in particular cuts with three-particle sub-amplitudes.  

We consider the triplet cut $(a\cdot i)=(a\cdot b)=(b\cdot i)=0$ in the double boxes, which separates out a three-point sub-amplitude and thus vanishes. Three-particle cut occurs in the topology of $I_{B,i,j}^{db}$, $I_{D,i,j}^{db}$, $I_{E,i,j}^{db}$,  and $I_{G,i,j}^{db}$. We add the double triangles $I^{dt}_{i,i+2;i+2,i}$, $I^{dt}_{i,i+4;i+4,i}$, $I_{i,i+2;i+2,i-2}^{dt}$ and $I_{i,1+2;i+2,i+4}^{dt}$ involving in this cut to cancel the ones in the double boxes.  The numerator for the double triangle is just as \eqref{eq:numerator_bt} we introduce in the previous subsection. As a consistency check,  we further verify that once we use the three-particle cut to fix the integrals, higher odd-particle cuts are automatically absent.

\paragraph{Double $\epsilon$-numerator:} We begin with canceling the tree-particle for the doubles boxes whose numerators are double $\epsilon$-tensor contract together. First, The double box $I_{B,i,j}^{db}$ (say $i=1,j=4$) have the triplet cut $(a\cdot 3)=(a\cdot b)=(b\cdot 3)=0$ cutting out the . It can be canceled by $I_{1,3;3,5}^{dt}$: 

\begin{equation}
\vcenter{\hbox{\scalebox{0.7}{


}}}
\end{equation}

For double box $I_{D,i,j}^{db}[n_b]$ (say $i=1,j=5$), there are two three-particle cuts, $(a\cdot 1)=(a\cdot b)=(b\cdot 1)=0$ and $(a\cdot 3)=(a\cdot b)=(b\cdot 3)=0$. The first one (red) can be canceled by adding $I^{dt}_{1,3;3,1}$, $I^{dt}_{1,3;5,1}$. The second one (blue) can be cancelled by  $I^{dt}_{1,3;3,1}$, $I^{dt}_{1,3;3,5}$. That is, 

\begin{equation}
\vcenter{\hbox{\scalebox{0.7}{
%

}}}
\end{equation}

For double box  $I_{E,i,j}^{db}[n_b]$ (say $i=1, j=7$), its has three-particle cut $(a\cdot 1)=(a\cdot b)=(b\cdot 1)=0$. Adding the double triangle $I_{1,3;5,1}^{dt}$ and $I_{1,3;7,1}^{dt}$,  we have

\begin{equation}
\vcenter{\hbox{\scalebox{0.7}{
%

}}}
\end{equation}

Finally, perhaps the most complicated case is the non-soft integral $I_{G,1}^{db}[n_a]$.  It has triple cut $(a\cdot 1)=(a\cdot b)=(b\cdot 1)=0$ and $(a\cdot 5)=(a\cdot b)=(b\cdot 5)=0$. The first one (red) is cancelled by $-I_{1,5;5,1}^{dt}+I_{1,5;7,1}^{dt}+I_{1,3;5,1}^{dt}-I_{1,3;7,1}^{dt}$. The second one (blue) is cancelled by $-I^{dt}_{1,5;5,1}+I^{dt}_{1,5;5,7}+I^{dt}_{3,5;5,1}-I^{dt}_{3,5;5,7}$. The combination is free of three-particle cut:

\begin{equation}
\vcenter{\hbox{\scalebox{0.7}{
%

}}}
\end{equation}

\paragraph{Single $\epsilon$-numerator:} The triple cut is rather trivial in the single $\epsilon$-numerator. The diagrams themselves vanish in the triple cut:

\begin{equation}
\vcenter{\hbox{\scalebox{0.7}{
%

}}}
\end{equation}


\subsection{The complete eight-point integrand}\label{sec:CompInt}
We have now fixed the two-loop eight-point integrand. It consists of three parts, reflecting the sequence of on-shell constraints used to fix the integrand: the soft-constructible, the maximal-cut, and double-triangle integrands. The latter is determined by the cancellation of soft-collinear divergences and vanishing cuts with three-point sub-amplitude. We write:
\begin{equation}
    \boxed{	A_{8}^{\text{$2$-loop}}=A_{8,soft}^{\text{$2$-loop}}+A_{8,max-cut}^{\text{$2$-loop}}+A_{8,tri}^{\text{$2$-loop}} .}
\end{equation}
The explicit form of $A_{8,soft}^{\text{$2$-loop}}$and $A_{8,max-cut}^{\text{$2$-loop}}$ are given in eq.(\ref{SoftAnsw}) and eq.(\ref{2LoopMaxCut}) respectively, while $A_{8,tri}^{\text{$2$-loop}}=A_{3\mathrm{pt}}^{dt}+A_{\mathrm{coll}}^{dt}$ is given as:
\begin{equation}\label{FinalTria}	\begin{aligned}
		A_{3\mathrm{pt}}^{dt}&= A_{8}^{\mathrm{tree}}\sum_{i=1}^8I_{i,i+2;i+2,i{+}4}^{dt}+ \sum_{i=1}^8 (-1)^i \mathcal{D}_{i,i+2,i+4} \bigg[ I^{dt}_{i,i+2;i+2,i}{-}I^{dt}_{i,i+2;i+2,i+4}{-}I^{dt}_{i,i+2;i+4,i}\\
		&\quad \ \  {+}I^{dt}_{i+2,i+4;i+4,i+2}{-}I^{dt}_{i+2,i+4;i+4,i}{-}I^{dt}_{i+2,i+4;i,i+2}{+}I^{dt}_{i+4,i-2;i,i+4} {-}I^{dt}_{i+4,i-2;i+2,i+4}\\
		&\quad  \ \ {+}I^{dt}_{i-2,i;i,i+4}-I^{dt}_{i-2,i;i,i+2}\bigg]{-}\sum_{i=1}^4 ({-}1)^i \mathcal{D}_{i,i+2,i+4}\bigg[{-}I^{dt}_{i,i+4;i+4,i}{+}I^{dt}_{i,i+4;i-2,i}\\
		&\quad \ \  {+}I^{dt}_{i,i+4;i+4,i-2}{+}I^{dt}_{i,i+2;i+4,i}{+}I^{dt}_{i+2,i+4;i+4,i}{-}I^{dt}_{i,i+2;i-2,i}{-}I^{dt}_{i+2,i+4;i+4,i-2}  \bigg].\\
		A_{\mathrm{coll}}^{dt}&= {-}A_{8}^{\mathrm{tree}}\sum_{i=1}^8I_{i,i+2;i+3,i+5}^{dt}{-}\sum_{i=1}^8 (-1)^i \mathcal{D}_{i,i+2,i+4}\left[I_{i{-}3,i{-}1;i,i+4}^{dt}{-} I_{i{-}3,i{-}1;i,i{+}2}^{dt}{-}I_{i{-}3,i{-}1;i+2,i+4}^{dt}\right].
	\end{aligned}
\end{equation}
In the above, we've organized the double triangles in subsets that manifest the vanishing three-point cut and soft-collinear divergences once combined with the double box integrals. 

The complete integrand is now determined. As further consistency checks, one can check whether other cuts with odd-point amplitude vanish. For example the following cut on $I^{db}_{E,5,3}$
\begin{center}
\scalebox{0.7}{
\begin{tikzpicture}[x=0.75pt,y=0.75pt,yscale=-1,xscale=1]

\draw  [line width=1.2]  (201,299.5) -- (251,299.5) -- (251,349.5) -- (201,349.5) -- cycle ;
\draw  [line width=1.2]  (251,299.5) -- (301,299.5) -- (301,349.5) -- (251,349.5) -- cycle ;
\draw [color={rgb, 255:red, 0; green, 0; blue, 0 }  ,draw opacity=1 ][line width=1.2]    (201,349.5) -- (183.25,367.25) ;
\draw [color={rgb, 255:red, 0; green, 0; blue, 0 }  ,draw opacity=1 ][line width=1.2]    (201,299.5) -- (196.88,295.38) -- (183.25,281.75) ;
\draw [color={rgb, 255:red, 0; green, 0; blue, 0 }  ,draw opacity=1 ][line width=1.2]    (301,299.5) -- (301,286.54) -- (301,279.25) ;
\draw [color={rgb, 255:red, 144; green, 19; blue, 254 }  ,draw opacity=1 ][line width=1.2]    (263.5,324) -- (249.5,324) -- (241.5,324) ;
\draw [color={rgb, 255:red, 0; green, 0; blue, 0 }  ,draw opacity=1 ][line width=1.2]    (301,299.5) -- (313.96,299.5) -- (321.25,299.5) ;
\draw [color={rgb, 255:red, 0; green, 0; blue, 0 }  ,draw opacity=1 ][line width=1.2]    (301,349.5) -- (313.96,349.5) -- (321.25,349.5) ;
\draw [color={rgb, 255:red, 0; green, 0; blue, 0 }  ,draw opacity=1 ][line width=1.2]    (301,369.75) -- (301,356.79) -- (301,349.5) ;
\draw [color={rgb, 255:red, 144; green, 19; blue, 254 }  ,draw opacity=1 ][line width=1.2]    (275.5,288) -- (275.5,302) -- (275.5,310) ;
\draw [color={rgb, 255:red, 0; green, 0; blue, 0 }  ,draw opacity=1 ][line width=1.2]    (251,299.5) -- (260.17,290.33) -- (265.32,285.18) ;
\draw [color={rgb, 255:red, 0; green, 0; blue, 0 }  ,draw opacity=1 ][line width=1.2]    (236.68,285.18) -- (245.85,294.35) -- (251,299.5) ;
\draw [color={rgb, 255:red, 144; green, 19; blue, 254 }  ,draw opacity=1 ][line width=1.2]    (223.5,288) -- (223.5,302) -- (223.5,310) ;

\draw (246.5,359.5) node [anchor=north west][inner sep=0.75pt]   [align=left] {$\displaystyle 5$};
\draw (314,316) node [anchor=north west][inner sep=0.75pt]   [align=left] {$\displaystyle 3$};
\draw (179,315.5) node [anchor=north west][inner sep=0.75pt]   [align=left] {$\displaystyle 6$};
\draw (218.5,265) node [anchor=north west][inner sep=0.75pt]   [align=left] {$\displaystyle 7$};
\draw (269.5,266) node [anchor=north west][inner sep=0.75pt]   [align=left] {$\displaystyle 1$};

\end{tikzpicture}}
\end{center}
contributes to a cut containing a five-point tree amplitude and must vanish. Indeed it cancels with contributions from 21 other double-box and double triangles. 

\begin{figure}[ht]   
\begin{center}
$\vcenter{\hbox{\scalebox{0.7}{
\begin{tikzpicture}[x=0.75pt,y=0.75pt,yscale=-1,xscale=1]

\draw  [line width=1.2]  (100,175.5) -- (150,175.5) -- (150,225.5) -- (100,225.5) -- cycle ;
\draw  [line width=1.2]  (150,175.5) -- (200,175.5) -- (200,225.5) -- (150,225.5) -- cycle ;
\draw [line width=1.2]    (100,225.5) -- (82.25,243.25) ;
\draw [color={rgb, 255:red, 0; green, 0; blue, 0 }  ,draw opacity=1 ][line width=1.2]    (100,175.5) -- (95.88,171.38) -- (82.25,157.75) ;
\draw [color={rgb, 255:red, 0; green, 0; blue, 0 }  ,draw opacity=1 ][line width=1.2]    (217.75,157.75) -- (200,175.5) ;
\draw [line width=1.2]    (217.75,243.25) -- (200,225.5) ;
\draw [line width=1.2]    (162.5,238.25) -- (150,225.5) ;
\draw [line width=1.2]    (150,225.5) -- (137.35,238.1) ;
\draw [color={rgb, 255:red, 144; green, 19; blue, 254 }  ,draw opacity=1 ][line width=1.2]    (122.5,165) -- (122.5,178.8) -- (122.5,183.4) ;
\draw [color={rgb, 255:red, 144; green, 19; blue, 254 }  ,draw opacity=1 ][line width=1.2]    (161,200.5) -- (147,200.5) -- (139,200.5) ;
\draw [line width=1.2]    (150,175.5) -- (137.5,162.75) ;
\draw [line width=1.2]    (162.65,162.9) -- (150,175.5) ;
\draw [color={rgb, 255:red, 144; green, 19; blue, 254 }  ,draw opacity=1 ][line width=1.2]    (175.5,218) -- (175.5,236.4) ;
\draw [color={rgb, 255:red, 144; green, 19; blue, 254 }  ,draw opacity=1 ][line width=1.2]    (174.1,165.4) -- (174.1,183.8) ;
\draw [color={rgb, 255:red, 144; green, 19; blue, 254 }  ,draw opacity=1 ][line width=1.2]    (120.5,216) -- (120.5,234.4) ;
\draw  [line width=1.2]  (532.6,176.5) -- (582.6,176.5) -- (582.6,226.5) -- (532.6,226.5) -- cycle ;
\draw  [line width=1.2]  (582.6,176.5) -- (632.6,176.5) -- (632.6,226.5) -- (582.6,226.5) -- cycle ;
\draw [color={rgb, 255:red, 0; green, 0; blue, 0 }  ,draw opacity=1 ][line width=1.2]    (532.6,226.5) -- (514.85,244.25) ;
\draw [color={rgb, 255:red, 0; green, 0; blue, 0 }  ,draw opacity=1 ][line width=1.2]    (532.6,176.5) -- (528.47,172.38) -- (514.85,158.75) ;
\draw [color={rgb, 255:red, 0; green, 0; blue, 0 }  ,draw opacity=1 ][line width=1.2]    (632.6,176.5) -- (632.6,163.54) -- (632.6,156.25) ;
\draw [color={rgb, 255:red, 144; green, 19; blue, 254 }  ,draw opacity=1 ][line width=1.2]    (593.6,201.5) -- (579.6,201.5) -- (571.6,201.5) ;
\draw [color={rgb, 255:red, 0; green, 0; blue, 0 }  ,draw opacity=1 ][line width=1.2]    (632.6,176.5) -- (645.56,176.5) -- (652.85,176.5) ;
\draw [color={rgb, 255:red, 0; green, 0; blue, 0 }  ,draw opacity=1 ][line width=1.2]    (632.6,226.5) -- (645.56,226.5) -- (652.85,226.5) ;
\draw [color={rgb, 255:red, 0; green, 0; blue, 0 }  ,draw opacity=1 ][line width=1.2]    (632.6,246.75) -- (632.6,233.79) -- (632.6,226.5) ;
\draw [color={rgb, 255:red, 0; green, 0; blue, 0 }  ,draw opacity=1 ][line width=1.2]    (582.6,176.5) -- (591.77,167.33) -- (596.92,162.18) ;
\draw [color={rgb, 255:red, 0; green, 0; blue, 0 }  ,draw opacity=1 ][line width=1.2]    (568.28,162.18) -- (577.45,171.35) -- (582.6,176.5) ;
\draw [color={rgb, 255:red, 144; green, 19; blue, 254 }  ,draw opacity=1 ][line width=1.2]    (645.1,201) -- (631.1,201) -- (623.1,201) ;
\draw [color={rgb, 255:red, 144; green, 19; blue, 254 }  ,draw opacity=1 ][line width=1.2]    (555.1,166.5) -- (555.1,180.3) -- (555.1,184.9) ;
\draw [color={rgb, 255:red, 144; green, 19; blue, 254 }  ,draw opacity=1 ][line width=1.2]    (610.6,168) -- (610.6,181.8) -- (610.6,186.4) ;
\draw [color={rgb, 255:red, 144; green, 19; blue, 254 }  ,draw opacity=1 ][line width=1.2]    (556.6,218) -- (556.6,231.8) -- (556.6,236.4) ;

\draw  [line width=1.2]  (317.4,175.2) -- (367.4,175.2) -- (367.4,225.2) -- (317.4,225.2) -- cycle ;
\draw  [line width=1.2]  (367.4,175.2) -- (417.4,175.2) -- (417.4,225.2) -- (367.4,225.2) -- cycle ;
\draw [color={rgb, 255:red, 0; green, 0; blue, 0 }  ,draw opacity=1 ][line width=1.2]    (317.4,225.2) -- (299.65,242.95) ;
\draw [color={rgb, 255:red, 0; green, 0; blue, 0 }  ,draw opacity=1 ][line width=1.2]    (317.4,175.2) -- (313.28,171.08) -- (299.65,157.45) ;
\draw [color={rgb, 255:red, 0; green, 0; blue, 0 }  ,draw opacity=1 ][line width=1.2]    (417.4,175.2) -- (417.4,162.24) -- (417.4,154.95) ;
\draw [color={rgb, 255:red, 144; green, 19; blue, 254 }  ,draw opacity=1 ][line width=1.2]    (429.9,199.7) -- (415.9,199.7) -- (407.9,199.7) ;
\draw [color={rgb, 255:red, 144; green, 19; blue, 254 }  ,draw opacity=1 ][line width=1.2]    (378.4,200.2) -- (364.4,200.2) -- (356.4,200.2) ;
\draw [color={rgb, 255:red, 0; green, 0; blue, 0 }  ,draw opacity=1 ][line width=1.2]    (417.4,175.2) -- (430.36,175.2) -- (437.65,175.2) ;
\draw [color={rgb, 255:red, 0; green, 0; blue, 0 }  ,draw opacity=1 ][line width=1.2]    (417.4,225.2) -- (430.36,225.2) -- (437.65,225.2) ;
\draw [color={rgb, 255:red, 0; green, 0; blue, 0 }  ,draw opacity=1 ][line width=1.2]    (417.4,245.45) -- (417.4,232.49) -- (417.4,225.2) ;
\draw [color={rgb, 255:red, 144; green, 19; blue, 254 }  ,draw opacity=1 ][line width=1.2]    (341.4,164.2) -- (341.4,178.2) -- (341.4,186.2) ;
\draw [color={rgb, 255:red, 0; green, 0; blue, 0 }  ,draw opacity=1 ][line width=1.2]    (366.9,226.38) -- (357.73,235.55) -- (352.58,240.7) ;
\draw [color={rgb, 255:red, 0; green, 0; blue, 0 }  ,draw opacity=1 ][line width=1.2]    (381.22,240.7) -- (372.05,231.53) -- (366.9,226.38) ;

\draw [color={rgb, 255:red, 144; green, 19; blue, 254 }  ,draw opacity=1 ][line width=1.2]    (342.4,214.2) -- (342.4,228.2) -- (342.4,236.2) ;
\draw [color={rgb, 255:red, 144; green, 19; blue, 254 }  ,draw opacity=1 ][line width=1.2]    (391.4,215.2) -- (391.4,229.2) -- (391.4,237.2) ;

\draw (118,234.5) node [anchor=north west][inner sep=0.75pt]   [align=left] {$\displaystyle 1$};
\draw (135,250) node [anchor=north west][inner sep=0.75pt]   [align=left] {$\displaystyle I_{C,1}^{db}$};
\draw (117,190.5) node [anchor=north west][inner sep=0.75pt]   [align=left] {$\displaystyle a$};
\draw (169.6,190.7) node [anchor=north west][inner sep=0.75pt]   [align=left] {$\displaystyle b$};
\draw (328.4,231.7) node [anchor=north west][inner sep=0.75pt]   [align=left] {$\displaystyle 1$};
\draw (438.8,192.5) node [anchor=north west][inner sep=0.75pt]   [align=left] {$\displaystyle 5$};
\draw (336.5,250) node [anchor=north west][inner sep=0.75pt]   [align=left] {$\displaystyle I_{E,1,5}^{db}\left[ n_{b}\right]$};
\draw (337.4,191.5) node [anchor=north west][inner sep=0.75pt]   [align=left] {$\displaystyle a$};
\draw (578.1,233) node [anchor=north west][inner sep=0.75pt]   [align=left] {$\displaystyle 1$};
\draw (649.
4,193.6) node [anchor=north west][inner sep=0.75pt]   [align=left] {$\displaystyle 7$};
\draw (549.3,250) node [anchor=north west][inner sep=0.75pt]   [align=left] {$\displaystyle I_{E,1,7}^{db}\left[ n_{b}\right]$};
\draw (550.2,191.5) node [anchor=north west][inner sep=0.75pt]   [align=left] {$\displaystyle a$};
\draw (606.4,192.5) node [anchor=north west][inner sep=0.75pt]   [align=left] {$\displaystyle b$};
\draw (52.5,190) node [anchor=north west][inner sep=0.75pt]  [font=\Large] [align=left] {$ A_{8}$};
\draw (251.5,190) node [anchor=north west][inner sep=0.75pt]  [font=\Large] [align=left] {$\mathcal{ D}_{3,5,7}$};
\draw (221.5,190) node [anchor=north west][inner sep=0.75pt]  [font=\Large] [align=left] {$\displaystyle +$};
\draw (456.6,190) node [anchor=north west][inner sep=0.75pt]  [font=\Large] [align=left] {$\displaystyle +$};
\draw (672.6,190) node [anchor=north west][inner sep=0.75pt]  [font=\Large] [align=left] {$\displaystyle =0$};
\draw (480.5,190) node [anchor=north west][inner sep=0.75pt]  [font=\Large] [align=left] {$\mathcal{ D}_{1,5,7}$};

\end{tikzpicture}

}}}$
\end{center}
\caption{The integral $I_{C,1}^{db}$, $I_{E,1,5}^{db}\left[n_b\right]$ and $I_{E,1,7}^{db}\left[n_b\right]$ will contribute to the elliptic cut $(a\cdot 1)=(a\cdot 3)=(a\cdot b)=(b\cdot 5)=(b\cdot 7)=0$. Due to isolating five-point sub-amplitude, the elliptic function will cancel in $A_8 I_{C,1}^{db}+\mathcal{D}_{1,5,7} I_{E,1,5}^{db}\left[n_b\right]+\mathcal{D}_{3,5,7} I_{E,1,7}^{db}\left[n_b\right]$. } \label{fig:ellptic_cut_1}
\end{figure}

The cancellation of such cuts actually plays an important role in the simplification of the integrated result. Consider cutting two more propagators leaving one degree of freedom of loop variables unfixed as in fig.\ref{fig:ellptic_cut_1}. This cut introduces a Jacobian factor of the form $J=\sqrt{Q(z)}$ . If the square root cannot be rationalized, then one has an elliptic integral. For example the cut $(a\cdot 1)=(a\cdot 3)=(a\cdot b)=(b\cdot 5)=(b\cdot 7)=0$ is elliptical. However, since such a cut isolates a five-point amplitude, it must vanish. Indeed the combination of $A_8 I_{C,1}^{db}+\mathcal{D}_{3,5,7} I_{E,1,5}^{db}\left[n_b\right]+\mathcal{D}_{1,5,7} I_{E,1,7}^{db}\left[n_b\right]$ under the cut is zero, illustrated in fig.\ref{fig:ellptic_cut_1}.

\begin{figure}[H]   
\begin{center}
$\vcenter{\hbox{\scalebox{0.7}{	
\begin{tikzpicture}[x=0.75pt,y=0.75pt,yscale=-1,xscale=1]

\draw  [line width=1.2]  (52,351) -- (102,351) -- (102,401) -- (52,401) -- cycle ;
\draw  [line width=1.2]  (102,351) -- (152,351) -- (152,401) -- (102,401) -- cycle ;
\draw [color={rgb, 255:red, 0; green, 0; blue, 0 }  ,draw opacity=1 ][line width=1.2]    (152,351) -- (152,338.04) -- (152,330.75) ;
\draw [color={rgb, 255:red, 144; green, 19; blue, 254 }  ,draw opacity=1 ][line width=1.2]    (62.5,375.5) -- (48.5,375.5) -- (44,375.5) ;
\draw [color={rgb, 255:red, 144; green, 19; blue, 254 }  ,draw opacity=1 ][line width=1.2]    (162.67,377) -- (151.5,377) -- (143.5,377) ;
\draw [color={rgb, 255:red, 0; green, 0; blue, 0 }  ,draw opacity=1 ][line width=1.2]    (152,351) -- (164.96,351) -- (172.25,351) ;
\draw [color={rgb, 255:red, 0; green, 0; blue, 0 }  ,draw opacity=1 ][line width=1.2]    (152,401) -- (164.96,401) -- (172.25,401) ;
\draw [color={rgb, 255:red, 0; green, 0; blue, 0 }  ,draw opacity=1 ][line width=1.2]    (152,421.25) -- (152,408.29) -- (152,401) ;
\draw  [line width=1.2]  (251,352) -- (301,352) -- (301,402) -- (251,402) -- cycle ;
\draw  [line width=1.2]  (301,352) -- (351,352) -- (351,402) -- (301,402) -- cycle ;
\draw [color={rgb, 255:red, 0; green, 0; blue, 0 }  ,draw opacity=1 ][line width=1.2]    (251,402) -- (233.25,419.75) ;
\draw [color={rgb, 255:red, 0; green, 0; blue, 0 }  ,draw opacity=1 ][line width=1.2]    (251,352) -- (246.88,347.88) -- (233.25,334.25) ;
\draw [color={rgb, 255:red, 0; green, 0; blue, 0 }  ,draw opacity=1 ][line width=1.2]    (351,352) -- (351,339.04) -- (351,331.75) ;
\draw [color={rgb, 255:red, 0; green, 0; blue, 0 }  ,draw opacity=1 ][line width=1.2]    (351,352) -- (363.96,352) -- (371.25,352) ;
\draw [color={rgb, 255:red, 0; green, 0; blue, 0 }  ,draw opacity=1 ][line width=1.2]    (351,402) -- (363.96,402) -- (371.25,402) ;
\draw [color={rgb, 255:red, 0; green, 0; blue, 0 }  ,draw opacity=1 ][line width=1.2]    (351,422.25) -- (351,409.29) -- (351,402) ;
\draw [color={rgb, 255:red, 0; green, 0; blue, 0 }  ,draw opacity=1 ][line width=1.2]    (301,352) -- (310.17,342.83) -- (315.32,337.68) ;
\draw [color={rgb, 255:red, 0; green, 0; blue, 0 }  ,draw opacity=1 ][line width=1.2]    (286.68,337.68) -- (295.85,346.85) -- (301,352) ;
\draw [color={rgb, 255:red, 0; green, 0; blue, 0 }  ,draw opacity=1 ][line width=1.2]    (31.75,351) -- (44.71,351) -- (52,351) ;
\draw [color={rgb, 255:red, 0; green, 0; blue, 0 }  ,draw opacity=1 ][line width=1.2]    (52,351) -- (52,338.04) -- (52,330.75) ;
\draw [color={rgb, 255:red, 0; green, 0; blue, 0 }  ,draw opacity=1 ][line width=1.2]    (52,421.25) -- (52,408.29) -- (52,401) ;
\draw [color={rgb, 255:red, 0; green, 0; blue, 0 }  ,draw opacity=1 ][line width=1.2]    (31.75,401) -- (44.71,401) -- (52,401) ;
\draw [color={rgb, 255:red, 144; green, 19; blue, 254 }  ,draw opacity=1 ][line width=1.2]    (343.25,376.58) -- (354.92,376.58) -- (361.58,376.58) ;
\draw [color={rgb, 255:red, 144; green, 19; blue, 254 }  ,draw opacity=1 ][line width=1.2]    (291.58,376.25) -- (303.25,376.25) -- (309.92,376.25) ;
\draw [color={rgb, 255:red, 144; green, 19; blue, 254 }  ,draw opacity=1 ][line width=1.2]    (93.25,376.25) -- (104.92,376.25) -- (111.58,376.25) ;
\draw  [line width=1.2]  (51,226) -- (101,226) -- (101,276) -- (51,276) -- cycle ;
\draw  [line width=1.2]  (101,226) -- (151,226) -- (151,276) -- (101,276) -- cycle ;
\draw [color={rgb, 255:red, 0; green, 0; blue, 0 }  ,draw opacity=1 ][line width=1.2]    (151,226) -- (151,213.04) -- (151,205.75) ;
\draw [color={rgb, 255:red, 144; green, 19; blue, 254 }  ,draw opacity=1 ][line width=1.2]    (61.5,250.5) -- (47.5,250.5) -- (43,250.5) ;
\draw [color={rgb, 255:red, 144; green, 19; blue, 254 }  ,draw opacity=1 ][line width=1.2]    (161.67,252) -- (150.5,252) -- (142.5,252) ;
\draw [color={rgb, 255:red, 0; green, 0; blue, 0 }  ,draw opacity=1 ][line width=1.2]    (151,226) -- (163.96,226) -- (171.25,226) ;
\draw [color={rgb, 255:red, 0; green, 0; blue, 0 }  ,draw opacity=1 ][line width=1.2]    (151,276) -- (163.96,276) -- (171.25,276) ;
\draw [color={rgb, 255:red, 0; green, 0; blue, 0 }  ,draw opacity=1 ][line width=1.2]    (151,296.25) -- (151,283.29) -- (151,276) ;
\draw  [line width=1.2]  (250,227) -- (300,227) -- (300,277) -- (250,277) -- cycle ;
\draw  [line width=1.2]  (300,227) -- (350,227) -- (350,277) -- (300,277) -- cycle ;
\draw [color={rgb, 255:red, 0; green, 0; blue, 0 }  ,draw opacity=1 ][line width=1.2]    (250,277) -- (232.25,294.75) ;
\draw [color={rgb, 255:red, 0; green, 0; blue, 0 }  ,draw opacity=1 ][line width=1.2]    (250,227) -- (245.88,222.88) -- (232.25,209.25) ;
\draw [color={rgb, 255:red, 0; green, 0; blue, 0 }  ,draw opacity=1 ][line width=1.2]    (350,227) -- (350,214.04) -- (350,206.75) ;
\draw [color={rgb, 255:red, 0; green, 0; blue, 0 }  ,draw opacity=1 ][line width=1.2]    (350,227) -- (362.96,227) -- (370.25,227) ;
\draw [color={rgb, 255:red, 0; green, 0; blue, 0 }  ,draw opacity=1 ][line width=1.2]    (350,277) -- (362.96,277) -- (370.25,277) ;
\draw [color={rgb, 255:red, 0; green, 0; blue, 0 }  ,draw opacity=1 ][line width=1.2]    (350,297.25) -- (350,284.29) -- (350,277) ;
\draw [color={rgb, 255:red, 0; green, 0; blue, 0 }  ,draw opacity=1 ][line width=1.2]    (285.68,291.32) -- (294.85,282.15) -- (300,277) ;
\draw [color={rgb, 255:red, 0; green, 0; blue, 0 }  ,draw opacity=1 ][line width=1.2]    (300,277) -- (309.17,286.17) -- (314.32,291.32) ;
\draw [color={rgb, 255:red, 0; green, 0; blue, 0 }  ,draw opacity=1 ][line width=1.2]    (30.75,226) -- (43.71,226) -- (51,226) ;
\draw [color={rgb, 255:red, 0; green, 0; blue, 0 }  ,draw opacity=1 ][line width=1.2]    (51,226) -- (51,213.04) -- (51,205.75) ;
\draw [color={rgb, 255:red, 0; green, 0; blue, 0 }  ,draw opacity=1 ][line width=1.2]    (51,296.25) -- (51,283.29) -- (51,276) ;
\draw [color={rgb, 255:red, 0; green, 0; blue, 0 }  ,draw opacity=1 ][line width=1.2]    (30.75,276) -- (43.71,276) -- (51,276) ;
\draw [color={rgb, 255:red, 208; green, 2; blue, 27 }  ,draw opacity=1 ][line width=1.2]    (271.83,287) -- (271.83,275.33) -- (271.83,268.67) ;
\draw [color={rgb, 255:red, 144; green, 19; blue, 254 }  ,draw opacity=1 ][line width=1.2]    (342.25,251.58) -- (353.92,251.58) -- (360.58,251.58) ;
\draw [color={rgb, 255:red, 144; green, 19; blue, 254 }  ,draw opacity=1 ][line width=1.2]    (290.58,251.25) -- (302.25,251.25) -- (308.92,251.25) ;
\draw [color={rgb, 255:red, 144; green, 19; blue, 254 }  ,draw opacity=1 ][line width=1.2]    (92.25,251.25) -- (103.92,251.25) -- (110.58,251.25) ;
\draw [color={rgb, 255:red, 144; green, 19; blue, 254 }  ,draw opacity=1 ][line width=1.2]    (124.5,286.33) -- (124.5,274.67) -- (124.5,268) ;
\draw [color={rgb, 255:red, 144; green, 19; blue, 254 }  ,draw opacity=1 ][line width=1.2]    (74.5,235.33) -- (74.5,223.67) -- (74.5,217) ;
\draw [color={rgb, 255:red, 144; green, 19; blue, 254 }  ,draw opacity=1 ][line width=1.2]    (273.5,236.33) -- (273.5,224.67) -- (273.5,218) ;
\draw [color={rgb, 255:red, 144; green, 19; blue, 254 }  ,draw opacity=1 ][line width=1.2]    (324.5,286.33) -- (324.5,274.67) -- (324.5,268) ;
\draw [color={rgb, 255:red, 144; green, 19; blue, 254 }  ,draw opacity=1 ][line width=1.2]    (75.83,410) -- (75.83,398.33) -- (75.83,391.67) ;
\draw [color={rgb, 255:red, 144; green, 19; blue, 254 }  ,draw opacity=1 ][line width=1.2]    (126.83,360) -- (126.83,348.33) -- (126.83,341.67) ;
\draw [color={rgb, 255:red, 144; green, 19; blue, 254 }  ,draw opacity=1 ][line width=1.2]    (271.83,361.33) -- (271.83,349.67) -- (271.83,343) ;
\draw [color={rgb, 255:red, 144; green, 19; blue, 254 }  ,draw opacity=1 ][line width=1.2]    (272.83,411.33) -- (272.83,399.67) -- (272.83,393) ;
\draw [color={rgb, 255:red, 144; green, 19; blue, 254 }  ,draw opacity=1 ][line width=1.2]    (328.83,360.33) -- (328.83,348.67) -- (328.83,342) ;

\draw (266.17,286.83) node [anchor=north west][inner sep=0.75pt]   [align=left] {$\displaystyle 3$};
\draw (373.5,243.33) node [anchor=north west][inner sep=0.75pt]   [align=left] {$\displaystyle 7$};
\draw (263.5,306) node [anchor=north west][inner sep=0.75pt]   [align=left] {$\displaystyle I_{E,3,7}^{db}\left[ n_{b}\right]$};
\draw (391,243.5) node [anchor=north west][inner sep=0.75pt]  [font=\Large] [align=left] {$\displaystyle =0,$};
\draw (96.5,282.5) node [anchor=north west][inner sep=0.75pt]   [align=left] {$\displaystyle 1$};
\draw (95.83,204.67) node [anchor=north west][inner sep=0.75pt]   [align=left] {$\displaystyle 5$};
\draw (73.5,308) node [anchor=north west][inner sep=0.75pt]   [align=left] {$\displaystyle I_{G,1}^{db}[ n_{a}]$};
\draw (191,243.5) node [anchor=north west][inner sep=0.75pt]  [font=\Large] [align=left] {$\displaystyle +$};
\draw (70,244) node [anchor=north west][inner sep=0.75pt]   [align=left] {$\displaystyle a$};
\draw (267,244) node [anchor=north west][inner sep=0.75pt]   [align=left] {$\displaystyle a$};
\draw (320,243) node [anchor=north west][inner sep=0.75pt]   [align=left] {$\displaystyle b$};
\draw (120.5,243) node [anchor=north west][inner sep=0.75pt]   [align=left] {$\displaystyle b$};
\draw (296.5,406.5) node [anchor=north west][inner sep=0.75pt]   [align=left] {$\displaystyle 5$};
\draw (371.83,368.67) node [anchor=north west][inner sep=0.75pt]   [align=left] {$\displaystyle 7$};
\draw (265.5,430) node [anchor=north west][inner sep=0.75pt]   [align=left] {$\displaystyle I_{E,5,7}^{db}\left[ n_{b}\right]$};
\draw (393,366.5) node [anchor=north west][inner sep=0.75pt]  [font=\Large] [align=left] {$\displaystyle =0.$};
\draw (97.5,406.5) node [anchor=north west][inner sep=0.75pt]   [align=left] {$\displaystyle 1$};
\draw (96.83,329.67) node [anchor=north west][inner sep=0.75pt]   [align=left] {$\displaystyle 5$};
\draw (76.5,431) node [anchor=north west][inner sep=0.75pt]   [align=left] {$\displaystyle I_{G,1}^{db}[ n_{a}]$};
\draw (192,366.5) node [anchor=north west][inner sep=0.75pt]  [font=\Large] [align=left] {$\displaystyle +$};
\draw (72,370) node [anchor=north west][inner sep=0.75pt]   [align=left] {$\displaystyle a$};
\draw (268,369) node [anchor=north west][inner sep=0.75pt]   [align=left] {$\displaystyle b$};
\draw (122,369) node [anchor=north west][inner sep=0.75pt]   [align=left] {$\displaystyle b$};
\draw (322,370) node [anchor=north west][inner sep=0.75pt]   [align=left] {$\displaystyle a$};

\end{tikzpicture}

}}}$
\end{center}
\caption{The integral $I_{G,1}^{db}\left[n_a\right]$ with contribute to the two elliptic cuts:  $(a\cdot 3)=(a\cdot 5)=(a\cdot b)=(b\cdot 7)=(b\cdot 1)=0$ and $(a\cdot 1)=(a\cdot 3)=(a\cdot b)=(b\cdot 5)=(b\cdot 7)=0$. These two cuts will isolate five-point sub amplitude so the elliptics will cancel with other integral.} \label{fig:ellptic_cut_2} 
\end{figure}

Some integrals contribute to more than one elliptic cut. For example, the integral $I_{G,1}^{db}\left[n_a\right]$ will contribute the two elliptic cut   $(a\cdot 3)=(a\cdot 5)=(a\cdot b)=(b\cdot 7)=(b\cdot 1)=0$ and $(a\cdot 1)=(a\cdot 3)=(a\cdot b)=(b\cdot 5)=(b\cdot 7)=0$.  Both of the cuts will isolate the five-point sub-amplitude so they will vanish in the end. In the first cut, the $I_{G,1}^{db}\left[n_a\right]+I_{E,3,7}^{db}\left[n_b\right]=0$ while in the second cut,  $I_{G,1}^{db}\left[n_a\right]+I_{E,5,7}^{db}\left[n_b\right]=0$ as shown in fig. \ref{fig:ellptic_cut_2}.

The above analysis tells us that while each individual integral will contain elliptical pieces, they will cancel when combined. This is indeed what we find in the next section.




\section{The computation of two-loop integrals}\label{Section5}

We now proceed to integrate the eight-point two-loop amplitude. Many of the integrands are kinematically equivalent to ones already computed for the six-point two-loop, and thus their result can be directly imported. For example, the double triangles at eight-point do not add the new topology, while $I^{db}_{B,i,j}$ is kinematically equivalent to the ``$I^{\rm crab}$'' integral in~\cite{Caron-Huot:2012sos}. The kissing triangles  $I_{G,i}^{kt}$  are the new topologies that can be straightforwardly integrated:  
\begin{equation}
	\begin{aligned}
		I_{G,1}^{kt}&=\int_{a,b}\frac{\sqrt{(1\cdot 3\cdot 5)}\sqrt{(5\cdot 7\cdot 1)}}{(a\cdot 1)(a\cdot 3)(a\cdot 5)(b\cdot 5)(b\cdot 7)(b\cdot 1)}=\frac{\pi^2}{4} \mathrm{sgn}_c [(1{\cdot} 3)] \mathrm{sgn}_c [(3{\cdot} 5)] \mathrm{sgn}_c [(5{\cdot} 7)] \mathrm{sgn}_c [(7{\cdot} 1)]\,.
	\end{aligned}
\end{equation}
Here we will focus on the remaining double-box and box-triangles.

\subsection{Generalities: kinematics, regularizations and all that}
Before proceeding, let us first recall the kinematics for eight-point amplitudes in ABJM. It is easy to see that we have twelve multiplicatively independent (dual) conformal cross-ratios, which we denote as:
\begin{equation}\label{crossratios}
	\begin{aligned}
		&u_{i}:=\frac{(i\cdot i+2)(i+3\cdot i+7)}{(i\cdot i+3)(i+2\cdot i+7)}, \qquad i=1, \ 2, \cdots, \ 8\\
		&v_{i}:=\frac{(i\cdot i+3)(i+4\cdot i+7)}{(i\cdot i+4)(i+3\cdot i+7)}, \qquad i=1,\ 2,\ 3,\ 4.
	\end{aligned}
\end{equation}
General cross-ratios are monomials of $u_i$ and $v_i$ variables, and as a shorthand notation, we define the product of $u$ and $v$:
\begin{equation}
    \begin{aligned}
        u_{I}:=\prod_{i\in I}u_i \quad \text{and} \quad  v_{I}:=\prod_{i\in I}v_i\,.
    \end{aligned}
\end{equation}

Note that these $12$ cross-ratios are not {\it functionally independent} as they satisfy $6$ Gram-determinant constraints, reflecting the fact that all the dual points live in $D=3$. This is equivalent to the requirement that embedding variables $y_{i=1,2,\dots, n}$ live in $D{+}2=5$ dimensions, or that any $6$ of them must be linearly dependent. They amount to the conditions that any $6\times 6$ Gram determinant of the form  $G_{i_1, \dots, i_6}:=\det \left[ (a\cdot b)|_{a,b=i_1, \dots, i_6}\right]$ must vanish. 
A convenient parametrization can be obtained using momentum twistor variables~\cite{Hodges:2009hk} subject to $D=3$ conditions~\cite{Elvang:2014fja}: as discussed in~\cite{He:2021eec}, we first parametrize the $n=8$ DCI kinematics in $D=4$ using $9$ variables from a {\it quiver} of $G_+(4,8)$,  and the reduction to $D=3$ amounts to certain ``folding” of the quiver which has $6$ variables. One such parametrization in $D=4$ gives a $4\times 8$ matrix ${\bf Z}^I_{i=1,2, \dots, 8}$ (for $I=1, 2, 3, 4$)~\cite{He:2021non}:
\begin{equation}
{\bf Z}=\left(
\begin{array}{cccccccc}
 1 & g_{7,8,9} & f_7 g'_{8,4,9,5} & f_4 f_7 f_8 g_{9,5,1} & f_1 f_4 f_5 f_7 f_8 f_9 & 0 & 0 & 0 \\
 0 & 1 & g_{4,5,6} & f_4 g'_{5,1,6,2} & f_1 f_4 f_5 \left(f_2 f_6+f_6+1\right) & f_1 f_2 f_4 f_5 f_6 & 0 & 0 \\
 0 & 0 & 1 & g_{1,2,3} & f_1 \left(f_3 f_2+f_2+1\right) & f_1 f_2 \left(f_3+1\right) & f_1 f_2 f_3 & 0 \\
 0 & 0 & 0 & 1 & 1 & 1 & 1 & 1 \\
\end{array}
\right)
\end{equation}
where we introduce shorthand notation $g_{i,j,k}:=f_i f_j f_k+f_i f_j+f_i+1$ and $g'_{i,j,k,l}:=f_i \left(f_j f_k f_l+f_j f_k+f_j+f_k+1\right)+1$. By reducing to $D=3$, we have $3$ additional constraints $$
f_7=\frac{1}{f_1 \left(f_3 f_2+f_2+1\right)}, \quad f_8=\frac{g_{1,2,3}}{f_2 \left(f_3+1\right)}, \quad  f_9=\frac{f_3 f_2+f_2+1}{f_3},$$ thus our eight-point kinematics is parametrized by $f_1, \cdots, f_6$. Though we will not explicitly use it, for any cross-ratio of \eqref{crossratios} we simply replace $(a\cdot b) \to \langle a{-}1, a, b{-}1 b\rangle$ where the bracket is defined to be $\langle i,j,k,l \rangle:=\det(Z_i Z_j Z_k Z_l)$.

Now we move to Feynman parametrization and regularizations for all the integrals. 
For example, the one-loop triangle can be parametrized as
\begin{equation}
	\Gamma(3) \int_a \frac{1}{(a\cdot A)^3}\rightarrow \frac{1}{4 (\frac{1}{2}A\cdot A)^{3/2}}\,,
\end{equation}
where $A$ is a sum of dual regions multiplied by Feynman parameters.
For two-loops, {\it e.g.}the double-triangle takes the form:
\begin{equation}
	\begin{aligned}
		\int_{a,b} \frac{1}{(a\cdot A)^2(a\cdot b)(b\cdot B)^2}&=\int_0^{\infty}\frac{dc}{4 \pi \sqrt{c}}\int\frac{d a_i db_i}{\text{vol}\left(\text{GL}(1)\right)}\frac{1}{\left((1+c)\frac{1}{2}A\cdot A+A\cdot B +\frac{1}{2}B\cdot B\right)^2}
	\end{aligned}
\end{equation}
where $A=\sum a_i y_i$ and $B=\sum b_i y_i $. Due to dual-conformal invariance, any box must be accompanied by loop-dependent tensor numerators. These can be readily rewritten as derivative operators acting on the above formula for double-triangle integral. For example, the double-box integral with double $\epsilon$-numerators can be written as 
\begin{equation}
	\begin{aligned}
	&\int_{a,b}\frac{\epsilon\left(a,i_1,i_2,i_3,\,^\mu \right)\epsilon\left(b,j_1,j_2,j_3,\,_\mu \right)}{(a\cdot i_1)(a\cdot i_2)(a \cdot i_3)(a\cdot b) (b\cdot j_1)(b \cdot j_2)(b \cdot j_3)}\\
		=&\int_{0}^{\infty}\frac{dc}{4\pi\sqrt{c}}\int\frac{da_i db_i}{\mathrm{vol}\left(\mathrm{GL}\left(1\right)\right)}\frac{\epsilon\left(\partial_{A},i_1,i_2,i_3,\,^\mu \right)\epsilon\left(\partial_{B},j_1,j_2,j_3,\,_\mu \right)}{\left(\left(c+1\right)\frac{1}{2}A\cdot A+A\cdot B+\frac{1}{2}B\cdot B\right)^{2}}\,.
	\end{aligned}
\end{equation}
The double-box integrals with single $\epsilon$-numerator, i.e. $\epsilon(a,i,j,k,b)$ for $I^{db}_D[n_t]$, $I^{db}_E[n_t]$, $I^{db}_F[n_t]$ and $I^{db}_{G}[n_{b/c}]$ integrates to zero. To see this, one simply needs to realize that in Feynman parameter space it can be written as:
\begin{equation}
    \frac{\epsilon(\partial_A,i,j,k,\partial_B)}{\left(\left(c+1\right)\frac{1}{2}A\cdot A+A\cdot B+\frac{1}{2}B\cdot B\right)^{2}} \quad  \propto\quad \epsilon(A,i,j,k,B)=0 \,,
\end{equation}
since for such case either $A$ or $B$ will be a sum of vectors $(y_i, y_{j}, y_{k})$. From now on we will only consider $n_b$ numerators for $I^{db}_D$ $I^{db}_E$  $I^{db}_F$, and $n_a$ for $I^{db}_G$. 

For divergent integrals, we will use mass regulator, which corresponds to letting the scalar fields obtain a vev. That is, we move the theory slightly onto the Higgs branch and analytically continue to the origin. This has the advantage of retaining the dual conformal symmetry, which was instrumental in constraining the integrand. In practice, this simply amounts to extending the  five-dimension vector to: $(\vec{x}, 1,x^2)\rightarrow (\vec{x}, 1,x^2{+}\mu_{\rm IR}^2)$, or $(i\cdot j)\rightarrow (i\cdot j){+}2\mu_{\rm IR}^2$.

Note that while we expect that the final answer has uniform transcendental weight $2$, one often encounters functions with higher transcendental weight in the intermediate steps. Thus it is often more convenient to evaluate these integrals at the level of symbol first. Most integrals will be {\it linearly reducible} (sometimes after some change of variable or subtraction of divergences), {\it i.e.} it can be decomposed into rational factors of the form $dx /(x-x_i)^n$ with $n\geq 1$, multiplied by logarithms or polylogarithms with arguments that are rational functions of $x$. The symbol of these integrals can be extracted in an automated manner as follows~(see~\cite{Caron-Huot:2012awx} for more detail). Suppose we have the following integral and  want to obtain its symbol
\begin{equation}
	\begin{aligned}
		\int_a^b d\log(x-x_i)\left(F(x)\otimes\omega(x)\right),
	\end{aligned}
\end{equation}
where $F(x)\otimes\omega(x)$ is an integrable linear reducible symbol in $x$. Since it's linearly reducible, here we will assume $\omega(x)$ is at most linear in $x$, while there's no restriction on its dependence on other kinematic invariants. Taking the total differential with respect to other variables gives two contributions:
\begin{enumerate}[label=(\arabic*)]
	\item differentiating with respect to other variables at the endpoints yield:
	\begin{equation}
		\left.d\log(x-x_i)(F(x)\otimes\omega(x))\right|_{x=a}^{x=b}\rightarrow (F(x)\otimes \omega(x)\otimes (b-x_i))-(F(x)\otimes \omega(x)\otimes (a-x_i)),
	\end{equation}
	\item differentiating with respect to other variables in $\omega(x)$: if $\omega(x)$ is a constant with respect to $x$ then
	\begin{equation}
		\left(\int_a^b d\log(x-x_i)F(x)\right)d\log\omega\;\rightarrow\;\left(\int_a^b d\log(x-x_i)F(x)\right)\otimes\omega
	\end{equation}
	while for $\omega(x)=x-x_j$, 
	\begin{equation}
	    \begin{aligned}
	        \left(\int_a^b d\log\frac{x-x_i}{x-x_j}F(x)\right)d\log(x_j-x_i)\;\rightarrow\;\left(\int_a^b d\log\frac{x-x_i}{x-x_j}F(x)\right)\otimes(x_j-x_i).
	    \end{aligned}
	\end{equation}
\end{enumerate}
Iteratively repeating the above procedure, we can obtain the symbol of the linearly reducible integrals. For the case where we encounter $\frac{dx}{(x-x_i)^n}$ with $n>1$, its result can be obtained by repeated differentiation with respect to $x_i$ on $\frac{dx}{(x-x_i)}$.

Those integrals that are not linear-reducible will, in general, integrate into elliptic functions. As discussed at the end of section \ref{sec:CompInt}, these elliptic pieces will cancel as they are associated with vanishing cuts.
In practice, when encountering these non-rationalizable terms, we will stop at the last $c$-integral, and leave them to be canceled with similar terms from other integrands. Finally, terms with $\pi$ that are missed by symbol methods can be obtained by either numerical integration or taking the double-soft limit such that the integrand reduces to a six-point integrand whose results are known. 


\subsection{Finite double box integrals}
We begin with finite double-boxes such as $I_{C,i}^{db}$, $I_{G,i}^{db}[n_b]$ as well as those with soft-collinear divergences that are canceled when combined with appropriate double triangles, such as $I_{A,1,5}^{db}$ and $I_{F,i}^{db}[n_b]$. Since these integrals, or their combinations, are convergent, we do not need to introduce a regulator.

First double box $I_{C,1}^{db}$ has the very simple expression in the $c$-integrand. 
\begin{equation}
	\begin{aligned}
		I_{C,1}^{db}&=\int_0^\infty \frac{dc}{4 \pi \sqrt{c}} \frac{1}{1+c}\times \bigg(\log(1+c)\log{u_{3,7}v_{2,3}}-2\log{v_{2}}\log{v_{3}}+\log{u_{1,5}v_{1,4}}\\
		&\qquad \cdot \log{u_{3,7}v_{2,3}} -2 \mathrm{Li}_2(1-u_3)-2 \mathrm{Li}_2(1-u_7)-2 \mathrm{Li}_2(1-v_{2})-2\mathrm{Li}_2(1-v_{3})\\
		&\qquad  +2\mathrm{Li}_2(1-u_{3}v_{2})+2\mathrm{Li}_2(1-u_{3}v_{3})+2\mathrm{Li}_2(1-u_{7}v_{2})+2\mathrm{Li}_2(1-u_{7}v_{3})\bigg)\\
		&- \text{Fmb}\left(\frac{u_{1,5}v_{1,4}}{u_{3,7}v_{2,3}},u_{1,5}v_{1,4}\right)\,.
	\end{aligned}
\end{equation}
The function $ \text{Fmb}\left(v,w\right)$ is, in fact, closely related to the one-loop four-mass box integral integrated over a square root, 
\begin{equation}
	\begin{aligned}
		\text{Fmb}\left(v,w\right)&:=\int \frac{dc}{4\pi \sqrt{c}} \frac{w-v}{(1+c)\sqrt{-4 v w+(v+w-(1+c)v w)^2}}\\
		&\times \Big(\log(z \bar{z}) \log{(1+z)}- \log(z \bar{z}) \log{(1+\bar{z})}+2 \mathrm{Li}_2(-z)-2\mathrm{Li}_2(-\bar{z})\Big)
	\end{aligned}
\end{equation}
where $z$ and $\bar{z}$ are 
 \begin{equation}
 	\begin{aligned}
 		 &z,\bar{z}=\frac{-v-w+v w(1+c) \pm \sqrt{-4 v w + (v+w - (1+c) v  w)^2} }{2 w}\,.
 	\end{aligned}
 \end{equation}
Due to the square root factor, this is an elliptic function. However, as argued previously, since the cut that leads to elliptic integral vanishes, this term will cancel against that arising from  $I^{db}_{E,1,5}$ and $I^{db}_{E,1,7}$. Thus we will leave it unintegrated with respect to $c$. 
Performing the final $c$ integral on the rest, we find: 
 \begin{equation}
	\begin{aligned}
		I_{C,1}^{db}=&\frac{1}{2}\log{2}\log{u_{3,7}v_{2,3}}-\frac{1}{2}\log{v_{2}}\log{v_{3}}+\frac{1}{4}\log{u_{1,5} v_{1,4}}\log{u_{3,7}  v_{2,3}}\\
		&-\frac{1}{2} \mathrm{Li}_2(1-u_3)-\frac{1}{2} \mathrm{Li}_2(1-u_7)-\frac{1}{2} \mathrm{Li}_2(1-v_2)-\frac{1}{2} \mathrm{Li}_2(1-v_{3})\\
		&+\frac{1}{2}\mathrm{Li}_2(1-u_3 v_{2})+\frac{1}{2}\mathrm{Li}_2(1-u_3 v_{3})+\frac{1}{2}\mathrm{Li}_2(1-u_7 v_{2})+\frac{1}{2}\mathrm{Li}_2(1-u_7 v_{3})\\
		&- \text{Fmb}\left(\frac{u_{1,5} v_{1,4}}{u_{3,7} v_{2,3}},u_{1,5}  v_{1,4}\right).
	\end{aligned}
\end{equation}
The $\log 2$ term will be canceled in the final answer after summing over all integrals. 

Similarly for $I^{db}_{G,1}[n_b]$, we obtain the following:
\begin{equation}
	\begin{aligned}
		I^{db}_{G,1}[n_b]
		&=\int_0^{\infty}\frac{dc}{4\pi\sqrt{c}}\frac{
		-\log{(1+c)}\log{u_{1,3,5,7} v_{1,2,3,4}}-2\,\text{Li}_2(-c)}{c(1+c)}\\
		&\quad \quad \quad+\text{Fmb}\left(\frac{u_{3,7} v_{2,3}}{u_{1,5} v_{1,4}},u_{3,7} v_{2,3}\right)+\text{Fmb}\left(\frac{u_{1,5} v_{1,4}}{u_{3,7} v_{2,3}},u_{1,5} v_{1,4}\right)\\
		&=-2+\frac{\pi^2}{6}+\frac{1}{2}(1-\log{2}) \log{u_{1,3,5,7}  v_{1,2,3,4}}\\
		&\quad+\text{Fmb}\left(\frac{u_{3,7} v_{2,3}}{u_{1,5} v_{1,4}},u_{3,7} v_{2,3}\right)+\text{Fmb}\left(\frac{u_{1,5} v_{1,4}}{u_{3,7} v_{2,3}},u_{1,5} v_{1,4}\right)\,.	\end{aligned}
\end{equation}
Note the presence of the same elliptic function. 

\noindent
For the double box $I_{A,1,5}^{db}$, we combine it with $I^{dt}_{1,3;4,6}$ and define the finite combination $\tilde{I}_{A,1,5}^{db}:=I_{A,1,5}^{db}+I^{dt}_{1,3;4,6}$. The integrated result is:

\begin{equation}\label{eq:IAresult}
	\begin{aligned}
		\tilde{I}_{A,1,5}^{db}&=2\int_0^{\infty}\frac{dc}{4\pi\sqrt{c}}\frac{1}{1+c}\times \Bigg( \frac{\pi^2}{6}-\mathrm{Li}_{2}(1-u_2)-\mathrm{Li}_{2}(1-u_3)-\mathrm{Li}_{2}(1-v_{2})\\
		&+\mathrm{Li}_{2}(1-u_2 v_{2})+\mathrm{Li}_{2}(1-u_3 v_{2})-\mathrm{Li}_{2}(1-(1+c)u_{1,4}  v_{4})-\log{u_2}\log{u_3}\Bigg)\\
		&=\frac{\pi^2}{4}-\frac{1}{2}\mathrm{Li}_{2}(1-u_2)-\frac{1}{2}\mathrm{Li}_{2}(1-u_3)-\frac{1}{2}\mathrm{Li}_{2}(1-v_{2})+\frac{1}{2}\mathrm{Li}_{2}(1-u_2 v_{2})\\
		&\quad +\frac{1}{2}\mathrm{Li}_{2}(1-u_3 v_{2})+\frac 14\log\left(\chi(u_{1,4} v_{4})\right)^{2}-\frac{1}{2}\log{u_2}\log{u_3},
	\end{aligned}
\end{equation}
where for convenience we have defined the variable 
\begin{equation}\label{chidef}
\chi(x):=\frac{1-\sqrt{1-x^{-1}}}{1+\sqrt{1-x^{-1}}}.
\end{equation}
We will discuss the significance of these variables when the final result is presented. As a consistency check, if we let $p_{6,7,8}^2$ go to zero, the integrated result~\eqref{eq:IAresult} will reduce to the critter integral at two-loop six-point.

Similarly for the double box $I_{F,1}^{db}[n_b]$, the finite combination is given as,  $\tilde{I}_{F,1}^{db}:=I_{F,1}^{db}[n_b]-I^{dt}_{1,3;4,6}+I^{dt}_{1,3;4,8}-I^{dt}_{1,3;6,8}$, we obtain:
\begin{equation}\label{eq:IFresult}
	\begin{aligned}
		\tilde{I}_{F,1}^{db}=&2\int_0^\infty \frac{dc}{4\pi \sqrt{c}}  \frac{1}{1+c}\times \bigg(-\log{u_2 v_{2}}\log{u_8 v_{3}}-\mathrm{Li}_2(1-u_2 v_{2})-\mathrm{Li}_2(1-u_8 v_{3})\\
		&-\mathrm{Li}_2(1-(1+c)u_1)+\mathrm{Li}_2 (1-(1+c) u_{1,6}  v_1)+\mathrm{Li}_2(1-(1+c)u_{1,4} v_{4})\bigg)\\
		&=-\frac{\pi^2}{6}+\frac 14\log(\chi(u_1))^2-\frac 14\log(\chi(u_{1,6} v_1))^2-\frac{1}{4}\log(\chi(u_{1,4} v_{4}))^2\\
		&\quad -\frac{1}{2}\log{u_2 v_{2}}\log{u_8 v_{3}}-\frac{1}{2} \mathrm{Li}_2(1-u_2 v_{2})-\frac{1}{2} \mathrm{Li}_2(1-u_8 v_{3}).
	\end{aligned}
\end{equation}
Upon taking the limit of $p_{4,5}^2$ and $p_{6,7}^2$ approaching zero, the result of integration~\eqref{eq:IFresult} will also become equivalent to the two-loop six-point critter integral.


\subsection{Divergent double box integrals}
\begin{figure}[H]
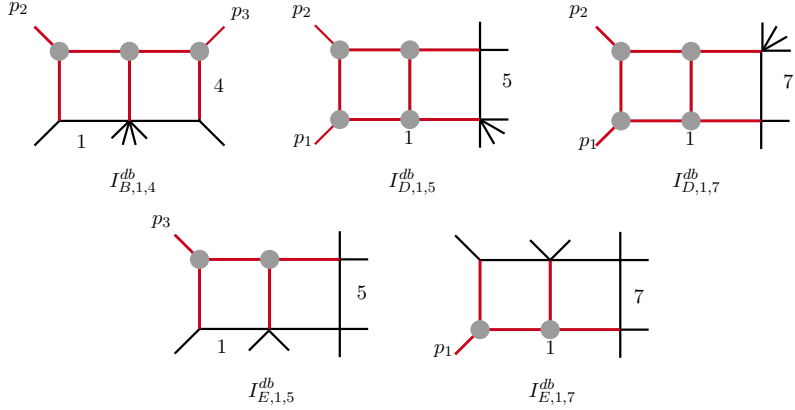
   
\begin{center}
$\vcenter{\hbox{\scalebox{0.7}{	



}}}$
\end{center}
\caption{ The five double boxes that have collinear or soft-collinear divergences.  We use the red line to indicate the loop momentum being collinear. The double boxes $I_{B,1,4}^{db}$ and $I_{D,1,j}^{db}$ have two collinear divergence regions. On the other hand, the double boxes $I_{E,1,5}^{db}(I_{E,1,7}^{db})$ only  diverge in one collinear region where both loop momenta are collinear to $p_{3}(p_1)$.} \label{fig:divergent_integral}
\end{figure}

We now consider the divergent integrals, where both loop momenta can become collinear with the external legs. This is kinematically allowed when an external massless leg is connected with two consecutive cubic vertices. If there are two such collinear regimes for a given graph, then they can overlap with certain propagators becoming soft. Such soft-collinear divergence is factorizable and leads to $\log^2 \mu^2_{\rm IR}$ times a number, while the previous case only leads to $\log \mu^2_{\rm IR}$. This is illustrated in fig.\ref{fig:divergent_integral}, wherein the first line of the integrand $I^{db}_{B,1,4}$ (or $I^{db}_{D,1,5}$) can have two collinear regions, both momentum proportional to leg $p_2$ or $p_3$ ($p_1$). On the other hand in the second line, there is only one collinear region.

Note that as we will be introducing mass regularization $\mu_{\text{IR}}^2$, in Feynman parameter space this introduces terms that are squares of Feynman parameters:  
\begin{equation}\label{eq:mu_containing_terms}
	\begin{aligned}
	\mu_{\text{IR}}^2 X=\mu_{\text{IR}}^2 \left(\left(\sum a+\sum b \right)^2+c \left(\sum a\right)^2\right)\,,
	\end{aligned}
\end{equation}
here $\sum a$ and $\sum b$ represent the sum over Feynman parameters of each loop. This spoils the linear reducibility of the integral. Following~\cite{Caron-Huot:2012sos} our strategy is to subtract something which has the same divergent behavior but which is simpler to integrate (in the sense that the $\mu_{\text{IR}}^2$ dependent term is the simplest). We then compute the correction that is the difference between the two, which is finite and can safely send $\mu^2_{\rm IR}\rightarrow0$. We illustrate this in an explicit example. 

\vskip 10 mm

\paragraph{Integrating $I_{B,1,4}^{db}$.} 

We start with simplifying the $\mu_{\text{IR}}^2$- term in  $I_{B,1,4}^{db}$. Since it is only relevant when the loop momenta are in the collinear regime, we can find a simpler expression by neglecting irrelevant Feynman parameters in that limit. This integral has two collinear regions, the loop momenta being proportional to $p_2$, where $(a_1, b_4,b_5)\rightarrow0$, or $p_3$ where $(a_1, a_2, b_5)\rightarrow0$. We can safely set $a_1, b_5\rightarrow0$ In both cases. Therefore, we can neglect $a_1, b_5$ in  $X$ and   integrate it out straightforwardly.

To proceed further, we replace $X$ with something simpler that has the same divergence behavior of $I_{B,1,4}^{db}$  but which is simpler to integrate. A good candidate is 

\begin{equation}
	\tilde{I}_{B,1,4}^{db}=I_{B,1,4}^{db}(X\longrightarrow (a_3+b_3)^2+c a_3^2)
\end{equation}
since this has identical soft and soft-collinear regions. But thanks to the simplified denominator, this can be integrated more easily.

The correction terms then correspond to restoring either $a_2$ or $b_4$, which modifies the logarithmic cut-off $X$. To capture this, it requires considering the difference between the regulator after simplification (i.e. in $\tilde{I}_{B,1,4}^{db}$) and the original regulator only drops $a_2(b_4)$ one at time. First, let's restore the correction of dropping $b_4$. We can first quickly integrate $a_1,a_2,b_5$ which regulator is irrelevant to  whether dropping $b_4$ in $X$, and we can obtain the function of the form
 \begin{equation}
     \log{X}\,\text{dependent term}+\text{remainder}.
 \end{equation}
Then the difference between the simplified regulator and the regulator only drops $a_2$ is the correction for $b_4$. Since the change is only sensitive to the small modification of the regulator, it turns out that only the  $\log{X}$ dependent term will survive the difference and the result is 
\begin{equation}
    \begin{aligned}
        &\frac{b_3+a_3 c}{b_3 b_4(b_3+a_3 c+b_4/y_1)}\log{X}\bigg|_{X=b_3^2+a_3^2c}^{X=(b_3+b_4)^2+a_3^2 c}\\
        =&\frac{b_3+a_3 c}{b_3 b_4(b_3+a_3 c+b_4/y_1)}\log{\frac{(b_3+b_4)^2+a_3^2 c}{b_3^2+a_3^2c}}
    \end{aligned}
\end{equation}
where $y_1=x_{1,3}^2/x_{1,4}^2$. By redefining $b_4$ to be $b_4 y_1$, the correction term for dropping $b_4$ is
\begin{equation}
	I_{B,1,4}^{\text{cor}}(y_1\equiv x_{1,3}^2/x_{1,4}^2)\big|_{a_2=0}=\int_0^{\infty}\frac{dc}{4\pi \sqrt{c}}\int_{a3<b3}\frac{d^2 [a_3 b_3 b_4]}{\text{vol}\left(\text{GL}(1)\right)} \frac{(b_3+a_3 c) \log{\frac{a_3^2 c+(b_3+b_4 y_1)^2}{b_3^2+a_3^2 c}}}{b_3 b_4 (b_3+b_4+a_3 c)^2}.
\end{equation}

\noindent
Similarly, for the $a_2$ correction is
\begin{equation}
	I_{B,1,4}^{\text{cor}}(y_2=x_{3,5}^2/x_{2,5}^2)\big|_{b_4=0}=\int_0^{\infty}\frac{dc}{4\pi \sqrt{c}}\int_{a3<b3}\frac{d^2 [a_2 a_3 b_3]}{\text{vol}\left(\text{GL}(1)\right)} \frac{b_3 y_2^2 \log{\frac{(a_2+b_3)^2+(a_2+b_3)^2 c}{b_3^2+a_3^2 c}}}{a_2(b_3+a_3 c)(a_2+b_3 y_2)^2}\,.
\end{equation}

\noindent
The final integrated result is

\begin{equation}
	\begin{aligned}
		I_{B,1,4}^{db} &=\tilde{I}_{B,1,4}^{db}+I^\text{cor}_{B,1,4}(x_{1,3}^2/x_{1,4}^2)\big|_{a_2=0}+I^\text{cor}_{B,1,5}(x_{3,5}^2/x_{2,5}^2)\big|_{b_4=0}\\
		&=-1+\frac{\pi^{2}}{4}+\frac{1}{2}\left(1+\log u_2\right)\log\frac{4\mu_{\mathrm{IR}}^{2}x_{1,5}^{2}}{x_{1,3}^{2}x_{3,5}^{2}}-\frac{1}{4}\log^{2}\frac{4\mu_{\mathrm{IR}}^{2}x_{1,5}^{2}}{x_{1,3}^{2}x_{3,5}^{2}}\\
	 &\quad-\mathrm{Li}_{2}\left(1-\frac{x_{1,3}^{2}}{x_{1,4}^{2}}\right)-\mathrm{Li}_{2}\left(1-\frac{x_{3,5}^{2}}{x_{2,5}^{2}}\right)+\frac{1}{2}\mathrm{Li}_{2}\left(1-\frac{1}{u_2}\right),
	\end{aligned}
\end{equation}
which is kinematically equivalent to the ``$I^{\rm crab}$'' integral in six-point.

\vskip 10 mm

\paragraph{Integrating $I_{D,1,j}^{db}[n_b]$.} 
The next divergent integral are $I_{D,1,j}^{db}[n_b]$($j=5,7$). Since the two integrals are related by relabelling, we can just focus on integrating $I_{D,1,5}^{db}$. The common behavior of the Feynman parameter for both collinear divergences is $b_5$ set to zero. Thus we can drop $b_5$ from $X_5$ and integrate out $b_5$, yielding
\begin{equation}
	I_{D,1,5}^{db}[n_b]=
 \int_0^\infty \frac{dc}{4\pi\sqrt{c}} \int \frac{[d^4a_1a_2a_3b_1b_3]}{\textrm{vol(GL(1))}} \frac{\big[a_2(2\cdot 5){-}2((A{+}B)\cdot 5)\big] (2\cdot 5)/(1\cdot 3)/((A{+}B)\cdot 5)^2}
 {\big((1+c)a_1a_3 + a_1b_3+a_3b_1+ b_1b_3 +\frac{\mu^2_\text{IR}}{(1\cdot 3)}X \big)^2}
\end{equation}
where $X:=(\sum a+\sum b)^2+c(\sum a)^2$.

The other Feynman parameters being set to zero in the collinear limits is $(a_3, b_3)$ and $(a_1, b_1)$. So the simpler form of the regulator is:
\begin{equation}
	\tilde{I}_{D,1,5}^{db}:= I_{D,1,5}^{db}[n_b]\big(X\longrightarrow a_2^2(1+c)\big) \,.
\end{equation}
Thanks to the simplified denominator, this can be integrated more easily.
Indeed after redefining $a_1+b_1\to b_1$, $a_3+b_3\to b_3$ together with a simple rescaling of the variables,
it can be seen to depend only on a single parameter $\epsilon:=\frac{4\mu_\text{IR}^2 x_{1,5}^2x_{3,5}^2}{x_{1,3}^2x_{2,5}^4}$:

\begin{equation}
	\begin{aligned}
		 \tilde{I}_{D,1,5}^{db}&= -\int_0^\infty \frac{dc}{4\pi \sqrt{c}} \int_{\substack{a_1<b_1\\a_3<b_3}}\frac{[d^4a_1a_2a_3b_1b_3]}{\textrm{vol(GL(1))}} \frac{a_2+2b_1+2b_3}{(a_2+b_1+b_3)^2(b_1b_3+a_1a_3 c+\epsilon a_2^2(1+c))^2}\\
		  &= 2-\frac{7\pi^2}{12}-\frac14\log^2 \epsilon. 
	\end{aligned}
\end{equation}

\noindent
Restoring $(a_3, b_3)$ and $(a_1, b_1)$ leads to the following change in the logarithmic cutoff

\begin{equation}
	\begin{aligned}
		I^\text{cor}_{D,1,5}(y_1{\equiv} x_{2,5}^2/x_{1,5}^2)\big|_{a_3=b_3=0}&= \int_0^\infty \frac{dc}{4\pi\sqrt{c}} \int_{a_1<b_1}\frac{[d^2a_1a_2b_1]}{\textrm{vol(GL(1))}} \frac{y_1(a_2y_1{+}2b_1)\log \frac{(a_2{+}b_1)^2{+}c (a_1{+}a_2)^2}{a_2^2(1+c)}}{b_1 (b_1+a_1c)(a_2y_1+b_1)^2}\\
 &=\frac{\pi^2}{6}-\text{Li}_2(1-y_1). 
	\end{aligned}
\end{equation}
\begin{equation}
    I^\text{cor}_{D,1,5}(y_2\equiv x_{2,5}^2/x_{3,5}^2)\big|_{a_1=b_1=0}=\frac{\pi^2}{6}-\text{Li}_2(1-y_2). 
\end{equation}
So that
\begin{equation}
	\begin{aligned}
			I_{D,1,5}^{db}[n_b]&=\tilde{I}_{D,1,5}^{db}+I^\text{cor}_{D,1,5}(x_{2,5}^2/x_{1,5}^2)\big|_{a_3=b_3=0}+ I^\text{cor}_{D,1,5}(x_{2,5}^2/x_{3,5}^2)\big|_{a_1=b_1=0}\\
			&=2-\frac{\pi^{2}}{4}-\frac{1}{4}\log^2\frac{4\mu_\text{IR}^2 x_{1,5}^2x_{3,5}^2}{x_{1,3}^2x_{2,5}^4}-\mathrm{Li}_{2}\left(1-\frac{x_{2,5}^{2}}{x_{1,5}^{2}}\right)-\mathrm{Li}_{2}\left(1-\frac{x_{2,5}^{2}}{x_{3,5}^{2}}\right),
	\end{aligned}
\end{equation}
which is kinematically equivalent to the ``$I^{\rm 2mh}$'' integral in six-point.
Similarly, 
\begin{equation}
	\begin{aligned}
			I_{D,1,7}^{db}[n_b]=2-\frac{\pi^{2}}{4}-\frac{1}{4}\log^2\frac{4\mu_{\mathrm{IR}}^{2}x_{1,7}^2x_{3,7}^{2}}{x_{1,3}^{2}x_{2,7}^{4}}-\mathrm{Li}_{2}\left(1-\frac{x_{2,7}^{2}}{x_{1,7}^{2}}\right)-\mathrm{Li}_{2}\left(1-\frac{x_{2,7}^{2}}{x_{3,7}^{2}}\right)\,.
	\end{aligned}
\end{equation}

\vskip 10 mm

\paragraph{Integrating $I_{E,1,j}^{db}[n_b]$.} 
Similarly, the two integrals of $I_{E,1,j}[n_b](j=5,7)$ are related by relabelling. We can just integrate out $I_{E,1,7}^{db}[n_b]$. A first observation is that in all divergent regions $a_3$, $b_5$, $b_7\rightarrow 0$. Thus we can drop $a_3$, $b_5$, $b_7$ from terms multiplying the mass in the denominator. We can easily integrate out $a_3$, $b_7$.

We simplify the regulator by taking the limit $a_1,b_1\rightarrow0$
\begin{equation}
	\tilde{I}_{E,1,7}^{db}:= I_{E,1,7}^{db}[n_b]\left(X\longrightarrow a_2^2 y^2(1+c)\right) 
\end{equation}
where $y\equiv\frac{x_{1,7}^{2}}{x_{2,7}^{2}}$. Simplifying the regulator will introduce the error. The finite correction term to compensate for the change of regulator is given by

\begin{equation}
	\begin{aligned}
		I_{E,1,7}^{\text{cor}}(u_7 v_2,y)=\int_{0}^{\infty}\frac{dc}{4\pi\sqrt{c}}\int_{a_{1}<b_{1}}&\frac{d^{2}a_{1}a_{2}b_{1}}{\mathrm{vol}\left(\mathrm{GL}\left(1\right)\right)}\frac{b_{1}\left(-u_7 v_2+1\right)\left(a_{2}u_7 v_2+a_{2}+2b_{1}\right)}{\left(a_{2}+b_{1}\right)^{2}\left(b_{1}+a_{1}c\right)\left(a_{2}u_7 v_2+b_{1}\right)^{2}}\\
		&\times \log {\frac{b_{1}^{2}+a_{1}^{2}c+2a_{2}\left(b_{1}+a_{1}c\right)y+a_2^2\left(1+c\right)y^2}{a_2^2 \left(1+c\right)y^2}}.
	\end{aligned}
\end{equation}

\noindent
After integrating out $I^{\text{cor}}_{E,1,7}(u_7 v_2,y)$, it gives

\begin{equation}
	\frac{1}{2}\log\frac{x_{1,7}^{2}}{x_{2,7}^{2}}\log (u_7 v_2)-\frac{1}{2}\log^{2} (u_7 v_2)-\mathrm{Li}_{2}\left(1-\frac{x_{1,5}^{2}}{x_{2,5}^{2}}\right)+\mathrm{Li}_{2}\left(1-\frac{x_{1,7}^{2}}{x_{2,7}^{2}}\right)\,.
\end{equation}

\noindent
The result of $I_{E,1,7}^{db}[n_b]$ integral is
\begin{equation}
	\begin{aligned}
		&\quad I_{E,1,7}^{db}[n_b]=\tilde{I}_{E,1,7}^{db}+I^{\text{cor}}_{E,1,7}(u,y)\\
		&=\frac{1}{4}\log{u_{3,7}  v_{2,3}}(2-\log{4}-\log{u_{1,5} v_{1,4}})+\frac{1}{2}\log{u_{7} v_{2}}\log{u_{1,5} v_{1,4}}\\
		&\ -\frac{1}{4}\log^2{u_7 v_{2}} 
		-\frac{1}{2}\log{u_7 v_{2}}\log{\frac{4 \mu_{\mathrm{IR}}^2 x_{5,7}^2}{x_{1,5}^2 x_{1,7}^2}} -\frac{1}{2}\mathrm{Li}_2(1-u_7 v_{2})-\frac{1}{2}\mathrm{Li}_2(1-u_3 v_{3})\\
		&\ -\mathrm{Li}_{2}\left(1-\frac{x_{1,5}^{2}}{x_{2,5}^{2}}\right) +\mathrm{Li}_{2}\left(1-\frac{x_{1,7}^{2}}{x_{2,7}^{2}}\right)+\text{Fmb}\left(\frac{u_{1,5} v_{1,4}}{u_{3,7} v_{2,3}},u_{1,5} v_{1,4}\right)\,.
	\end{aligned}
\end{equation}

\vskip 10mm

Similar, the integrating result of $I_{E,1,5}^{db}[n_b]$ is given by

\begin{equation}
	\begin{aligned}
		I_{E,1,5}^{db}[n_b]
		&=\frac{1}{4}\log{u_{3,7} v_{2,3}}(2-\log4-\log u_{1,5} v_{1,4} )+\frac{1}{2}\log{u_3 v_{3}}\log{u_{1,5} v_{1,4}}\\
		&-\frac{1}{4}\log^2{u_3 v_{3}}-\frac{1}{2}\log{u_3 v_{3}}\log{\frac{4 \mu_{\mathrm{IR}}^2 x_{5,7}^2}{x_{3,5}^2 x_{3,7}^2} }-\frac{1}{2}\mathrm{Li}_2(1-u_3 v_3)-\frac{1}{2}\mathrm{Li}_2(1-u_7 v_{2})\\
		& +\mathrm{Li}_{2}\left(1-\frac{x_{3,5}^{2}}{x_{2,5}^{2}}\right)-\mathrm{Li}_{2}\left(1-\frac{x_{3,7}^{2}}{x_{2,7}^{2}}\right)+\text{Fmb}\left(\frac{u_{1,5} v_{1,4} }{u_{3,7} v_{2,3}},u_{1,5} v_{1,4}\right)\,.
	\end{aligned}
\end{equation}


\subsection{Box-triangle integrals}\label{sec:parity_odd}
Finally, we consider the  box-triangle integrals. We pay special attention to the integrals that are proportional to $\mathcal{B}$, as they must integrate to functions that have non-trivial little group properties to restore the deficiency in $\mathcal{B}$. There are three box-triangles $I^{bt}_{A,i,j}$, $I^{bt}_{B,i}+I_{C,i}^{bt}[n_{\beta}]$ where the latter are combined such that their non-factorizable soft-collinear divergence cancels. 

Recall that we have defined $\chi(x)$ in \eqref{chidef}, and note that $\chi(x)$ and $\chi(1-x)$ are not multiplicative independent since 
\[
\chi(1-x)=
\begin{cases}
-\chi(x)&\text{if $x<0$ or $x>1$};\\
-\chi(x)^{-1}&\text{if $0<x<1$};
\end{cases}
\]
we usually take the simpler one as the letter in the symbol. 
In order to avoid the problem arising from branches of the square root in $\chi$, we assume that $u_i$, $v_i>1$ in this subsection. Moreover, we define two functions
\begin{equation}\label{eq:f-ftn}
	\begin{aligned}
		f(a,b)\equiv &\frac{\pi^2}{8}-\frac{1}{8}\log^2{\frac{1-a}{1+a}}+\frac{1}{4}\log{\frac{1-a}{1+a}}\log{\frac{b-a}{a+b}}\\
		&-\frac{1}{2}\mathrm{Li}_2\left(\frac{1-b}{1-a}\right)-\frac{1}{2}\mathrm{Li}_2\left(\frac{1-b}{1+a}\right)+\frac{1}{4}\mathrm{Li}_2\left(\frac{1-b^2}{1-a^2}\right),\\
		F(x,y)\equiv &f(\sqrt{1-x^{-1}},\sqrt{1-y^{-1}}),
	\end{aligned}
\end{equation}
they are building blocks of box-triangle integrals in this subsection, and their symbol are fairly simple,
\begin{equation}
	\begin{aligned}
		\mathcal S[f(a,b)]&=\frac{1}{4} \left(\frac{b^2-a^2}{a^2-1}\otimes \frac{1-b}{1+b}+\frac{1-a}{1+a}\otimes\frac{b-a}{b+a}\right),\\
		\mathcal S[F(x,y)]&=\frac{1}{4} \left(\bigl(1-\frac{x}{y}\bigr)\otimes 
		\chi(y)+\chi(x)\otimes\chi\biggl(\frac{(x-1)y}{x-y}\biggr)\right)\quad \text{if $x>1$ or $x<0$}.
	\end{aligned}
\end{equation}

\paragraph{Integrating $I_{A,1,j}^{bt}$.} There are two integrals that need to be evaluated in this topology ($j=5,7$). We begin with integrating $I_{A,1,5}^{bt}$, and then relabel it to obtain $I_{A,1,7}^{bt}$. The Feynman integral of $I_{A,1,5}^{bt}$ read in this case:
\begin{equation}
	\begin{aligned}
	&-\int_0^{\infty}\frac{dc}{4\pi\sqrt{c}}\int \frac{da_1 a_2 a_3 b_4 b_6}{\text{vol}(\text{GL}(1))} \epsilon(\partial_A,1,2,3,4)
		\frac{1}{\left((c+1)\frac{1}{2}A^2+A\cdot B+\frac{1}{2}B^2\right)^2}\\
		&=\int_0^{\infty}\frac{dc}{4\pi\sqrt{c}} \int \frac{da_1 a_2 a_3 b_4 b_6}{\text{vol}(\text{GL}(1))} \frac{2 b_6\, \epsilon(6,1,2,3,4)}{\left((c+1)\frac{1}{2}A^2+A\cdot B+\frac{1}{2}B^2\right)^3}\\
	\end{aligned}
\end{equation}
where $A=a_1 y_1+a_2 y_2+a_3 y_3$, $B=b_4 y_4+b_6 y_6$. The integrand can be straightforwardly integrated, yielding
\begin{equation}\label{eq:result_bt_A}
	\begin{aligned}
		& \frac{\epsilon(6,1,2,3,4)\sqrt{(2\cdot 4\cdot 6)}}{\sqrt{2}(2\cdot 4)}\int_0^\infty\frac{dc}{4\pi \sqrt{c}}\frac{1}{1+c-\frac{1-u_2v_2}{u_{1,4}v_4}}\times \bigg(\frac{\pi^2}{6}-\log{u_2 v_2}\log{u_{1 ,4} v_4}\\
	&\qquad -\log{(1+c)}\log{u_2 v_{2}}-\mathrm{Li}_2(1-u_2 v_{2})-\mathrm{Li}_2(1-(1+c)u_1 u_4 v_{4})\bigg)\\
		=&\frac{\epsilon(6,1,2,3,4)\sqrt{(2\cdot 4\cdot 6)}}{\sqrt{2}(2\cdot 4)(1\cdot3)(2\cdot6)(4\cdot6)\sqrt{1-\frac{1-u_2v_2}{u_{1,4}v_4}}}
		F\biggl(u_{1,4}v_4,\frac{u_{1,4}v_4}{1-u_2v_2}\biggr),
	\end{aligned}
\end{equation}
and its symbol is
\begin{equation}
    \frac{1}{4}\left(u_2 v_2 \otimes \chi\left(\frac{u_{1,4} v_4}{1-u_2 v_2}\right)+\chi\left(u_{1,4} v_4\right)\otimes \chi\left(\frac{1-u_{1,4} v_4}{u_2 v_2}\right)\right).
\end{equation}

\noindent

Remarkably, the prefactor is simply a product of sign functions thanks to the identity \eqref{eq:signftn idenity 1} and \eqref{eq:signftn idenity 2} (see appendix~\ref{sec:identity} for details):
\begin{equation}\label{Sign1}
	\begin{aligned}
		\frac{\epsilon(6,1,2,3,4)\sqrt{(2\cdot4\cdot6)}}{\sqrt{2}(2\cdot4)(1\cdot3)(2\cdot6)(4\cdot6)\sqrt{1-\frac{1-u_2v_2}{u_{1,4}v_4}}}=&\frac{1}{i} \mathrm{sign}( \langle12\rangle\langle45\rangle)\mathrm{sign} \left(\langle23\rangle\langle1|p_{3,4,5}|2\rangle+\langle13\rangle (2\cdot 6) \right)\,,
	\end{aligned}
\end{equation}
where ${\rm sign}(x)\equiv x/\sqrt{x^2}$ which differs from ${\rm sgn}_c$ in eq.(\ref{SignDef}). Note that this product of sign functions actually takes little group weight at legs $2,3,4,5$.

Similarly, we can obtain the integrated result of box-triangle $I_{A,1,7}^{bt}$ by shifting the  legs $y_1\leftrightarrow y_3, \, y_4\rightarrow y_8$, and left $y_2$, $y_6$ unchanged of the $I_{A,1,5}^{bt}$:

\begin{equation}
	\begin{aligned}
	\frac{1}{i} \, \mathrm{sign}\langle67\rangle\, \mathrm{sign}\langle12\rangle\, \mathrm{sign}\left(\langle18\rangle \langle2|p_{6,7,8}|1\rangle+\langle82\rangle(2\cdot6)\right) F\biggl(u_{1,6}v_1,\frac{u_{1,6}v_1}{1-u_8v_3}\biggr)
	\end{aligned}
\end{equation}
and its symbols
\begin{equation}
	\begin{aligned}
		\frac{1}{4} \left((u_8 v_{3})\otimes \chi\biggl(\frac{u_{1,6}v_1}{1-u_8v_3}\biggr)+\chi(u_{1,6}v_1)\otimes\chi\biggl(\frac{1-u_{1,6} v_1 }{u_8 v_3}\biggr)\right).
	\end{aligned}
\end{equation}

\paragraph{Integrating $I^{bt}_{B,1}+I_{C,1}^{bt}[n_{\beta}]$:}  Let's define the following rescaled combination:
\begin{equation}\label{eq:normalize_bt_2}
	\begin{aligned}
		&I^{bt}_{B+C,1}:=\left(\frac{\epsilon(1,2,3,4,8) \sqrt{(2\cdot 4\cdot 8)}}{\sqrt{2}(2\cdot 4)}\right)^{-1}\times \left(I^{bt}_{B,1}+I_{C,1}^{bt}[n_{\beta}]\right)\\
		&= \int_{a,b} 
		\frac{\epsilon(a,1,2,3,4)/\epsilon(1,2,3,4,8)}{(a\cdot1)(a\cdot 2)(a\cdot3)(a\cdot b)(b\cdot4)(b\cdot 8)}{-}\frac{\big((1{\cdot} 6)\epsilon(b,3,4,6,8){+}(6{\cdot} 8)\epsilon(b,1,3,4,6)\big){/}\epsilon(1,3,4,6,8)}{(2\cdot 8)(a\cdot1)(a\cdot3)(a\cdot b)(b\cdot4)(b\cdot6)(b\cdot 8)} \\
	&=\int_0^\infty \frac{dc}{4 \pi \sqrt{c}}  \int \frac{d^5 a_1 a_2 a_3 b_4 b_6 b_8}{\text{vol}(\text{GL}(1))} \bigg(\frac{\epsilon(\partial_A,1,2,3,4)}{\epsilon(1,2,3,4,8)}(\partial_B\cdot 6)\\
		 &\qquad  -(2\cdot \partial_A)\frac{(1\cdot 6)\epsilon(\partial_B,3,4,6,8)+(6\cdot 8)\epsilon(\partial_B,1,3,4,6)}{(2\cdot 8)\epsilon(1,3,4,6,8)}\Bigg) \frac{1}{\left((c+1)\frac{1}{2}A^2+A\cdot B+\frac{1}{2}B^2\right)^2}
	\end{aligned}
\end{equation}
where $A=a_1 y_1+a_2 y_2+a_3 y_3$, $B=b_4 y_4+b_6 y_6+b_8 y_8$.

We do integral in the order $a_2$, $a_3$, $b_4$, $a_1$, and $b_8$, setting $b_6=1$. In the process, the $\epsilon$-tensor will neatly cancel out:

\begin{equation}\label{eq:result_bt_B+C}
	\begin{aligned}
		&I^{bt}_{B+C,1}=\frac{1}{(1\cdot 3)(2\cdot 8)(4\cdot 8)} \int \frac{dc}{4 \pi \sqrt{c}} \frac{u_1}{(1+c)u_1-1}\Big( -\log\left((1+c)u_1\right)\log \frac{u_2 v_2}{u_8 v_3}\\
		& -\mathrm{Li}_2\left(1-u_6 v_1\right)+\mathrm{Li}_2\left(1-u_4 v_4\right)+\mathrm{Li}_2\left(1-(1+c)u_{1,6} v_1\right)-\mathrm{Li}_2\left(1-(1+c)u_{1,4} v_4\right) \Big)\\
		&=\frac{1}{(1\cdot 3)(2\cdot 8)(4\cdot 8)\sqrt{1{-}u_1^{-1}}} \left( F(u_{1,4}v_4,u_1){-}F(u_{1,6}v_1,u_1){+}\frac{1}{4}\log{\chi(u_1)}\log{\frac{u_2 v_2(1{-}u_6 v_1)}{u_8 v_3(1{-}u_4 v_4)}}\right)
	\end{aligned}
\end{equation}
here again the answer is expressed in term of function $F$ in~\eqref{eq:f-ftn}, and its symbol reads 
\begin{equation}
	\begin{aligned}
		&\frac{1}{4}\times \left(\frac{u_2 v_2}{u_8 v_3}\otimes \chi(u_1)+\chi(u_1)\otimes\frac{u_2 v_2}{u_8 v_3}+\chi(u_{1,4}v_4)\otimes\chi\biggl(\frac{1-u_{1,4} v_4}{1-u_4 v_4}\biggr) \right.\\
		&\qquad \quad \left.-\chi(u_{1,6} v_1)\otimes\chi\biggl(\frac{1-u_{1,6} v_1}{1-u_6 v_1}\biggr)+\chi(u_1)\otimes\frac{1-u_6 v_1}{1-u_4 v_4}\right)\,.
	\end{aligned}
\end{equation}
The factors in front of the transcendental function in~\eqref{eq:result_bt_B+C} can again be combined with the scaling factor in eq.(\ref{eq:normalize_bt_2}) to produce a product of sign functions: 

\begin{equation}\label{Sign2}
	\frac{\epsilon(8,1,2,3,4)\sqrt{(2\cdot4\cdot8)}}{\sqrt{2}(2\cdot4)(1\cdot3)(2\cdot8)(4\cdot8) \sqrt{1-u_1^{-1}}}=\frac{1}{i}  \mathrm{sign} \langle12\rangle\,  \mathrm{sign} \langle8|p_{1,2}|3\rangle.
\end{equation}
Once again this sign function takes little group at legs $8,1,2,3$.

\paragraph{Integrating $I_{C,1}^{bt}[n_{\alpha}]$:} The last one box-triangle in Feynman parameters space is by striping off kinematic prefactor $2\langle12\rangle\langle45\rangle\langle67\rangle\langle3|p_{1,2}|8\rangle/\epsilon(1,3,4,6,8)$:
\begin{equation}
	\begin{aligned}
	&-\int_0^{\infty}\frac{dc}{4\pi\sqrt{c}}\int \frac{da_1  a_3 b_4 b_6 b_8}{\text{vol}(\text{GL}(1))} \sum_{i=1,3,4,6,8} \alpha_i (\partial_A\cdot i)
		\frac{1}{\left((c+1)\frac{1}{2}A^2+A\cdot B+\frac{1}{2}B^2\right)^2}\\
		&=\int_0^{\infty}\frac{dc}{4\pi\sqrt{c}} \int \frac{da_1 a_3 b_4 b_6 b_8}{\text{vol}(\text{GL}(1))} \sum_{i=1,3,4,6,8} \frac{2 \alpha_i (A+B\cdot i)}{\left((c+1)\frac{1}{2}A^2+A\cdot B+\frac{1}{2}B^2\right)^3}\\
	\end{aligned}
\end{equation}
where $A=a_1 y_1+a_3 y_3$, $B=b_4 y_4+b_6 y_6+b_8 y_8$. 
The Feynman parameters of this integral can be straightforwardly integrated,  then the integral arriving
\begin{equation}\label{eq:result_bt_C2}
\begin{aligned}
		&\int_0^{\infty} \frac{dc}{4 \pi \sqrt{c}} \frac{2(1{-}u_6 v_1{-}u_4 v_4(1{-}u_{1,6} v_1))}{1{-}u_6 v_1 {+}u_4 v_4((1{+}c)u_{1,6} v_1{-}1)}\times \Big(\log{u_6 v_1}\log{u_4 v_4}{+}\mathrm{Li}_2\left(1{-}(1{+}c)u_1\right)\\
		&+\mathrm{Li}_2\left(1{-}u_6 v_1\right){-}\mathrm{Li}_2\left(1-(1{+}c)u_1 u_6 v_1\right){+}\mathrm{Li}_2\left(1{-}u_4 v_4\right){-}\mathrm{Li}_2\left(1{-}(1{+}c)u_1 u_4 v_4\right)\Big)\\
		= & 2\sqrt{\frac{1{-}u_6 v_1{-}u_4 v_4 (1{-}u_{1,6}v_1)}{u_{1,4,6}v_{1,4}}} \times \left( {-}F\left(u_1,\frac{u_{1,4,6}v_{1,4}}{u_6 v_1{+}u_4 v_4{-}1}\right){+}F\left(u_{1,4}v_4,\frac{u_{1,4,6}v_{1,4}}{u_6 v_1{+}u_4 v_4{-}1}\right)\right.\\
		&\qquad \qquad \qquad \qquad \qquad \qquad  \qquad\left.{+}F\left(u_{1,6}v_1,\frac{u_{1,4,6}v_{1,4}}{u_6 v_1{+}u_4 v_4{-}1}\right)\right).
	\end{aligned}
\end{equation}
The kinematic prefactor also combines into a sign function with little group weight
\begin{equation}
    \frac{2\langle12\rangle\langle45\rangle\langle67\rangle\langle3|p_{1,2}|8\rangle}{\epsilon(1,3,4,6,8)\left(2\sqrt{\frac{1-u_6 v_1-u_4 v_4 (1-u_{1,6}v_1)}{u_{1,4,6}v_{1,4}}}\right)^{-1}}= \frac{1}{i} \mathrm{sign} \langle12\rangle  \mathrm{sign} \langle45\rangle  \mathrm{sign} \langle67\rangle \mathrm{sign} \left(\langle8|p_{1,2}|3\rangle\right).
\end{equation}
In terms of the symbols, the integrated result is simplified to
\begin{equation}
    \begin{gathered}
     \frac{1}{4}\left({-}\chi(u_1)\otimes \chi\left(\frac{(1{-}u_1)u_{4,6} v_{1,4}}{(1{-}u_6 v_1)(1{-}u_4 v_4)}\right){+}\chi(u_{1,4} v_4)\otimes \chi\left(\frac{u_6 v_1(1{-}u_{1,4} v_4)}{1{-}u_4 v_4}\right)\right.
     \\\left.{+}\chi(u_{1,6} v_1)\otimes \chi\left(\frac{u_4 v_4(1{-}u_{1,6} v_1)}{1{-}u_6 v_1}\right)\right).
    \end{gathered}
\end{equation}


\section{The two-loop eight-point amplitude}\label{Section6}
We now return to the complete two-loop integrand,
\begin{equation}
	A_{8}^{\text{$2$-loop}}=A_{8,soft}^{\text{$2$-loop}}+A_{8,max-cut}^{\text{$2$-loop}}+A_{8,tri}^{\text{$2$-loop}}
\end{equation}
where each set is defined in eq.(\ref{SoftAnsw}), eq.(\ref{2LoopMaxCut}) and eq.(\ref{FinalTria}). Schematically, the result is organized as combinations of leading singularities dressed with dual conformal integrals. The leading singularities are organized into the $(A^{\rm tree}_8, \mathcal{B}_{i,j,k}, \mathcal{D}_{i,j,k}, \bar{\mathcal{D}}_{i,j,k})$, where $(\mathcal{B}_{i,j,k}, \mathcal{D}_{i,j,k})$ can be identified as the sum and difference of one-loop maximal cut respectively. Since the one-loop amplitude, eq.(\ref{OneLoopResult}), is given in terms of $\mathcal{B}_{i,j,k}$, its dressing at two-loop can be considered as the correction to the one-loop amplitude. 
\subsection{The integrated result}

Let us first begin with terms proportional to $A^{\rm tree}$ and $\mathcal{D}_{i,j,k}$, since they are linearly dependent (see eq.(\ref{TreeRel})). This entails part of $A_{8,soft}^{\text{$2$-loop}}$ and all of $A_{8,tri}^{\text{$2$-loop}}$. That they should be considered in combination can also be seen from the cancellation of elliptical cuts, as discussed at the beginning of section \ref{Section5}. We can further separate these terms into the part that contains infrared divergences while at the same time free of elliptical and three-point amplitude sub-cut, and the remaining part. For the former, we have
\begin{equation}
	\begin{aligned}
		&A^{\rm tree}_8{\times} \left[\sum_{i=1}^8 I_{B,i,i+3} +\frac{1}{2} I_{C,i}^{db}[n_b]\right]{+}\sum_{i=1}^8 (-1)^i \mathcal{D}_{i,i+2,i+4}{\times} \bigg[ I_{D,i,i+4}+I_{D,i+2,i} \\
		& \qquad \qquad \qquad \qquad +I_{E,i+4,i+2}+I_{E,i+4,i}-\frac{1}{2}I_{G,i}\bigg].
	\end{aligned}
\end{equation} 
where
\begin{equation}
    \begin{aligned}
        &I_{B,i,i+3}:=I^{db}_{B,i,i+3}[n_b]+I^{dt}_{i,i+2;i+2,i+4}\\
        &I_{D,i,i+4}:=I^{db}_{D,i,i+4}[n_b]+I^{dt}_{i,i+2;i+2,i}-I^{dt}_{i,i+2;i+2,i+4}-I^{dt}_{i,i+2;i+4,i}\\
        &I_{D,i+2,i}:=I^{db}_{D,i+2,i}[n_b]+I^{dt}_{i+2,i+4;i+4,i+2}-I^{dt}_{i+2,i+4;i+4,i}-I^{dt}_{i+2,i+4;i,i+2}\\
        &I_{E,i+4,i+2}:=I^{db}_{E,i+4,i+2}[n_b]+I^{dt}_{i+4,i-2;i-2,i+4}-I^{dt}_{i+4,i-2;i+2,i+4}\\
        &I_{E,i+4,i}:=I^{db}_{E,i+4,i}[n_b]+I^{dt}_{i-2,i;i,i+4}-I^{dt}_{i-2,i;i,i+2}\\
        &I_{G,i}:=I^{db}_{G,i}[n_a]-I^{dt}_{i,i+4;i+4,i}+I^{dt}_{i,i+4;i-2,i}+I^{dt}_{i,i+4;i+4,i-2}+I^{dt}_{i,i+2;i+4,i}\\
        &\qquad \qquad    +I^{dt}_{i+2,i+4;i+4,i}-I^{dt}_{i,i+2;i-2,i}-I^{dt}_{i+2,i+4;i+4,i-2}.
    \end{aligned}
\end{equation}

Substituting the integrated result the above collapses into an universal $A_8^{\rm tree}$ prefactor multiplied by the eight-point BDS ansatz of $\mathcal{N}=4$ SYM:
\begin{equation}
    \begin{aligned}
    {\rm BDS}_8+\pi^2=&\, -\frac 13\pi^2+\frac{1}{2} \sum _{i=1}^8 \left(\log ^2\left(\frac{(i\cdot i{+}3)}{(i{+}1\cdot i{+}3)}\right)-\log ^2\left(\frac{4 \mu ^2 (i\cdot i{+}3)}{(i\cdot i{+}2) (i{+}1\cdot i{+}3)}\right)\right)\\
    &\ -\sum _{i=1}^8 \left(\log \left(u_i\right) \log \left(\frac{4 \mu ^2}{(i{-}1\cdot i{+}2)}\right){+}\text{Li}_2\left(1-u_i\right)\right)\\
    &\ -\sum _{i=1}^4 \left(\log \left(v_i\right) \log \left(\frac{4 \mu ^2}{(i{-}1\cdot i{+}3)}\right){+}\text{Li}_2\left(1-v_i\right)\right).
    \end{aligned}\,
\end{equation}
Remarkably the BDS ansatz can be identified with a set of integrands that captures the infrared divergences accompanied by those necessary to cancel unphysical cuts. We will return to this observation at the conclusion. The remaining terms can be organized as  
\begin{equation}
	\begin{aligned}
		&A^{\rm tree}_8  R^{\text{even},A}_8+\sum_{i=1}^8(-1)^i \mathcal{D}_{i,i+2,i+4}  R^{\text{even}, \mathcal{D}}_{i,i+2,i+4}\\
		=&A^{\rm tree}_8   \sum_{i=1}^8 \left[-I^{db}_{A,i,i+4}-I^{dt}_{i,i+2;i+3,i+5}\right]\\
		+&\sum_{i=1}^8(-1)^i \mathcal{D}_{i,i+2,i+4}\left[-I_{F,i-3}^{db}[n_b]+I^{dt}_{i-3,i-1;i,i-4}-I^{dt}_{i-3,i-1;i,i+2}-I^{dt}_{i-3,i-1;i+2,i-4}\right]\,,
	\end{aligned}
\end{equation}
where the superscript $R^{\text{even}}$ indicates that the integrand involves even number of Levi-Cevita tensors. The integrated result is then given as:
\begin{equation}\label{RA}
	\begin{aligned}
    R^{\text{even},A}_8=&{-}\sum_{i=1}^8 \left.\frac{\pi^2}{4}{-}\frac{1}{2}\mathrm{Li}_{2}(1{-}u_{i+1}){-}\frac{1}{2}\mathrm{Li}_{2}(1{-}u_{i+2}){-}\frac{1}{2}\mathrm{Li}_{2}(1{-}v_{i+1}){+}\frac{1}{2}\mathrm{Li}_{2}(1{-}u_{i+1} v_{i+1})\right. \\
    &\qquad \quad \left. +\frac{1}{2}\mathrm{Li}_{2}(1-u_{i+2} v_{i+1})+\frac{1}{4}\log\left(\chi\left({u_{i,i+3} v_{i+3}}\right)\right)^{2}-\frac{1}{2}\log{u_{i+1}}\log{u_{i+2}}\right. ,
    \end{aligned}
    \end{equation}
    
\begin{equation}\label{RD}
    \begin{aligned}
	R^{\text{even}, \mathcal{D}}_{i,i+2,i+4}&= \frac{\pi^2}{6}{-}\frac{1}{4}\log\left(\chi \left(u_{i-3}\right)\right)^2{+}\frac{1}{4}\log\left(\chi\left({u_{i-3,i+2} v_{i+1}}\right)\right)^2{+}\frac{1}{4}\log\left(\chi\left({u_{i-3,i} v_{i}}\right)\right)^2\\
		&{+}\frac{1}{2}\log{u_{i-2} v_{i+2}}\log{u_{i-4} v_{i-1}}{+}\frac{1}{2} \mathrm{Li}_2(1-u_{i-2} v_{i+2}){+}\frac{1}{2} \mathrm{Li}_2(1-u_{i-4} v_{i-1})\,.
			\end{aligned}
\end{equation}
We now consider the remaining terms, which are proportional to $\mathcal{B}_{i,j,k}$. Recall that the double box integrals $I^{db}_{D|E|F}[n_t]$ and $I^{db}_G[n_{b/c}]$ integrate to zero. Thus terms proportional to $\mathcal{B}_{i,j,k}$ are simply box-triangle integrals:  
\begin{equation}\label{Bdressing}
	\begin{aligned}
	&\sum_{i=1}^8 (-1)^i\mathcal{B}_{i,i+2,i+4}	R^{\text{odd}}_{i,i+2,i+4}\\
	=&\sum_{i=1}^8 (-1)^i\mathcal{B}_{i,i+2,i+4} \left[I^{bt}_{A,i-1,i+3}+I^{bt}_{A,i+3,i+1}+I^{bt}_{B,i+1}+I^{bt}_{C,i+1}[n_\beta]\right]\,,
	\end{aligned}
\end{equation}
where the superscript $R^{\text{odd}}$ represents odd number (one) of Levi-Cevita tensors. Using the results in eq.(\ref{eq:result_bt_A}), eq.(\ref{Sign1}), eq.(\ref{eq:result_bt_B+C}) and eq.(\ref{Sign2}), we have:
\begin{equation}\label{RB}
	\begin{aligned}
		R^{\text{odd}}_{i,i{+}2,i{+}4}&=\frac{1}{i}\,  \mathrm{sign} \langle i{-}1\,i\rangle\, \mathrm{sign} \langle i{+}2\,i{+}3\rangle\mathrm{sign} \left(\langle i\,i{+}1\rangle\langle i{-}1|p_{i{+}1,i{+}2,i{+}3}|i\rangle{+}\langle i{-}1\,i{+}1\rangle (i\cdot i{+}4) \right)\\
		&\quad \times F\biggl(u_{i-1,i+2}v_{i+2},\frac{u_{i-1,i+2}v_{i+2}}{1{-}u_iv_i}\biggr)+\frac{1}{i}\,\mathrm{sign}\langle i\,i{+}1\rangle\, \mathrm{sign}\langle i{+}3\,i{+}4\rangle\\
		&\mathrm{sign}\left(\langle i{+}3\,i{+}2\rangle \langle i{+}4|p_{i,i+1,i+2}|i{+}3\rangle{+}\langle i{+}2\, i{+}4\rangle(i\cdot i{+}4)\right){\times}F\biggl(u_{i,i+3}v_{i+3},\frac{u_{i,i+3}v_{i+3}}{1{-}u_{i+2}v_{i+2}}\biggr)\\
		&+\frac{1}{i}\,  \mathrm{sign} \langle i{+}1\,i{+}2\rangle\,    \mathrm{sign} \langle i|p_{i+1,i+2}|i{+}3\rangle
		\times\bigg( F(u_{i+1,i+4}v_i,u_{i+1}){-}F(u_{i-2,i+1}v_{i+1},u_{i+1})\\
		&\left.{+}\frac{1}{4}\log{\chi(u_{i+1})}\log{\frac{u_{i+2} v_{i+2}(1{-}u_{i-2} v_{i+1})}{v_i v_{i+3}(1{-}u_{i+4} v_i)}}\right)\,.
	\end{aligned}
\end{equation}

Finally we have the terms proportional to $\bar{\mathcal{D}}$ is:
\begin{equation}\label{barD}
\sum_{i=1}^8\bar{\mathcal{D}}_{i,i{+}2,i{+}4}R^{\text{even}, \bar{\mathcal{D}}}_{i,i{+}2,i{+}4}=\sum_{i=1}^8\bar{\mathcal{D}}_{i,i{+}2,i{+}4}\left[ \frac{1}{2}I^{kt}_{G,i}+ I^{bt}_{C,i{+}5}[n_\alpha] \right].	
\end{equation}
We find: 
\begin{equation}\label{RDbar}
\begin{aligned}
    &R^{\text{even}, \bar{\mathcal{D}}}_{i,i{+}2,i{+}4}{=} \frac{\pi^2}{2} \mathrm{sign}\langle i\, i{+}1\rangle \mathrm{sign}\langle i{+}2\, i{+}3\rangle\mathrm{sign}\langle i{+}4\, i{+}5\rangle \mathrm{sign}\langle i{+}6\, i{+}7\rangle {+} \frac{1}{i} \mathrm{sign} \langle i{+}5\,i{+}6\rangle\\
    & \qquad \qquad  \mathrm{sign} \langle i\,i{+}1\rangle  \mathrm{sign} \langle i{+}2\,i{+}3\rangle \mathrm{sign} \langle i{+}4|p_{i{+}5,i{+}6}|i{+}7\rangle\left[F\left(u_{i+5},\frac{u_{i,i+2,i+5}v_{i,i+1}}{u_{i+2} v_{i+1}{+}u_i v_i{-}1}\right)\right.\\
&\qquad \qquad  \left.{-}F\left(u_{i,i+5}v_i,\frac{u_{i,i+2,i+5}v_{i,i+1}}{u_{i+2} v_{i+1}{+}u_i v_i{-}1}\right){-}F\left(u_{i+2,i+5}v_{i+1},\frac{u_{i,i+2,i+5}v_{i,i+1}}{u_{i+2} v_{i+1}{+}u_i v_i{-}1}\right)\right].
\end{aligned}
\end{equation}

Putting everything together, we obtain
\begin{equation}\label{FinalAnsw}
\boxed{
\begin{aligned}
		A^{\rm 2-loop}_8&=A^{\rm tree}_8 \times \left[ {\rm BDS}_8+\pi^2 +R^{\text{even},A}_8\right]\\
	&+\sum_{i=1}^8(-1)^i \left[\mathcal{D}_{i,i+2,i+4} R^{\text{even}, \mathcal{D}}_{i,i+2,i+4} + \mathcal{B}_{i,i+2,i+4}	R^{\text{odd}}_{i,i+2,i+4}\right]+\sum_{i=1}^8 \bar{\mathcal{D}}_{i,i+2,i+4}	R^{\text{even}, \bar{\mathcal{D}}}_{i,i{+}2,i{+}4}
	\end{aligned}}
\end{equation}
where $(R^{\text{even},A}_8, R^{\text{even}, \mathcal{D}, \bar{\mathcal{D}}}_{i,i+2,i+4},  R^{\text{odd}}_{i,i+2,i+4})$ are given in eq.(\ref{RA}), (\ref{RD}), (\ref{RDbar}) and (\ref{RB}) respectively.

\subsection{Consistency checks and analytic structure}
The construction of the integrand has already utilized several non-trivial constraints, including the matching of maximal and soft-cuts, the correct soft-collinear divergences, and the absence of un-physical cuts. For the integrated result, we can further check that the amplitude has the correct little-group parity, and the reflection behavior eq.(\ref{ReflectionRule}). As we will see, the integrated result realizes these properties in a non-trivial fashion.      

\paragraph{Little-group parity} We begin with considering $Z_2$ little group transformation of external leg $\Lambda_i\rightarrow -\Lambda_i$. The full amplitude should pick up a sign depending on the fermion number $F_i$ of the particle on leg $i$:
\begin{equation}\label{eq:parity_of_am}
    A_8\left(\Lambda_1,\ldots,-\Lambda_i,\ldots,\Lambda_n\right)=(-1)^{F_i} A_8\left(\Lambda_1,\ldots,\Lambda_i,\ldots,\Lambda_n\right).
\end{equation}
Since the combination of leading singularities appearing in two-loop amplitude have distinct little group parity compared to the tree-amplitude~\eqref{eq:little_gp_parity}, the functions dressing these combinations must provide compensating L
little group weight. \\ 

According to~\eqref{eq:little_gp_parity}, $\mathcal{D}_{i,j,k}$ has the same parity of the amplitude so the functions proportional to  $\mathcal{D}_{i,j,k}$ should be little group neutral. The non-trivial ones are  $\mathcal{B}_{i,j,k}$ and  $\bar{\mathcal{D}}_{i,j,k}$. The former has half number of legs with the same parity and half with the opposite parity, while the latter is totally opposite parity of amplitude. The compensating functions are comprised of sign functions. Let's first look at the function proportional to $\mathcal{B}_{a,a+2,a+4}$. The coefficient $\mathcal{B}_{a,a+2,a+4}$ has opposite little group at legs $i=a,a{+}1,a{+}2,a{+}3$. We can see that the three box-triangles have a sign function that exactly carries non-trivial little group weight at these legs:
\begin{equation}
    \begin{aligned}
        &I^{bt}_{A,i-1,i+3}:\, \mathrm{sign} \langle i{-}1\,i\rangle\, \mathrm{sign} \langle i{+}2\,i{+}3\rangle\mathrm{sign} \left(\langle i\,i{+}1\rangle\langle i{-}1|p_{i{+}1,i{+}2,i{+}3}|i\rangle{+}\langle i{-}1\,i{+}1\rangle (i\cdot i{+}4) \right)\\
        &I^{bt}_{A,i+3,i+1}:\, \mathrm{sign}\langle i\,i{+}1\rangle\, \mathrm{sign}\langle i{+}3\,i{+}4\rangle \mathrm{sign}\left(\langle i{+}3\,i{+}2\rangle \langle i{+}4|p_{i,i+1,i+2}|i{+}3\rangle{+}\langle i{+}2\, i{+}4\rangle(i\cdot i{+}4)\right)\\
        &I^{bt}_{B,i+1}+I^{bt}_{C,i+1}[n_\beta]:\,\mathrm{sign} \langle i{+}1\,i{+}2\rangle\,    \mathrm{sign} \langle i|p_{i+1,i+2}|i{+}3\rangle\,,
    \end{aligned}
\end{equation}
where we set $i=a$. 

For the functions proportional to $\bar{\mathcal{D}}_{a,a+2,a+4}$, the sign functions need to have little group weight at all legs to restore the correct parity of $\bar{\mathcal{D}}_{a,a+2,a+4}$. We can see the that the sign functions in $I^{kt}_{G,i}$ and $I^{bt}_{C,i{+}5}[n_\alpha]$ satisfy this property:
\begin{equation}
    \begin{aligned}
        &I^{kt}_{G,i}:\, \frac{\langle i, i{+}1\rangle}{\sqrt{-(i\cdot i{+}2)-i \epsilon}} \frac{\langle i{+}2, i{+}3\rangle}{\sqrt{-(i{+}2\cdot i{+}4)-i \epsilon}}\frac{\langle i{+}4, i{+}5\rangle}{\sqrt{-(i{+}4\cdot i{+}6)-i \epsilon}} \frac{\langle i{+}6, i{+}7\rangle}{\sqrt{-(i{+}6\cdot i)-i\epsilon}},\\
        &I^{bt}_{C,i{+}5}[n_\alpha]: \, \mathrm{sign} \langle i{+}5,i{+}6\rangle  \mathrm{sign} \langle i,i{+}1\rangle  \mathrm{sign} \langle i{+}2,i{+}3\rangle \mathrm{sign} \langle i{+}4|p_{i{+}5,i{+}6}|i{+}7\rangle.
    \end{aligned}
\end{equation}
Therefore, the full two-loop amplitude has the correct little weight.

\paragraph{Reflection symmetry} We now turn to the reflection symmetry, where the two-loop amplitude should be  invariant under the reflection $\{\Lambda_1,\Lambda_2,\ldots, \Lambda_8 \}\rightarrow\{\Lambda_1,\Lambda_8,\ldots, \Lambda_2 \}$. The reflection rule of the coefficients appearing in the two-loop amplitude is summarized in~\eqref{eq:reflection_rule}. The non-trivial reflection rules imply that the functions proportional to these coefficients have non-trivial properties under reflection. First, let's consider the function proportional to $\mathcal{D}_{i,j,k}$. According to eq.(\ref{eq:reflection_rule}), $\mathcal{D}_{i,j,k}\leftrightarrow-\mathcal{D}_{i',j',k'}$, where $\{i,j,k\}$ and $\{i',j',k'\}$ are reflection pairs and consists of totally even or totally odd legs respectively. The extra sign then exactly matches the alternating sign when cycling shifting by odd sites for the part proportional to $\mathcal{D}_{i,j,k}$ in the final answer eq.(\ref{FinalAnsw}). One can trivially check that the functions proportional to $\mathcal{D}_{i,j,k}$ map to functions proportional to $\mathcal{D}_{i',j',k'}$. The mapping becomes non-trivial when sign functions are involved, which we now turn to.\\

Similar analysis can be applied to the $\bar{\mathcal{D}}_{i,j,k}$ sector. Under reflection $\bar{\mathcal{D}}_{i,j,k}\leftrightarrow \bar{\mathcal{D}}_{i',j',k'}$, where the pair $\{i,j,k\}$ and $\{i',j',k'\}$ are reflection pairs and consists of totally even or totally odd legs respectively. This is consistent with the fact that the cyclic sum for  $\bar{\mathcal{D}}_{i,j,k}$ in eq.(\ref{FinalAnsw}) does not pick up a sign upon cyclic shifts by odd sites. Now we just need to show that the function proportional to $\bar{\mathcal{D}}_{i, j, k}$ maps to $\bar{\mathcal{D}}_{i',j',k'}$  under reflection. Consider for example the reflection pair $\bar{\mathcal{D}}_{1,3,5}\leftrightarrow \bar{\mathcal{D}}_{2,6,8}$. From eq.(\ref{barD}) the functions dressing $\bar{\mathcal{D}}_{1,3,5}$ come from kissing triangle $I^{kt}_{G,1}$ and box-triangle $I^{bt}_{C,6}[n_\alpha]$. We find
\begin{equation}
    \begin{aligned}
        &I^{kt}_{G,1}=\frac{\langle 1 2\rangle}{\sqrt{-(1\cdot 3)-i \epsilon}} \frac{\langle 34\rangle}{\sqrt{-(3\cdot 5)-i \epsilon}}\frac{\langle 5 6\rangle}{\sqrt{-(5\cdot 7)-i \epsilon}} \frac{\langle 78\rangle}{\sqrt{-(7\cdot 1)-i\epsilon}} \pi^2\\
        &\mapsto  \frac{\langle 1 8\rangle}{\sqrt{-(2\cdot 8)-i \epsilon}} \frac{\langle 76\rangle}{\sqrt{-(6\cdot 8)-i \epsilon}}\frac{\langle 5 4\rangle}{\sqrt{-(4\cdot 6)-i \epsilon}} \frac{\langle 32\rangle}{\sqrt{-(2\cdot 4)-i\epsilon}} \pi^2\\
        &\qquad = I^{kt}_{D,2},
    \end{aligned}
\end{equation}
and 
\begin{equation}
\begin{aligned}
    &I^{bt}_{C,6}[n_\alpha]=\frac{1}{i} \mathrm{sign} \langle 12\rangle  \mathrm{sign} \langle 34\rangle \mathrm{sign} \langle 67\rangle \mathrm{sign} \langle 5|p_{6,7}|8\rangle\bigg(F\left(u_{6},\frac{u_{1,3,6}v_{1,2}}{u_{3} v_{2}{+}u_1 v_1{-}1}\right)\\
&\qquad \qquad \qquad  {-}F\left(u_{1,6}v_1,\frac{u_{1,3,6}v_{1,2}}{u_{3} v_{2}{+}u_1 v_1{-}1}\right){-}F\left(u_{3,6}v_{2},\frac{u_{1,3,6}v_{1,2}}{u_{3} v_{2}{+}u_1 v_1{-}1}\right)\bigg)\\
&\mapsto-\frac{1}{i} \mathrm{sign} \langle 18\rangle  \mathrm{sign} \langle 76\rangle \mathrm{sign} \langle 43\rangle \mathrm{sign} \langle 5|p_{3,4}|2\rangle\bigg(F\left(u_{3},\frac{u_{3,6,8}v_{2,3}}{u_{8} v_{3}{+}u_6 v_2{-}1}\right)\\
& \qquad \qquad  {-}F\left(u_{3,6}v_2,\frac{u_{3,6,8}v_{2,3}}{u_{8} v_{3}{+}u_6 v_2{-}1}\right){-}F\left(u_{8,3}v_{3},\frac{u_{3,6,8}v_{2,3}}{u_{8} v_{3}{+}u_6 v_2{-}1}\right)\bigg)\\
&\qquad =I^{bt}_{C,3}[n_\alpha].
\end{aligned}
\end{equation}
Note that the argument in each sign function is actually reversed under the map, since there is an even number of them, the combination is invariant.   \\

Finally, we consider the terms proportional to $\mathcal{B}_{i,j,k}$, which are dressed by box-triangles $I_{A}^{bt}, I_{B}^{bt}, I_C^{bt}$. Without loss of generality, we  focus on the sector $\mathcal{B}_{1,3,5}$. According to~\eqref{eq:reflection_rule}, the sector $\mathcal{B}_{1,3,5}$ will be reflected to the sector $\mathcal{B}_{2,6,8}$. We see explicitly:
\begin{equation}
    \begin{aligned}
        &I_{A,8,4}^{bt}=\frac{1}{i} \mathrm{sign} \langle81\rangle \mathrm{sign} \langle34 \rangle \mathrm{sign}\left( \langle12\rangle \langle8|p_{2,3,4}|1\rangle+\langle82\rangle(1\cdot 5)\right) \times F\left(u_{3,8} v_3,1{-}\frac{u_1 v_1}{u_{3,8} v_3}\right)\\
        &\quad\mapsto\ \frac{1}{i} \mathrm{sign}\langle21\rangle \mathrm{sign} \langle76\rangle \mathrm{sign} \left({-}\langle18\rangle \langle2|p_{6,7,8}|1\rangle{+}\langle28\rangle(2\cdot6)\right) \times F\left(u_{1,6} v_1,1{-}\frac{u_8 v_3}{u_{1,6} v_1}\right)\\
        &\quad   =-I_{A,1,7}^{bt}.
    \end{aligned}
\end{equation}
Here the reflection of $I_{A,8,4}^{bt}$ picks up an extra minus compared to $I_{A,1,7}^{bt}$, due to the odd number of sign-functions. Similarly, 
\begin{equation}
    \begin{aligned}
        &I_{B,2}^{bt}{+}I_{C,2}^{bt}[n_{\beta}]=\frac{1}{i}    \mathrm{sign} \langle23\rangle  \mathrm{sign} \langle1|p_{2,3}|4\rangle\\
        &\quad  \times \left(F(u_{2,5} v_1,u_2)-F(u_{2,7} v_2,u_2)+\frac{1}{4}\log(\chi(u_2))\log{\frac{u_3 v_3(1-u_7 v_2)}{u_1 v_4 (1-u_5 v_1)}}\right)\\
        \mapsto &\frac{1}{i} \mathrm{sign }\langle 87\rangle \mathrm{sign }\langle1|p_{7,8}|6\rangle\\
        &\times \left(F(u_{4,7} v_3,u_7)-F(u_{2,7} v_2,u_7)+\frac{1}{4}\log(\chi(u_7))\log{\frac{u_6 v_1 (1-u_2 v_2)}{u_8 v_4(1-u_4 v_3)}}\right)\\
        =&-\left(I_{B,7}^{bt}{+}I_{C,7}^{bt}[n_{\beta}]\right)\,.
    \end{aligned}
\end{equation}
Again, we see that reflection picks up a minus sign. The overall minus sign then  exactly compensates the minus sign when cyclic rotates from $\mathcal{B}_{1,3,5}$ to $\mathcal{B}_{2,6,8}$.

\paragraph{Symbol letters and analytic structure} Let us summarize all symbol letters that appear in the two-loop eight-point amplitudes. As we have mentioned that the $12$ cross-ratios, $u_{1\leq i\leq 8}$ and $v_{1\leq j\leq 4}$, satisfy constraints thus there are only $6$ of them that are functionally independent. The symbol letters can be divided into two classes: those letters that are simple polynomials of cross-ratios, and those that involve square roots which correspond to pure phases. For the former, there are $40$ letters:
\begin{equation}\label{evenlett}
\{u_i,v_j,1-u_i,1-v_j,1-u_jv_{j},1-u_{j+1}v_{j},1-u_{j+4}v_{j},1-u_{j+5}v_{j}\}_{1\leq i\leq 8,1\leq j\leq 4}\,.
\end{equation}
They are nothing but the symbol letters from functions of the form ${\rm Li}_2\left(1-\frac{(a\cdot c)(b\cdot d)}{(a \cdot b)(c \cdot d)}\right)$ where in the argument, the cross-ratio with $a<b<c<d$ is a monomial of $u_i$, $v_j$ variables. 

As discussed in~\cite{He:2021mme}, these letters belong to the one-loop alphabet of $n=8$ amplitudes in ${\cal N}=4$ SYM; they also give weight-two functions satisfying first-entry conditions and Steinmann relations. Since these letters are parity invariant, they stay unchanged under ``folding'' , or the dimensional reduction to $D=3$ (except that they satisfy more constraints, such that only $6$ are independent). We expect that all parity-invariant symbol letters of ABJM amplitudes can be obtained from dimensional reducing those in sYM.

Finally for the letters $\chi(x)$, these represent phases. To see this, note that 
\begin{equation}
\chi(x)=\frac{\sqrt{x}-\sqrt{x-1}}{\sqrt{x}+\sqrt{x-1}}=e^{-i\theta}.
\end{equation}
where we have parameterize $x=\cos\frac{\theta}{2}$. Thus $\log \chi(x)=-i\theta$ gives a phase. For $0<x<1$ the phase is real and imaginary otherwise.  In our case, $x$ can either be products of cross-ratios, 
\begin{equation}\label{1stchi}
\{\chi (u_i),\chi (u_i u_{i+5} v_{i})\}_{1\leq i\leq 8}
\end{equation}
or rational functions of cross-ratios:
\begin{equation}\label{2ndchi}
\begin{aligned}
\biggl\{&\frac{u_{i,i+5} v_i}{1-u_{i-2} v_{i+2}},\frac{u_{i,i+5} v_i}{1-u_{i-1} v_{i+2}},
\frac{1-u_{i,i+5} v_i}{u_{i-2} v_{i+2}},\frac{1-u_{i, i+5} v_i}{u_{i-1} v_{i+2}},\frac{1-u_{i,i+5} v_i}{1-u_{i+5} v_i},\frac{1-u_{i,i+5} v_i}{1-u_i v_i}
,\\
&\frac{(1-u_i)u_{i+3,i+5}v_{i,i+3}}{(1-u_{i+5}v_i)(1-u_{i+3}v_{i+3})},
\frac{u_{i+5}v_i(1-u_{i,i+3}v_{i+3})}{1-u_{i+3}v_{i+3}},
\frac{u_{i+3}v_{i+3}(1-u_{i,i+5}v_{i})}{1-u_{i+5}v_{i}}\biggr\}_{1\leq i\leq 8}\,.
\end{aligned}
\end{equation}

Recall that for $n=6$, there are only $6$ parity-invariant letters and $3$ phases, $u_i, 1-u_i$ and $\chi(u_i)$ for $i=1,2,3$, thus we have seen a dramatic proliferation of symbol letters from $n=6$ to $n=8$. However, these letters only appear in a few simple functions. We already mentioned that rational letters in \eqref{evenlett} only come from ${\rm Li}_2 \left(1-\frac{(a\cdot c)(b\cdot d)}{(a \cdot b)(c \cdot d)}\right)$ and similar $\log \log$ functions; similarly all the $\chi(x)$ only appear in $\log \log$ terms and $F$ function defined in~\eqref{eq:f-ftn}. To be more precise, both terms of the form $\arccos(\sqrt{x})^2 \sim \log^2 \chi(x)$ in the even part, and those of the form $\log \chi(x) \log \left(\frac{(a\cdot c)(b\cdot d)}{(a \cdot b)(c \cdot d)}\right)$ in the odd part contain only simpler $\chi(x)$ of \eqref{1stchi}; the more complicated ones of \eqref{2ndchi} exclusively appear in $F$ functions defined in \eqref{eq:f-ftn}, which are accompanied with sign functions as prefactors. Therefore, the amplitude only contains three types of weight-two functions. Furthermore, it is straightforward to check that all these functions satisfy physical-discontinuity conditions~\cite{Gaiotto:2010fk}: the first entries are either $x$ or $\chi(x)$ with $x$ being products of $u_i$ and $v_i$ variables. 

Since the phase switches between real and imaginary at $x=0,1$, these are the associated branch points. As we will see, these correspond to collinear and soft limits. Without loss of generality, we set $i=1$ in eq.(\ref{1stchi}) and eq.(\ref{2ndchi}). We find
\begin{equation}
\begin{aligned}
    u_1&=\frac{(1\cdot 3)(4\cdot 8)}{(1\cdot 4)(3\cdot8)}\quad\left\{\begin{array}{cc} =0& \quad p_1\propto p_2 \\ =1& \quad y_1=y_8\; {\textrm or} \;y_3=y_4\end{array}\right. \\
    u_{1,6} v_1&=\frac{(1\cdot 3)(6\cdot 8)}{(1\cdot 6)(3\cdot8)}\quad\left\{\begin{array}{cc} =0& \quad p_1\propto p_2\; {\textrm or}\; p_6 \propto p_7\\ =1& \quad y_1=y_8\end{array}\right. .
    \end{aligned}
\end{equation}
For rational arguments we instead have 
\begin{equation}
\begin{aligned}
    \frac{u_{1,6}v_1}{1-u_7v_3}&=\frac{\frac{(1\cdot 3)(6\cdot 8)}{(1\cdot 6)(3\cdot8)}}{1-\frac{(7\cdot 1)(3\cdot 6)}{(7\cdot 3)(1\cdot 6)}}\quad\left\{\begin{array}{cc} =0& \quad p_1\propto p_2 \\ =1& \quad y_1=y_8, 
    \end{array}\right. \\
    \frac{1-u_{1,6}v_1}{u_7v_3}&=\frac{1-\frac{(1\cdot 3)(6\cdot 8)}{(1\cdot 6)(3\cdot8)}}{\frac{(7\cdot 1)(3\cdot 6)}{(7\cdot 3)(1\cdot 6)}}\quad\left\{\begin{array}{cc} =0& \quad p_7\propto p_8 \\ =1& \quad 
    y_6=y_7\end{array}\right.\\
    \frac{1-u_{1,6}v_1}{(1-u_6v_1)}&=\frac{1-\frac{(1\cdot 3)(6\cdot 8)}{(1\cdot 6)(3\cdot8)}}{1-\frac{(1\cdot 4)(6\cdot 8)}{(1\cdot 6)(4\cdot8)}}\quad\left\{\begin{array}{cc} =0& \quad y_1=y_8 \quad (y_1=y_8+\epsilon y_5) \\ =1& \quad y_6=y_7\; {\textrm or} \; y_3=y_4\end{array}\right. \\
    \frac{(1-u_1)u_{4,6}v_{1,4}}{(1-u_6v_1)(1-u_4 v_4)}&=\frac{(1-\frac{(1\cdot 3)(4\cdot 8)}{(1\cdot 4)(3\cdot8)}) \frac{(1\cdot 4)(6\cdot 8)}{(1\cdot 6)(4\cdot8)}\frac{(4\cdot 6)(8\cdot 3)}{(4\cdot 8)(6\cdot 3)}}{(1-\frac{(1\cdot 4)(6\cdot 8)}{(1\cdot 6)(4\cdot8)})(1-\frac{(4\cdot 6)(8\cdot 3)}{(4\cdot 8)(6\cdot 3)})}\quad\left\{\begin{array}{cc} =0& \quad p_4 \propto p_5\;{\textrm or}\; p_6 \propto p_7 \\ =1& \quad y_1=y_8\quad (y_1=y_8+\epsilon y_7) \end{array}\right. \\
    \frac{u_6v_1(1-u_{1,4}v_4)}{(1-u_4 v_4)}&=\frac{(6\cdot 8)(1\cdot 4)}{(6\cdot 1)(4\cdot 8)}\frac{1-\frac{(1\cdot 3)(4\cdot 6)}{(1\cdot 4)(3\cdot 6)}}{1-\frac{(4\cdot 6)(8\cdot 3)}{(4\cdot 8)(6\cdot 3)}}\quad\left\{\begin{array}{cc} =0& \quad p_6\propto p_7 \\ =1& \quad y_1=y_8\end{array}\right. .
    \end{aligned}
\end{equation}
For limits where the numerator and denominator are approaching zero, we've indicated how it is approached that leads to the limiting value. Note that for some cases, there can be overlapped between collinear and soft limits. It is interesting that these limits invariantly lead to odd-point kinematics.


\section{Conclusion and outlook}
In this paper, we have computed the two-loop eight-point amplitude of ABJM theory. The integrand was determined by constraints involving soft cuts, maximal cuts, the absence of collinear-soft divergences, and the vanishing of cuts involving three-point sub-amplitudes. Further checks were done where  unphysical cuts are absent, i.e., cuts where the tree amplitude vanishes. The complete integration was integrated using Higgs regularization, and the integrated result was checked to satisfy the correct little-group parity and reflection symmetry. 

Interestingly, we can identify a collection of integrals that reproduce the four-dimensional BDS piece just as observed at six points \cite{Caron-Huot:2012sos}. Here the set includes IR divergent integrals as well as those that are needed to cancel unphysical cuts. It is natural to conjecture that this defines the set of integrals that reproduce BDS$_n$ for arbitrary multiplicity. We leave its verification to future work~\cite{progress1}. Note that already at the eight-point, several integrals  contain elliptic pieces, as seen in fig.~\ref{fig:ellptic_cut_1} and fig.~\ref{fig:ellptic_cut_2}. These elliptic pieces cancel since they correspond to unphysical cuts. At ten points, we will have the first non-vanishing elliptic cut:
\begin{figure}[H]   
\begin{center}
$\vcenter{\hbox{\scalebox{1}{	

\begin{tikzpicture}[x=0.75pt,y=0.75pt,yscale=-1,xscale=1]

\draw  [line width=1.2]  (350,150.5) -- (400,150.5) -- (400,200.5) -- (350,200.5) -- cycle ;
\draw  [line width=1.2]  (400,150.5) -- (450,150.5) -- (450,200.5) -- (400,200.5) -- cycle ;
\draw [line width=1.2]    (350,200.5) -- (332.25,218.25) ;
\draw [color={rgb, 255:red, 0; green, 0; blue, 0 }  ,draw opacity=1 ][line width=1.2]    (350,150.5) -- (345.88,146.38) -- (332.25,132.75) ;
\draw [color={rgb, 255:red, 0; green, 0; blue, 0 }  ,draw opacity=1 ][line width=1.2]    (467.75,132.75) -- (450,150.5) ;
\draw [line width=1.2]    (467.75,218.25) -- (450,200.5) ;
\draw [line width=1.2]    (412.5,213.25) -- (400,200.5) ;
\draw [line width=1.2]    (400,200.5) -- (387.35,213.1) ;
\draw [color={rgb, 255:red, 144; green, 19; blue, 254 }  ,draw opacity=1 ][line width=1.2]    (372.5,140) -- (372.5,153.8) -- (372.5,158.4) ;
\draw [color={rgb, 255:red, 144; green, 19; blue, 254 }  ,draw opacity=1 ][line width=1.2]    (411,175.5) -- (397,175.5) -- (389,175.5) ;
\draw [line width=1.2]    (400,150.5) -- (387.5,137.75) ;
\draw [line width=1.2]    (412.65,137.9) -- (400,150.5) ;
\draw [color={rgb, 255:red, 144; green, 19; blue, 254 }  ,draw opacity=1 ][line width=1.2]    (425.5,193) -- (425.5,211.4) ;
\draw [color={rgb, 255:red, 144; green, 19; blue, 254 }  ,draw opacity=1 ][line width=1.2]    (424.1,140.4) -- (424.1,158.8) ;
\draw [color={rgb, 255:red, 144; green, 19; blue, 254 }  ,draw opacity=1 ][line width=1.2]    (370.5,191) -- (370.5,209.4) ;
\draw [line width=1.2]    (400.03,132.65) -- (400,150.5) ;
\draw [line width=1.2]    (400,200.5) -- (399.97,218.35) ;

\draw (366,212) node [anchor=north west][inner sep=0.75pt]   [align=left] {$\displaystyle 1$};
\draw (367,170) node [anchor=north west][inner sep=0.75pt]   [align=left] {$\displaystyle a$};
\draw (419.6,168) node [anchor=north west][inner sep=0.75pt]   [align=left] {$\displaystyle b$};
\draw (330.4,166.3) node [anchor=north west][inner sep=0.75pt]   [align=left] {$\displaystyle 2$};
\draw (367.8,125) node [anchor=north west][inner sep=0.75pt]   [align=left] {$\displaystyle 3$};
\draw (419.4,125) node [anchor=north west][inner sep=0.75pt]   [align=left] {$\displaystyle 6$};
\draw (458.4,167.1) node [anchor=north west][inner sep=0.75pt]   [align=left] {$\displaystyle 7$};
\draw (420.4,212) node [anchor=north west][inner sep=0.75pt]   [align=left] {$\displaystyle 8$};
\draw (470.4,200) node [anchor=north west][inner sep=0.75pt]   [align=left] {$\displaystyle .$};

\end{tikzpicture}

}}}$
\end{center}
\end{figure}
\noindent Thus we expect that the results for two-loop $n\geq 10$ can no longer be expressed in terms of multiple polylogarithms only. 

It is possible to push the frontier to higher loops via a bootstrap program based on perturbative data and various physical constraints. For $n=6$, as a starting point, we may use the symbol alphabet of $9$ letters $\{u_i, 1-u_i, \chi(u_i)\}$, and it is straightforward to construct the space of higher-weight functions satisfying physical discontinuity conditions and Steinmann relations. However, we find that all known constraints so far are insufficient to determine the three-loop amplitude, thus more constraints are needed already there. One possibility is to look for the analog of ${\bar Q}$ anomaly equations~\cite{Caron-Huot:2011dec}, which have played a crucial role in sYM, for ABJM amplitudes; our results up to two-loop eight points provide rich data for ``deriving'' such equations, which in turn can greatly facilitate future perturbative computations and bootstrap in the theory. Starting at $n=8$, the ``folding” of the sYM alphabet~\cite{He:2021eec} contains many more letters (polynomials of cross-ratios similar to those in~\eqref{evenlett}) which we expect to appear for higher-loop ABJM amplitudes. What makes the bootstrap much more difficult, however, is the proliferation of $\chi$-type letters, and the possibility of elliptic symbol letters for higher loops and multiplicities. We leave the study of higher-loop amplitudes and their analytic structure to future works. 

Just as observed at two-loop six points and one-loop general points, the non-analyticity of the amplitude is not limited to poles and branch cuts due to the ubiquitous sign functions. Their presence is crucial for the result to satisfy little-group parity and reflection symmetry. At one loop these sign functions are directly mapped to the kinematics of maximal cut, {\it i.e.} the arguments are the square of external momenta at each corner of the triangle cut. At two-loop things are more complicated. It will be desirable to have a systematic understanding of the structure of the arguments for these sign functions. 

Note that the one-loop integrand in eq. (\ref{OneLoopX}) contains a reference point $X$. The fact that the integrand is independent of this reference point can be verified by a series of tedious Schouten identities. This highly suggests that different choices of $X$ correspond to different ``triangulations'' of a fundamental geometric object, similar to the geometric formulation of the planar integrand of ${\cal N}=4$ sYM, {\it i.e.} the amplituhedron~\cite{Arkani-Hamed:2013jha}. This highly suggests a geometric definition of all-loop planar integrands of ABJM theory. So far, the tree-level ~\cite{Huang:2021jlh, He:2021llb, Lukowski:2021fkf} and four-point multi-loop~\cite{He:2022cup} has been successfully defined. This indicates that a full definition is within reach~\cite{New}.

Finally, the result here provides non-trivial data for a potential pentagon program for ABJM amplitudes~\cite{Basso:2018tif}. In particular, our two-loop eight-point analysis would provide explicit data to constrain the spinor pentagons, which are flux-tube excitations belonging to the bi-fundamental representation.


\acknowledgments
We would like to thank Chi Zhang and Marco Bianchi for useful discussions, as well as Ryota Kojima for collaborations in the initial stage of this project. We also thank Andrei Belitsky for his useful comments on the draft. The research of SH is supported in part by Key Research
Program of CAS, Grant No. XDPB15 and National Natural Science Foundation of China
under Grant No. Grant No. 11935013, 11947301, 12047502, 12047503, 12225510. Y.-t. Huang and C.-K. Kuo are supported by Taiwan Ministry of Science and Technology Grant No. 109-2112-M-002 -020 -MY3 as well as MOST 110-2923-M-002 -016 -MY3.

\appendix

\section{Several identities}\label{sec:identity}
In this appendix, we aim to prove a series of identities used in the text. We prove that the coefficients of the transcendental functions in the box-triangle can be expressible in terms of the product of the sign function.

\noindent
First we consider the coefficient in front of box-triangle $I_{A,1,5}^{bt}$  and prove the following identity:

\begin{equation}\label{eq:coef_IbtA16}
	\begin{aligned}
		\frac{\epsilon(6,1,2,3,4)\sqrt{(2\cdot4\cdot6)}}{\sqrt{2}(2\cdot4)(1\cdot3)(2\cdot6)(4\cdot6)\sqrt{1{-}\frac{1{-}u_2 v_2}{u_{1,4} v_4}}}=\frac{1}{i} \mathrm{sign} \langle12\rangle\, \mathrm{sign} \langle45\rangle\, \mathrm{sign} \left(\langle23\rangle\langle1|p_{3,4,5}|2\rangle{+}\langle13\rangle (2\cdot 6) \right).
	\end{aligned}
\end{equation}
Our strategy is to express the five-dimensional $\epsilon$ symbol in terms of angle brackets.  To do so, we start from the definition of the $\epsilon$-symbol as a determinant and use the translation invariant properties to set $x_2=0$. In doing so, we must remember to normalize the determinant such that $\epsilon(i,j,k,l,m)^2$ agrees with the Gram determinant formula $\epsilon(i_1,\ldots,i_5)\epsilon(j_1,\ldots,j_5):=\mathrm{det}\left[(i_i\cdot j_j)\right]$, since this is the convention used in the main text; this requires an extra factor of $2i \sqrt{2}$. Thus

\begin{equation}
	\begin{aligned}
		\epsilon(6,1,2,3,4)&=2\sqrt{2}\, i\, \text{det}(y_6,y_1,y_2,y_3,y_4)= 2\sqrt{2}\, i\, \text{det}
		\begin{pmatrix}
		\vec{p}_2+\vec{p}_3+\vec{p}_4+\vec{p}_5 & -\vec{p}_1 & 0 & \vec{p}_2 & \vec{p}_2+\vec{p}_3\\
		1 & 1 & 1 &1 &1\\
		(2\cdot6) & 0 & 0 & 0 & (2\cdot 4) &
		\end{pmatrix}\\
		&= 2\sqrt{2}\, i \left(-(2\cdot6) \,\text{det}(p_1,p_2,p_3)-(2\cdot 4) \sum_{i=3,4,5} \, \text{det}(p_1,p_2,p_i) \right).
	\end{aligned}
\end{equation}

\noindent
This determinant can now be evaluated in terms of three-dimensional ones, which in turn give two-brackets: $\text{det}(\vec{p}_i,\vec{p}_j,\vec{p}_k):=\frac{1}{2}\langle ij \rangle \langle jk\rangle\langle ki \rangle  $. This way, we obtain

\begin{equation}\label{eq:signftn idenity 1}
	\epsilon(6,1,2,3,4)= i \sqrt{2} \langle12\rangle \langle23\rangle \left(\langle23\rangle\langle1|p_{3,4,5}|2\rangle+\langle13\rangle(2\cdot6)\right).
\end{equation}

\noindent
The cross-ratio $\sqrt{1-\frac{1-u_2 v_2}{u_{1,4} v_4}}$ also can be expressed in terms of the angle brackets and its form is very similar to $\epsilon(6,1,2,3,4)$
\begin{equation}\label{eq:signftn idenity 2}
	\begin{aligned}
		\sqrt{1-\frac{1-u_2 v_2}{u_{1,4} v_4}}=\sqrt{\frac{\left(\langle23\rangle\langle1|p_{3,4,5}|2\rangle+\langle13\rangle(2\cdot6)\right)^2}{\langle12\rangle^2\langle45\rangle^2(2\cdot6)}}.
	\end{aligned}
\end{equation}
By combining all the ingredients together, we can derive the identity~\eqref{eq:coef_IbtA16}.

\vskip 10mm

\noindent
Performing a similar computation, we  find that the coefficient of transcendental functions in the box-triangle $I_{A,1,7}^{bt}$ is equal

\begin{equation}
	\frac{\epsilon(6,1,2,3,8)\sqrt{(2\cdot6\cdot8)}}{\sqrt{2}(2\cdot8)(1\cdot3)(2\cdot6)(6\cdot8) \sqrt{1{-}\frac{1{-}u_8 v_3}{u_{1,6} v_1}}}=\frac{1}{i} \, \mathrm{sign}\langle67\rangle\, \mathrm{sign}\langle12\rangle\, \mathrm{sign}\left(\langle18\rangle \langle2|p_{6,7,8}|1\rangle{+}\langle82\rangle(2\cdot6)\right),
\end{equation}
the one in front of the combination box-triangle $I^{bt}_{B,1}+I_{C,1}^{bt}[n_{\beta}]$ is as 
\begin{equation}
	\frac{\epsilon(8,1,2,3,4)\sqrt{(2\cdot4\cdot8)}}{\sqrt{2}(2\cdot4)(1\cdot3)(2\cdot8)(4\cdot8) \sqrt{1-u_1^{-1}}}=\frac{1}{i}  \mathrm{sign} \langle12\rangle\,   \mathrm{sign} \langle8|p_{1,2}|3\rangle,
\end{equation}
and the one in front of the  box-triangle $I_{C,1}^{bt}[n_{\alpha}]$ is
\begin{equation}
    \frac{2\langle12\rangle\langle45\rangle\langle67\rangle\langle3|p_{1,2}|8\rangle}{\epsilon(1,3,4,6,8)\left(2\sqrt{\frac{1-u_6 v_1-u_4 v_4 (1-u_{1,6}v_1)}{u_{1,4,6}v_{1,4}}}\right)^{-1}}= \frac{1}{i} \mathrm{sign} \langle12\rangle  \mathrm{sign} \langle45\rangle  \mathrm{sign} \langle67\rangle \mathrm{sign} \langle8|p_{1,2}|3\rangle.
\end{equation}

\section{Leading singularities for eight points}\label{LSApp}
 In this appendix, we will derive the explicit form of the leading singularities at the eight-point. We evaluate the  Grassmannian integral~\eqref{eq:grass} by solving the orthogonal condition ($\pm$ labels two branches) and localizing the minors ($1,2$ labels two solutions), leading to a set of four independent leading singularities for a given minor. 

Here, we choose to localize the Grassmannian on the cell $M_4=0$. On the positive branch, where all ordered minors are positive, the 2 solutions are given as:

\begin{equation}
	\begin{aligned}
		C^{+}_{1(2)}=\begin{pmatrix}
-\lambda_1 & \lambda_2 & -\lambda_3 & \lambda_4 & -\lambda_5 & \lambda_6 & -\lambda_7 & \lambda_8\\
c_{3,1} & c_{3,2} & c_{3,3} &0 &0 &0 &0 &c_{3,8}\\
c_{4,1} & c_{4,2} & c_{4,3} & c_{4,4} & c_{4,5} & c_{4,6} & c_{4,7} & c_{4,8}
\end{pmatrix}
	\end{aligned}
\end{equation}
where
\begin{eqnarray}
    c_{3,1}&=&p_{4,5,6,7}^2+(\langle1|{+}\langle8|)p_{4,5,6,7}|1\rangle\mp\langle23\rangle\sqrt{-p_{4,5,6,7}^2}\nonumber\\
    c_{3,8}&=&p_{4,5,6,7}^2-(\langle1|{+}\langle8|)p_{4,5,6,7}|8\rangle\mp\langle23\rangle\sqrt{-p_{4,5,6,7}^2}\nonumber\\
    c_{3,2}&=&-(\langle1|{+}\langle8|)p_{4,5,6,7}|2\rangle\pm (\langle1|{+}\langle8|)|3\rangle \sqrt{-p_{4,5,6,7}^2}\nonumber\\
    c_{3,3}&=&(\langle1|{+}\langle8|)p_{4,5,6,7}|3\rangle\mp (\langle1|{+}\langle8|)|2\rangle\sqrt{-p_{4,5,6,7}^2}\nonumber\\
    c_{4,1}&=&-\langle46\rangle c_{3,8},\quad c_{4,8}=-\langle 46\rangle c_{3,1},\quad c_{4,2}=-\langle57\rangle c_{3,3},\quad c_{4,3}=-\langle57\rangle c_{3,2}\\
    c_{4,4}&=&p_{4,5,6,7}^2 (\langle1|{+}\langle8|)|6\rangle {-}\langle18\rangle (\langle1|{+}\langle8|) p_{4,5,6,7}|6\rangle \mp \langle23\rangle (\langle1|{+}\langle8|)|6\rangle \sqrt{-p_{4,5,6,7}^2}\nonumber\\
    c_{4,6}&=&-p_{4,5,6,7}^2(\langle1|{+}\langle8|)|4\rangle{+}\langle18\rangle (\langle1|{+}\langle8|)p_{4,5,6,7}|4\rangle \pm \langle23\rangle (\langle1|{+}\langle8|)|4\rangle \sqrt{-p_{4,5,6,7}^2}\nonumber\\
    c_{4,5}&=&\langle23\rangle(\langle1|{+}\langle8|)p_{4,5,6,7}|7\rangle\pm\big (\langle1|{+}\langle8|)p_{4,5,6,7}|7\rangle{+}\langle18\rangle (\langle1|{+}\langle8|)|7\rangle \big)\sqrt{-p_{4,5,6,7}^2}\nonumber\\
    c_{4,7}&=&-\langle23\rangle(\langle1|{+}\langle8|)p_{4,5,6,7}|5\rangle\mp \big((\langle1|{+}\langle8|)p_{4,5,6,7}|5\rangle{+}\langle18\rangle(\langle1|{+}\langle8|)|5\rangle\big) \sqrt{-p_{4,5,6,7}^2}\,.\nonumber
\end{eqnarray}
In the above we've defined solution 1 to be the upper of $\pm, \mp$, and solution 2 to be the lower.  For the solutions in the negative branch, we just need to flip the sign of the matrix element $c_{4,j}\rightarrow -c_{4,j}$ ($j=1,4,6,8$), i.e.

\begin{equation}
	\begin{aligned}
		C^-_{1(2)}=\begin{pmatrix}
-\lambda_1 & \lambda_2 & -\lambda_3 & \lambda_4 & -\lambda_5 & \lambda_6 & -\lambda_7 & \lambda_8\\
c_{3,1} & c_{3,2} & c_{3,3} &0 &0 &0 &0 &c_{3,8}\\
-c_{4,1} & c_{4,2} & c_{4,3} & -c_{4,4} & c_{4,5} & -c_{4,6} & c_{4,7} & -c_{4,8}
\end{pmatrix}.
	\end{aligned}
\end{equation}
The leading singularities arising from $M_4=0$ are then given as:
\begin{equation}
	\begin{aligned}
		&\left.{\rm LS}_{\pm,1}[4]\equiv\frac{1}{4\sqrt{-p_{4,5,6,7}^2}}\frac{\delta^3(p)\delta^{(12)}(C \cdot \eta_I)}{M_1M_2M_3}\right|_{C=C^{\pm}_1}\\
		&\left.{\rm LS}_{\pm,2}[4]\equiv-\frac{1}{4\sqrt{-p_{4,5,6,7}^2}}\frac{\delta^3(p)\delta^{(12)}(C \cdot \eta_I)}{M_1M_2M_3}\right|_{C=C^{\pm}_2}\,,
	\end{aligned}
\end{equation}
where the factor $\frac{1}{4\sqrt{-p_{4,5,6,7}^2}}$ stems from the Jacobian factor for solving the orthogonal constraint and localizing on $M_4=0$. Remaining leading singularities ${\rm LS}_{\pm,1(2)}[i]$ arising from localizing on $M_i=0$, for $i=1,2,3$,  can be obtained from cyclic shifting the indices $\Lambda_i\rightarrow\Lambda_{i{-}1}$, and shifting column $i$ of $C$ to $i{-}1$,  such that instead of $M_4=0$, one has $M_3=0$, $M_2=0$ and $M_1=0$ under each successive shift.
The leading singularities ${\rm LS}_{\pm,1(2)}[i]$ for $i=1,2,3$ are then identified as 
\begin{equation}
    \begin{aligned}
        &{\rm LS}_{\pm,1}[1]\equiv-\left. \frac{1}{4\sqrt{-p_{1,2,3,4}^2}}\frac{\delta^3(p)\delta^{(12)}(C \cdot \eta_I)}{M_2M_3M_4}\right|_{C=C^{\mp}_1|_{i\rightarrow i{-}3}}\\
        &{\rm LS}_{\pm,2}[1]\equiv\left. \frac{1}{4\sqrt{-p_{1,2,3,4}^2}}\frac{\delta^3(p)\delta^{(12)}(C \cdot \eta_I)}{M_2M_3M_4}\right|_{C=C^{\mp}_2|_{i\rightarrow i{-}3}}
    \end{aligned}
\end{equation}

\begin{equation}
    \begin{aligned}
        &{\rm LS}_{\pm,1}[2]\equiv\left. \frac{1}{4\sqrt{-p_{2,3,4,5}^2}}\frac{\delta^3(p)\delta^{(12)}(C \cdot \eta_I)}{M_1M_3M_4}\right|_{C=C^{\pm}_1|_{i\rightarrow i{-}2}}\\
        &{\rm LS}_{\pm,2}[2]\equiv\left. -\frac{1}{4\sqrt{-p_{2,3,4,5}^2}}\frac{\delta^3(p)\delta^{(12)}(C \cdot \eta_I)}{M_1M_3M_4}\right|_{C=C^{\pm}_2|_{i\rightarrow i{-}2}}
    \end{aligned}
\end{equation}

\begin{equation}
    \begin{aligned}
        &{\rm LS}_{\pm,1}[3]\equiv\left. -\frac{1}{4\sqrt{-p_{3,4,5,6}^2}}\frac{\delta^3(p)\delta^{(12)}(C \cdot \eta_I)}{M_1M_2M_4}\right|_{C=C^{\mp}_1|_{i\rightarrow i{-}1}}\\
        &{\rm LS}_{\pm,2}[3]\equiv\left. \frac{1}{4\sqrt{-p_{3,4,5,6}^2}}\frac{\delta^3(p)\delta^{(12)}(C \cdot \eta_I)}{M_1M_2M_4}\right|_{C=C^{\mp}_2|_{i\rightarrow i{-}1}}\,.
    \end{aligned}
\end{equation}
Trivially from construction, under cyclic shit of the on-shell data $\Lambda_i\rightarrow\Lambda_{i-1}$, they are related as:
\begin{equation}\label{ShiftSym}
    \begin{aligned}
        &{\rm LS}_{+,1(2)}[4]\rightarrow{\rm LS}_{-,1(2)}[3]\rightarrow-{\rm LS}_{+,1(2)}[2]\rightarrow-{\rm LS}_{-,1(2)}[1]\rightarrow{\rm LS}_{+,2(1)}[4],\\
        &{\rm LS}_{-,1(2)}[4]\rightarrow-{\rm LS}_{+,1(2)}[3]\rightarrow-{\rm LS}_{-,1(2)}[2]\rightarrow{\rm LS}_{+,1(2)}[1]\rightarrow{\rm LS}_{-,1(2)}[4].
    \end{aligned}
\end{equation}
Another useful symmetry often discussed in ABJM theory is reflection symmetry:
\begin{equation}
    \begin{aligned}
        \{\Lambda_1,\Lambda_2,\ldots, \Lambda_8 \}\rightarrow\{\Lambda_1,\Lambda_8,\ldots, \Lambda_2 \}.
    \end{aligned}
\end{equation}
Under reflection, these leading singularities are related by
\begin{equation}\label{ReflectSym}
    \begin{aligned}
        &{\rm LS}_{+,1(2)}[1]\leftrightarrow  - {\rm LS}_{-,2(1)}[2], \quad  {\rm LS}_{-,1(2)}[1]\leftrightarrow   -{\rm LS}_{+,1(2)}[2],
    \end{aligned}
\end{equation}
\begin{equation}
    \begin{aligned}
        {\rm LS}_{\pm,1(2)}[3]\leftrightarrow  - {\rm LS}_{\mp,2(1)}[4].
    \end{aligned}
\end{equation}
As well, for parity $\Lambda_i\rightarrow-\Lambda_i$:
\begin{equation}
    \begin{aligned}
        {\rm LS}_{+,1(2)}[1]\leftrightarrow (-)^{F_i}{\rm LS}_{-,2(1)}[1] \quad \text{for $i=1,2,3,4$},\\
        {\rm LS}_{+,1(2)}[1]\leftrightarrow (-)^{F_i}{\rm LS}_{-,1(2)}[1] \quad \text{for $i=5,6,7,8$};\\
    \end{aligned}
\end{equation}
\begin{equation}
    \begin{aligned}
        {\rm LS}_{+,1(2)}[2]\leftrightarrow (-)^{F_i}{\rm LS}_{-,2(1)}[2] \quad \text{for $i=2,3,4,5$},\\
        {\rm LS}_{+,1(2)}[2]\leftrightarrow (-)^{F_i}{\rm LS}_{-,1(2)}[2] \quad \text{for $i=1,6,7,8$};\\
    \end{aligned}
\end{equation}
\begin{equation}
    \begin{aligned}
        {\rm LS}_{+,1(2)}[3]\leftrightarrow (-)^{F_i}{\rm LS}_{-,2(1)}[3] \quad \text{for $i=3,4,5,6$},\\
        {\rm LS}_{+,1(2)}[3]\leftrightarrow (-)^{F_i}{\rm LS}_{-,1(2)}[3] \quad \text{for $i=1,2,7,8$};\\
    \end{aligned}
\end{equation}
\begin{equation}
    \begin{aligned}
        {\rm LS}_{+,1(2)}[4]\leftrightarrow (-)^{F_i}{\rm LS}_{-,2(1)}[4] \quad \text{for $i=4,5,6,7$},\\
        {\rm LS}_{+,1(2)}[4]\leftrightarrow (-)^{F_i}{\rm LS}_{-,1(2)}[4] \quad \text{for $i=1,2,3,8$};\\
    \end{aligned}
\end{equation}
where $F_i$ is the fermion number of the particle on leg $i$.

Component amplitudes are obtained from the superamplitude by integrating out appropriate $\eta_{i,I}$s. For example, $A_{\bar{\phi} \phi \bar{\psi} \phi \bar{\psi} \psi \bar{\phi} \psi }$ can be obtain by integrating out $\eta_{i,I}$ with $i=1,6,7,8$. This specific choice is convenient in the sense that only ${\rm LS}_{\pm,1}[4]$ contributes while ${\rm LS}_{\pm,1}[2]$ vanishes. The explicit contribution from ${\rm LS}_{\pm,1}[4]$ is given as:
\begin{equation}
	\begin{aligned}
		&{\rm LS}_{\pm,1}[4]|_{\eta_{2}\eta_{3}\eta_{4}\eta_{5}}=\frac{1}{4\sqrt{-p_{4,5,6,7}^2}}\frac{\delta^3(p)}{p_{1,2,3}^2 p_{4,5,6}^2  p_{5,6,7}^2p_{8,1,2}^2} n_{\pm,1} \tilde{n}_{\pm,1}\\
		&{\rm LS}_{\pm,2}[4]|_{\eta_{2}\eta_{3}\eta_{4}\eta_{5}}=-\frac{1}{4\sqrt{-p_{4,5,6,7}^2}}\frac{\delta^3(p)}{p_{1,2,3}^2 p_{4,5,6}^2  p_{5,6,7}^2p_{8,1,2}^2} n_{\pm,2} \tilde{n}_{\pm,2}
	\end{aligned}
\end{equation}
where the subscript indicates selecting the term proportional to $\prod_{I=1,2,3}\eta_{2,I}\eta_{3,I}\eta_{4,I}\eta_{5,I}$, and
\begin{equation}
	\begin{aligned}
		&n_{+,1(2)}=\mp \sqrt{-p_{4,5,6,7}^2}(\langle67\rangle \langle\!\langle 14 \rangle\!\rangle+\langle23\rangle  \langle\!\langle 58 \rangle\!\rangle)+ \langle\!\langle 48 \rangle\!\rangle(-\langle45\rangle\langle81\rangle+\langle23\rangle\langle67\rangle)-p_{4,5,6,7}^2  \langle\!\langle 15 \rangle\!\rangle, \\
	&\tilde{n}_{-,1(2)}=\pm \sqrt{-p_{4,5,6,7}^2}(\langle81\rangle \langle\!\langle 36 \rangle\!\rangle+\langle45\rangle \langle\!\langle 72 \rangle\!\rangle)+\langle\!\langle 73 \rangle\!\rangle (\langle45\rangle \langle81\rangle-\langle23\rangle \langle67\rangle)+p_{4,5,6,7}^2 \langle\!\langle 62 \rangle\!\rangle,\\
	&\qquad \quad  n_{-,1(2)}=n_{+,1(2)}(\lambda_4\rightarrow-\lambda_4), \quad  \text{and} \quad  \tilde{n}_{-,1(2)}=\tilde{n}_{+,1(2)}(\lambda_4\rightarrow-\lambda_4),
	\end{aligned}
\end{equation}
where the bracket represents $\langle\!\langle ij \rangle\!\rangle:=\langle i|p_{i,i+1,\cdots,j}|j\rangle$. The tree amplitude is then given by the sum of the four leading singularities
\begin{equation}
	\begin{aligned}
		&A_8^{\text{tree}}(\bar{\phi} \phi \bar{\psi} \phi \bar{\psi} \psi \bar{\phi} \psi )=\frac{\delta^3(p)}{p_{1,2,3}^2 p_{4,5,6}^2  p_{5,6,7}^2p_{8,1,2}^2} \times \bigg( -p_{4,5,6,7}^2 \left(\langle23\rangle \langle\!\langle 58 \rangle\!\rangle \langle\!\langle 62 \rangle\!\rangle  +\langle81 \rangle\langle\!\langle 36 \rangle\!\rangle \langle\!\langle 15 \rangle\!\rangle \right)\\
		& +\langle45\rangle \langle67\rangle \left(\langle23\rangle\langle\!\langle 48 \rangle\!\rangle \langle\!\langle 72 \rangle\!\rangle+\langle81\rangle \langle\!\langle 37 \rangle\!\rangle \langle\!\langle 14 \rangle\!\rangle\right)-\langle45\rangle \langle81\rangle^2 \langle\!\langle 48 \rangle\!\rangle \langle\!\langle 36 \rangle\!\rangle-\langle67\rangle \langle23\rangle^2 \langle\!\langle 37 \rangle\!\rangle \langle\!\langle 58 \rangle\!\rangle \bigg)\,.
	\end{aligned}
\end{equation}


\section{Cyclicity, reflection, and parity of $\mathcal{B}_{i,j,k}$, $\mathcal{D}_{i,j,k}$, $\bar{\mathcal{D}}_{i,j,k}$}
In this appendix, we will discuss how $\mathcal{B}_{i,j,k}$, $\mathcal{D}_{i,j,k}$, and  $\bar{\mathcal{D}}_{i,j,k}$, which we define in our main text, transform under the various symmetries.\\

First, we consider the cyclic by one site $\Lambda_i\rightarrow\Lambda_{i-1}$:
\begin{equation}
    \begin{gathered}
    \mathcal{B}_{1,3,5}\rightarrow \mathcal{B}_{4,6,8}\rightarrow -\mathcal{B}_{1,3,7}\rightarrow -\mathcal{B}_{2,4,6}\rightarrow \mathcal{B}_{1,5,7}\rightarrow\mathcal{B}_{2,4,8}\rightarrow-\mathcal{B}_{3,5,7}\rightarrow-\mathcal{B}_{2,6,8}\rightarrow\mathcal{B}_{1,3,5};\\
    \mathcal{D}_{1,3,5}(\mathcal{D}_{1,5,7})\rightarrow -\bar{\mathcal{D}}_{2,4,8}(\bar{\mathcal{D}}_{4,6,8})\rightarrow-\mathcal{D}_{1,3,7}(\mathcal{D}_{3,5,7})\rightarrow\bar{\mathcal{D}}_{2,4,6}(\bar{\mathcal{D}}_{2,6,8})\rightarrow\mathcal{D}_{1,3,5}(\mathcal{D}_{1,5,7}),\\
     \bar{\mathcal{D}}_{1,3,5} (\bar{\mathcal{D}}_{1,5,7})\rightarrow \mathcal{D}_{2,4,8}(\mathcal{D}_{4,6,8})\rightarrow-\bar{\mathcal{D}}_{1,3,7}(\bar{\mathcal{D}}_{3,5,7})\rightarrow-\mathcal{D}_{2,4,6}(\mathcal{D}_{2,6,8})\rightarrow\bar{\mathcal{D}}_{1,3,5} (\bar{\mathcal{D}}_{1,5,7}).
    \end{gathered}
\end{equation}
For the reflection symmetry $\{\Lambda_1,\Lambda_2, \ldots, \Lambda_8 \} \rightarrow \{\Lambda_1,\Lambda_8, \ldots, \Lambda_2 \} $:
\begin{equation}\label{eq:reflection_rule}
      \begin{gathered}
       \mathcal{B}_{1,3,5}\leftrightarrow\mathcal{B}_{2,6,8}, \quad \mathcal{B}_{1,3,7}\leftrightarrow\mathcal{B}_{2,4,8}, \quad \mathcal{B}_{1,5,7}\leftrightarrow  \mathcal{B}_{2,4,6}, \quad \mathcal{B}_{3,5,7}\leftrightarrow  \mathcal{B}_{4,6,8};\\
        \mathcal{D}_{1,3,5}(\mathcal{D}_{1,5,7})\leftrightarrow-\mathcal{D}_{2,6,8}(\mathcal{D}_{2,4,6}), \quad \mathcal{D}_{1,3,7}(\mathcal{D}_{3,5,7})\leftrightarrow-\mathcal{D}_{2,4,8}(\mathcal{D}_{4,6,8});\\
       \bar{ \mathcal{D}}_{1,3,5}(\bar{\mathcal{D}}_{1,5,7})\leftrightarrow\bar{\mathcal{D}}_{2,6,8}(\bar{\mathcal{D}}_{2,4,6}), \quad \bar{\mathcal{D}}_{1,3,7}(\bar{\mathcal{D}}_{3,5,7})\leftrightarrow\bar{\mathcal{D}}_{2,4,8}(\bar{\mathcal{D}}_{4,6,8}).
      \end{gathered}
\end{equation}
Finally, for the parity transformation $\Lambda_i\rightarrow -\Lambda_i$:
\begin{equation}\label{eq:little_gp_parity}
    \begin{gathered}
     \mathcal{B}_{a,a{+}2,a{+}4}\rightarrow (-)^{F_i+1} \mathcal{B}_{a,a{+}2,a{+}4}\ \text{for $i=a,a{+}1,a{+}2,a{+}3$},\\
        \mathcal{B}_{a,a{+}2,a{+}4}\rightarrow (-)^{F_i} \mathcal{B}_{a,a{+}2,a{+}4}\ \text{for $i=a{+}4,a{+}5,a{+}6,a{+}7$};\\ 
        \mathcal{D}_{a,a{+}2,a{+}4}\rightarrow (-)^{F_i} \mathcal{D}_{a,a{+}2,a{+}4}\ \text{for $i=1,2,\ldots,8$};\\
        \bar{\mathcal{D}}_{a,a{+}2,a{+}4}\rightarrow (-)^{F_i+1} \bar{\mathcal{D}}_{a,a{+}2,a{+}4}\  \text{for $i=1,2,\ldots,8$}\,,
    \end{gathered}
\end{equation}
where the legs that transforms as $(-)^{F_i{+}1}$ has opposite parity.

\section{Numerators of box-triangle integrals}\label{BoxTriaApp}
In this section, we demonstrate how the numerators of box-triangles $I^{bt}_{C,i}[n_{\alpha(\beta)}]$ are constructed. This topology will contribute to the two-loop maximal cut $\mathcal{C}^{i+1}_{\includegraphics[scale=0.13]{Cut2Sym.pdf} }$, and on the cut the numerators $n^{bt}_{C,i,\alpha(\beta)}$ are required to reproduce the sign patterns in \eqref{eq:unity2}. Furthermore, this topology will contribute to the collinear-soft divergence, and the numerator is required to vanish at these kinematic points. We use the $i=1$ as an example.\\ 

We begin with the numerator $n^{bt}_{C,i=1,\alpha}$. Since $n^{bt}_{C,1,\alpha}$ is evaluated to $+1$ under all four maximal cut solutions, its numerator can only be inner product $(b\cdot j)$s. We write the ansatz for the  numerator as $\alpha_1 (b\cdot1){+}\alpha_3(b\cdot 3){+}\alpha_4(b\cdot 4){+}\alpha_6(b\cdot 6){+}\alpha_8(b\cdot 8)$. Now, imposing that our ansatz vanishes under collinear region $y_a=a_1 y_1+a_8 y_8$, $y_b=b_1 y_1+b_8 y_8$ gives:
\begin{equation}\label{eq:scalar_collinear_1}
    \alpha_3 \left(b_1 (1\cdot3)+b_8(3\cdot8)\right)+\alpha_4 \left(b_1 (1\cdot4)+b_8(4\cdot8)\right)+\alpha_6 \left(b_1 (1\cdot6)+b_8(6\cdot8)\right)=0.
\end{equation}
while vanishing in the region $y_a=a_3 y_3+a_4 y_4$, $y_b=b_3 y_3+b_4 y_4$ gives:
\begin{equation}\label{eq:scalar_collinear_2}
    \alpha_1 \left(b_3(1\cdot3)+b_4(1\cdot4)\right)+\alpha_6 \left(b_3(3\cdot6)+b_4(4\cdot6)\right)+\alpha_8 \left(b_3(3\cdot8)+b_4(4\cdot8)\right)=0.
\end{equation}
Solving eq.~\eqref{eq:scalar_collinear_1} and eq.~\eqref{eq:scalar_collinear_2}, the coefficients can be determined up to a normalization factor:
\begin{equation}
    \begin{aligned}
    		 \left\{\begin{array}{l}
\alpha_1=N\left((3\cdot6)(4\cdot8)-(3\cdot8)(4\cdot6)\right) \\
\alpha_3=-N\left((1\cdot4)(6\cdot8)-(1\cdot6)(4\cdot8)\right) \\
\alpha_4=N\left((1\cdot3)(6\cdot8)-(1\cdot6)(3\cdot8)\right) \\
\alpha_6=-N\left((1\cdot3)(4\cdot8)-(1\cdot4)(3\cdot8)\right) \\
\alpha_8=N\left((1\cdot3)(4\cdot6)-(1\cdot4)(3\cdot6)\right) 
\end{array}\right.
    \end{aligned}
\end{equation}
and $N$ can be determined to be $\frac{2\langle12\rangle\langle45\rangle\langle67\rangle\langle3|p_{1,2}|8\rangle}{\epsilon(1,3,4,6,8)}$ by requiring that the integrand evaluate to +1 on the maximal cut.\\

We turn to fix the numerator $n^{bt}_{C,i=1,\beta}$. As discussed in sec~\ref{sec:maximal_cut}, the soft-colinear divergence of box-triangle $I^{bt}_{C,1}[n_\beta]$ should cancel against $I^{bt}_{B,1}$. Choosing the numerator of $I^{bt}_{B,1}$ to be $\epsilon(b,1,2,3,4)\sqrt{(2\cdot4\cdot8)}/\sqrt{2}(2\cdot4)$, it evaluates to zero in the soft-collinear region $y_a=a_3 y_3+a_4 y_4$, $y_b=b_3 y_3+b_4 y_4$. Therefore, the numerator $n_\beta$ should also vanish in this region. According to the sign pattern~\eqref{eq:unity2} of maximal cut~$\mathcal{C}^{2}_{\includegraphics[scale=0.13]{Cut2Sym.pdf} }$, the numerator should reflect the sign change which switch $b_+$ to  $b_-$. The possible numerator $n^{bt}_{C,1,\beta}$ is
\begin{equation}
    n^{bt}_{C,1,\beta}=\beta_1 \epsilon(b,3,4,6,8)+\beta_8 \epsilon(b,1,3,4,6).
\end{equation}
Now imposing that two box-triangle $I^{bt}_{B,1}$ and $I^{bt}_{C,1}[n_\beta]$ cancels each other in the other collinear region $y_a=a_1 y_1+a_8 y_8$, $y_b=b_1 y_1+b_8 y_8$ yields:
\begin{equation}
    \frac{\epsilon(1,2,3,4,8)\sqrt{(2\cdot4\cdot8)}}{\sqrt{2}(2\cdot4)(2\cdot8)(a\cdot3)(b\cdot4)}+\frac{\left(b_1 \beta_1+b_8 \beta_8\right)\epsilon(1,3,4,6,8)}{(b_1 (1\cdot6)+b_8 (6\cdot8))(a\cdot3)(b\cdot4)}=0.
\end{equation}
The solution of above equation is
\begin{equation}\label{eq:soln_of_tensor_bt}
\left\{\begin{aligned}
\beta_1&=-\frac{(1\cdot6)\epsilon(1,2,3,4,8)\sqrt{(2\cdot4\cdot8)}}{\sqrt{2}(2\cdot4)(2\cdot8)\epsilon(1,3,4,6,8)}\\
\beta_8&=-\frac{(6\cdot8)\epsilon(1,2,3,4,8)\sqrt{(2\cdot4\cdot8)}}{\sqrt{2}(2\cdot4)(2\cdot8)\epsilon(1,3,4,6,8)}
\end{aligned}\right.\ .
\end{equation}
One can check that the solution~\eqref{eq:soln_of_tensor_bt} under the maximal cut~$\mathcal{C}^{2}_{\includegraphics[scale=0.13]{Cut2Sym.pdf} }$ is unity.

Note that the above solution stems from a particular choice for the numerator of $I^{bt}_{B,1}$,  $\epsilon(b,1,2,3,4)\sqrt{(2\cdot4\cdot8)}/\sqrt{2}(2\cdot4)$. One could have started with another viable choice $\epsilon(b,1,2,3,8)\sqrt{(2\cdot4\cdot8)}/\sqrt{2}(2\cdot8)$, and proceed with the same procedure to determine another corresponding $n_\beta$. Here we show that the two choices are equivalent. We apply Schouten identity on the box-triangle $I_{B,1}^{bt}$:
\begin{equation}
    \begin{aligned}
        &\frac{\epsilon(a,1,2,3,4)\sqrt{(2\cdot4\cdot8)}}{\sqrt{2}(2\cdot4)(a\cdot1)(a\cdot2)(a\cdot3)(a\cdot b)(b\cdot 4)(b\cdot 8)}\\
        =& \frac{\epsilon(1,2,3,4,8)\sqrt{(2\cdot4\cdot8)}}{\sqrt{2}(2\cdot4)(2\cdot8)(a\cdot1)(a\cdot3)(a\cdot b)(b\cdot4)(b\cdot8)}\\
        +&\frac{\epsilon(a,1,2,3,8)\sqrt{(2\cdot4\cdot8)}}{\sqrt{2}(2\cdot8)(a\cdot1)(a\cdot2)(a\cdot3)(a\cdot b)(b\cdot 4)(b\cdot 8)}.
    \end{aligned}
\end{equation}
It produces the other choice of numerator for $I^{bt}_{B,1}$ and an extra double triangle with the tensor numerator. On the other hand if we apply Schouten identity on $(1\cdot6)\epsilon(b,3,4,6,8)$ of the box-triangle $I_{C,1}^{bt}[n_\beta]$:
\begin{equation}
    \begin{aligned}
        &-\frac{(1\cdot6)\epsilon(b,3,4,6,8)+(6\cdot8)\epsilon(b,1,3,4,6)}{\sqrt{2}(2\cdot4)(2\cdot8)(a\cdot1)(a\cdot3)(a\cdot b)(b\cdot4)(b\cdot6)(b\cdot8)}\frac{\epsilon(1,2,3,4,8)\sqrt{(2\cdot4\cdot8)}}{\epsilon(1,3,4,6,8)}\\
        =&-\frac{(3\cdot6)\epsilon(b,6,8,1,4)+(4\cdot6)\epsilon(b,6,8,1,3)}{\sqrt{2}(2\cdot4)(2\cdot8)(a\cdot1)(a\cdot3)(a\cdot b)(b\cdot4)(b\cdot6)(b\cdot8)}\frac{\epsilon(1,2,3,4,8)\sqrt{(2\cdot4\cdot8)}}{\epsilon(1,3,4,6,8)}\\
        &-\frac{\epsilon(1,2,3,4,8)\sqrt{(2\cdot4\cdot8)}}{\sqrt{2}(2\cdot4)(2\cdot8)(a\cdot1)(a\cdot3)(a\cdot b)(b\cdot4)(b\cdot8)}.
    \end{aligned}
\end{equation}
It creates another choice of numerator of $I^{bt}_{C,1}[n_\beta]$ and the same double triangle with opposite sign. Hence, the two choices of tensor numerators are equivalent.


\section{The $+$ of $\delta^+(P^2)$ in maximal cuts}\label{AppKin}
In section~\ref{sec:maximal_cut}, we've constructed a set of numerators that evaluate to $\pm1$ on the maximal cut, with the sign depending on the cut solution. To check that the numerators do the job, sometimes we need to resort to numerics. However special care is needed when selecting the numerical cut solutions, since the solution needs to be forward-pointing, i.e., we have $\delta^+(\ell^2)$ where $\ell^0>0$.  This means that we should only consider solutions where along each close loop, we can identify a direction of the loop where the momenta are always pointing in the future direction. We will use the numerator $n_{C,i=1,\beta}^{bt}$ of the box-triangle to illustrate this subtlety. 

First, we consider the rule of reading out the momentum flow between the regions. For each vertex, we consider a clockwise orientation of (red) arrows connecting different regions. If an arrow points from region $i$ to $j$, it represents that the momentum flow between the two regions is given by $y_j-y_i$, with the direction given by the sign of the first component of $y_j-y_i$. If the sign is positive, then the momentum flow is pointing outward from the vertex. A negative sign represents pointing inward toward the vertex. For example, let's consider the quartic vertex below, where the sign of the difference in clockwise directions is: 
\begin{figure}[H]   
\begin{center}
\scalebox{1}{	

\begin{tikzpicture}[x=0.75pt,y=0.75pt,yscale=-1,xscale=1]

\draw [line width=1.2]    (148.2,107.2) -- (251,106.6) ;
\draw [line width=1.2]    (200.6,149.8) -- (200.2,70.6) ;
\draw [color={rgb, 255:red, 208; green, 2; blue, 27 }  ,draw opacity=1 ]   (174.2,115.6) -- (173.86,98.4) ;
\draw [shift={(173.8,95.4)}, rotate = 88.87] [fill={rgb, 255:red, 208; green, 2; blue, 27 }  ,fill opacity=1 ][line width=1.2]  [draw opacity=0] (8.93,-4.29) -- (0,0) -- (8.93,4.29) -- cycle    ;
\draw [color={rgb, 255:red, 208; green, 2; blue, 27 }  ,draw opacity=1 ]   (221.8,115.8) -- (221.8,97.2) ;
\draw [shift={(221.8,118.8)}, rotate = 270] [fill={rgb, 255:red, 208; green, 2; blue, 27 }  ,fill opacity=1 ][line width=1.2]  [draw opacity=0] (8.93,-4.29) -- (0,0) -- (8.93,4.29) -- cycle    ;
\draw [color={rgb, 255:red, 208; green, 2; blue, 27 }  ,draw opacity=1 ]   (210.8,91.35) -- (189,91) ;
\draw [shift={(213.8,91.4)}, rotate = 180.92] [fill={rgb, 255:red, 208; green, 2; blue, 27 }  ,fill opacity=1 ][line width=1.2]  [draw opacity=0] (8.93,-4.29) -- (0,0) -- (8.93,4.29) -- cycle    ;
\draw [color={rgb, 255:red, 208; green, 2; blue, 27 }  ,draw opacity=1 ]   (211.6,122.2) -- (189.8,121.85) ;
\draw [shift={(186.8,121.8)}, rotate = 0.92] [fill={rgb, 255:red, 208; green, 2; blue, 27 }  ,fill opacity=1 ][line width=1.2]  [draw opacity=0] (8.93,-4.29) -- (0,0) -- (8.93,4.29) -- cycle    ;

\draw (169.1,120.2) node [anchor=north west][inner sep=0.75pt]   [align=left] {$\displaystyle 1$};
\draw (169.5,77.6) node [anchor=north west][inner sep=0.75pt]   [align=left] {$\displaystyle 2$};
\draw (217.5,77.8) node [anchor=north west][inner sep=0.75pt]   [align=left] {$\displaystyle 3$};
\draw (218.3,119.4) node [anchor=north west][inner sep=0.75pt]   [align=left] {$\displaystyle 4$};
\draw (268.2,68.4) node [anchor=north west][inner sep=0.75pt]   [align=left] {$\displaystyle  \begin{aligned}
y^0_{2} -y^0_{1} &=(+)\\
y^0_{3} -y^0_{2} &=(-)\\
y^0_{4} -y^0_{3} &=(+)\\
y^0_{1} -y^0_{4} &=(-)
\end{aligned}$};
\end{tikzpicture}
}
\end{center}
\end{figure}
\noindent
Following our rule, momentum $p_1$ and $p_3$ are outgoing and $p_2$ and $p_4$ are incoming:
\begin{center}
\scalebox{1}{	

\begin{tikzpicture}[baseline={([yshift=-.5ex]current bounding box.center)},x=0.75pt,y=0.75pt,yscale=-1,xscale=1]

\draw [line width=1.0]    (148.2,107.2) -- (251,106.6) ;
\draw [line width=1.2]    (200.6,149.8) -- (200.2,70.6) ;
\draw [color={rgb, 255:red, 0; green, 0; blue, 0 }  ,draw opacity=1 ]   (200.6,139.8) -- (200.26,122.6) ;
\draw [shift={(200.2,119.6)}, rotate = 88.87] [fill={rgb, 255:red, 0; green, 0; blue, 0 }  ,fill opacity=1 ][line width=0.08]  [draw opacity=0] (8.93,-4.29) -- (0,0) -- (8.93,4.29) -- cycle    ;
\draw [color={rgb, 255:red, 0; green, 0; blue, 0 }  ,draw opacity=1 ]   (200.6,93.2) -- (200.6,74.6) ;
\draw [shift={(200.6,96.2)}, rotate = 270] [fill={rgb, 255:red, 0; green, 0; blue, 0 }  ,fill opacity=1 ][line width=0.08]  [draw opacity=0] (8.93,-4.29) -- (0,0) -- (8.93,4.29) -- cycle    ;
\draw [color={rgb, 255:red, 0; green, 0; blue, 0 }  ,draw opacity=1 ]   (231.2,106.95) -- (209.4,106.6) ;
\draw [shift={(234.2,107)}, rotate = 180.92] [fill={rgb, 255:red, 0; green, 0; blue, 0 }  ,fill opacity=1 ][line width=0.08]  [draw opacity=0] (8.93,-4.29) -- (0,0) -- (8.93,4.29) -- cycle    ;
\draw [color={rgb, 255:red, 0; green, 0; blue, 0 }  ,draw opacity=1 ]   (190.6,107.4) -- (168.8,107.05) ;
\draw [shift={(165.8,107)}, rotate = 0.92] [fill={rgb, 255:red, 0; green, 0; blue, 0 }  ,fill opacity=1 ][line width=0.08]  [draw opacity=0] (8.93,-4.29) -- (0,0) -- (8.93,4.29) -- cycle    ;

\draw (125.5,98.2) node [anchor=north west][inner sep=0.75pt]   [align=left] {$\displaystyle p_{1}$};
\draw (194.7,47.6) node [anchor=north west][inner sep=0.75pt]   [align=left] {$\displaystyle p_{2}$};
\draw (259.5,99) node [anchor=north west][inner sep=0.75pt]   [align=left] {$\displaystyle p_{3}$};
\draw (195.1,153.4) node [anchor=north west][inner sep=0.75pt]   [align=left] {$\displaystyle p_{4}$};

\end{tikzpicture}

\ \  .
}
\end{center}

For the box-triangle $I^{bt}_{C,1}$, we have the following region differences
\begin{equation}\label{eq:sign_region}
        \{y_1-y_a,\,y_3-y_a, \,y_b-y_a,\, y_4-y_b,\, y_6-y_b, \,y_8-y_b\}\,, 
\end{equation}
and we label them with red arrows in the following diagram, each being clockwise to a unique vertex,
\begin{figure}[H]   
\begin{center}
$\vcenter{\hbox{\scalebox{1}{	

\begin{tikzpicture}[x=0.75pt,y=0.75pt,yscale=-1,xscale=1]

\draw  [line width=1.2]  (291,339.5) -- (341,339.5) -- (341,389.5) -- (291,389.5) -- cycle ;
\draw [color={rgb, 255:red, 0; green, 0; blue, 0 }  ,draw opacity=1 ][line width=1.2]    (249.5,364.33) -- (231.75,382.08) ;
\draw [color={rgb, 255:red, 0; green, 0; blue, 0 }  ,draw opacity=1 ][line width=1.2]    (249.5,364.33) -- (245.38,360.21) -- (231.75,346.58) ;
\draw [color={rgb, 255:red, 0; green, 0; blue, 0 }  ,draw opacity=1 ][line width=1.2]    (341,339.5) -- (341,326.54) -- (341,319.25) ;
\draw [color={rgb, 255:red, 0; green, 0; blue, 0 }  ,draw opacity=1 ][line width=1.2]    (341,339.5) -- (353.96,339.5) -- (361.25,339.5) ;
\draw [color={rgb, 255:red, 0; green, 0; blue, 0 }  ,draw opacity=1 ][line width=1.2]    (341,389.5) -- (353.96,389.5) -- (361.25,389.5) ;
\draw [color={rgb, 255:red, 0; green, 0; blue, 0 }  ,draw opacity=1 ][line width=1.2]    (341,409.75) -- (341,396.79) -- (341,389.5) ;
\draw [color={rgb, 255:red, 0; green, 0; blue, 0 }  ,draw opacity=1 ][line width=1.2]    (291,319.25) -- (291,332.21) -- (291,339.5) ;
\draw [color={rgb, 255:red, 0; green, 0; blue, 0 }  ,draw opacity=1 ][line width=1.2]    (291,389.5) -- (291,402.46) -- (291,409.75) ;
\draw  [line width=1.2]  (249.5,364.33) -- (291,339.5) -- (291,389.17) -- cycle ;
\draw [color={rgb, 255:red, 208; green, 2; blue, 27 }  ,draw opacity=1 ]   (260.78,342.8) -- (275,357.17) ;
\draw [shift={(258.67,340.67)}, rotate = 45.29] [fill={rgb, 255:red, 208; green, 2; blue, 27 }  ,fill opacity=1 ][line width=1.0]  [draw opacity=0] (8.93,-4.29) -- (0,0) -- (8.93,4.29) -- cycle    ;
\draw [color={rgb, 255:red, 208; green, 2; blue, 27 }  ,draw opacity=1 ]   (265.69,384.85) -- (267.88,381.63) -- (275.67,370.17) ;
\draw [shift={(264,387.33)}, rotate = 304.2] [fill={rgb, 255:red, 208; green, 2; blue, 27 }  ,fill opacity=1 ][line width=1.0]  [draw opacity=0] (8.93,-4.29) -- (0,0) -- (8.93,4.29) -- cycle    ;
\draw [color={rgb, 255:red, 208; green, 2; blue, 27 }  ,draw opacity=1 ]   (283,365.5) -- (299.67,365.5) ;
\draw [shift={(302.67,365.5)}, rotate = 180] [fill={rgb, 255:red, 208; green, 2; blue, 27 }  ,fill opacity=1 ][line width=1.0]  [draw opacity=0] (8.93,-4.29) -- (0,0) -- (8.93,4.29) -- cycle    ;
\draw [color={rgb, 255:red, 208; green, 2; blue, 27 }  ,draw opacity=1 ]   (333.67,365.83) -- (350.33,365.83) ;
\draw [shift={(353.33,365.83)}, rotate = 180] [fill={rgb, 255:red, 208; green, 2; blue, 27 }  ,fill opacity=1 ][line width=1.0]  [draw opacity=0] (8.93,-4.29) -- (0,0) -- (8.93,4.29) -- cycle    ;
\draw [color={rgb, 255:red, 208; green, 2; blue, 27 }  ,draw opacity=1 ]   (317,345.83) -- (316.44,330.5) ;
\draw [shift={(316.33,327.5)}, rotate = 87.92] [fill={rgb, 255:red, 208; green, 2; blue, 27 }  ,fill opacity=1 ][line width=1.0]  [draw opacity=0] (8.93,-4.29) -- (0,0) -- (8.93,4.29) -- cycle    ;
\draw [color={rgb, 255:red, 208; green, 2; blue, 27 }  ,draw opacity=1 ]   (320.56,398.5) -- (320,383.17) ;
\draw [shift={(320.67,401.5)}, rotate = 267.92] [fill={rgb, 255:red, 208; green, 2; blue, 27 }  ,fill opacity=1 ][line width=1.0]  [draw opacity=0] (8.93,-4.29) -- (0,0) -- (8.93,4.29) -- cycle    ;

\draw (253.5,387.83) node [anchor=north west][inner sep=0.75pt]   [align=left] {$\displaystyle 1$};
\draw (254,325) node [anchor=north west][inner sep=0.75pt]   [align=left] {$\displaystyle 3$};
\draw (219.5,355.5) node [anchor=north west][inner sep=0.75pt]   [align=left] {$\displaystyle 2$};
\draw (311.33,310.33) node [anchor=north west][inner sep=0.75pt]   [align=left] {$\displaystyle 4$};
\draw (355.33,356.33) node [anchor=north west][inner sep=0.75pt]   [align=left] {$\displaystyle 6$};
\draw (316,401.67) node [anchor=north west][inner sep=0.75pt]   [align=left] {$\displaystyle 8$};
\draw (270,360) node [anchor=north west][inner sep=0.75pt]   [align=left] {$\displaystyle a$};
\draw (315,355) node [anchor=north west][inner sep=0.75pt]   [align=left] {$\displaystyle b$};

\end{tikzpicture}

}}}$.
\end{center}
\end{figure}

Now let's consider two sets of sign patterns for the first component of the differences in eq.\eqref{eq:sign_region} :
\begin{equation}
    (A)\; \{-,-,-,+,+,+\}, \quad (B)\; \{+,+,+,+,+,+\}\,.
\end{equation}
For kinematics $(A)$ the momentum flow is illustrated in the following: 

\begin{figure}[H]   
\begin{center}
$\vcenter{\hbox{\scalebox{1}{	

\begin{tikzpicture}[x=0.75pt,y=0.75pt,yscale=-1,xscale=1]

\draw  [line width=1.2]  (271,319.5) -- (321,319.5) -- (321,369.5) -- (271,369.5) -- cycle ;
\draw [color={rgb, 255:red, 0; green, 0; blue, 0 }  ,draw opacity=1 ][line width=1.2]    (229.5,344.33) -- (211.75,362.08) ;
\draw [color={rgb, 255:red, 0; green, 0; blue, 0 }  ,draw opacity=1 ][line width=1.2]    (229.5,344.33) -- (225.38,340.21) -- (211.75,326.58) ;
\draw [color={rgb, 255:red, 0; green, 0; blue, 0 }  ,draw opacity=1 ][line width=1.2]    (321,319.5) -- (321,306.54) -- (321,299.25) ;
\draw [color={rgb, 255:red, 0; green, 0; blue, 0 }  ,draw opacity=1 ][line width=1.2]    (321,319.5) -- (333.96,319.5) -- (341.25,319.5) ;
\draw [color={rgb, 255:red, 0; green, 0; blue, 0 }  ,draw opacity=1 ][line width=1.2]    (321,369.5) -- (333.96,369.5) -- (341.25,369.5) ;
\draw [color={rgb, 255:red, 0; green, 0; blue, 0 }  ,draw opacity=1 ][line width=1.2]    (321,389.75) -- (321,376.79) -- (321,369.5) ;
\draw [color={rgb, 255:red, 0; green, 0; blue, 0 }  ,draw opacity=1 ][line width=1.2]    (271,299.25) -- (271,312.21) -- (271,319.5) ;
\draw [color={rgb, 255:red, 0; green, 0; blue, 0 }  ,draw opacity=1 ][line width=1.2]    (271,369.5) -- (271,382.46) -- (271,389.75) ;
\draw  [line width=1.2]  (229.5,344.33) -- (271,319.5) -- (271,369.17) -- cycle ;
\draw    (245.58,354.04) -- (258.67,361.83) ;
\draw [shift={(243,352.5)}, rotate = 30.78] [fill={rgb, 255:red, 0; green, 0; blue, 0 }  ][line width=1.0]  [draw opacity=0] (8.93,-4.29) -- (0,0) -- (8.93,4.29) -- cycle    ;
\draw    (232.67,342.5) -- (250.78,331.4) ;
\draw [shift={(253.33,329.83)}, rotate = 148.5] [fill={rgb, 255:red, 0; green, 0; blue, 0 }  ][line width=1.0]  [draw opacity=0] (8.93,-4.29) -- (0,0) -- (8.93,4.29) -- cycle    ;
\draw    (270.78,348.17) -- (271.33,333.17) ;
\draw [shift={(270.67,351.17)}, rotate = 272.12] [fill={rgb, 255:red, 0; green, 0; blue, 0 }  ][line width=1.0]  [draw opacity=0] (8.93,-4.29) -- (0,0) -- (8.93,4.29) -- cycle    ;
\draw    (299,369.21) -- (278.67,369.5) ;
\draw [shift={(302,369.17)}, rotate = 179.18] [fill={rgb, 255:red, 0; green, 0; blue, 0 }  ][line width=1.0]  [draw opacity=0] (8.93,-4.29) -- (0,0) -- (8.93,4.29) -- cycle    ;
\draw    (314,319.17) -- (293.67,319.46) ;
\draw [shift={(290.67,319.5)}, rotate = 359.18] [fill={rgb, 255:red, 0; green, 0; blue, 0 }  ][line width=1.0]  [draw opacity=0] (8.93,-4.29) -- (0,0) -- (8.93,4.29) -- cycle    ;
\draw    (320.67,359.17) -- (320.95,342.83) ;
\draw [shift={(321,339.83)}, rotate = 90.99] [fill={rgb, 255:red, 0; green, 0; blue, 0 }  ][line width=1.0]  [draw opacity=0] (8.93,-4.29) -- (0,0) -- (8.93,4.29) -- cycle    ;

\draw (235.5,364.5) node [anchor=north west][inner sep=0.75pt]   [align=left] {$\displaystyle 1$};
\draw (234,310) node [anchor=north west][inner sep=0.75pt]   [align=left] {$\displaystyle 3$};
\draw (199.5,335.5) node [anchor=north west][inner sep=0.75pt]   [align=left] {$\displaystyle 2$};
\draw (291,300) node [anchor=north west][inner sep=0.75pt]   [align=left] {$\displaystyle 4$};
\draw (328,333) node [anchor=north west][inner sep=0.75pt]   [align=left] {$\displaystyle 6$};
\draw (291,380) node [anchor=north west][inner sep=0.75pt]   [align=left] {$\displaystyle 8$};
\draw (250,338) node [anchor=north west][inner sep=0.75pt]   [align=left] {$\displaystyle a$};
\draw (295,335) node [anchor=north west][inner sep=0.75pt]   [align=left] {$\displaystyle b$};

\end{tikzpicture}

}}}$.
\end{center}
\end{figure}

\noindent
We see that loop $a$ forms a clockwise loop while loop $b$ forms a counterclockwise loop. For kinematics $(B)$, as shown below, loop $b$ does not form a closed loop,  
\begin{figure}[H]   
\begin{center}
$\vcenter{\hbox{\scalebox{1}{	

\begin{tikzpicture}[x=0.75pt,y=0.75pt,yscale=-1,xscale=1]

\draw  [line width=1.2]  (291,339.5) -- (341,339.5) -- (341,389.5) -- (291,389.5) -- cycle ;
\draw [color={rgb, 255:red, 0; green, 0; blue, 0 }  ,draw opacity=1 ][line width=1.2]    (249.5,364.33) -- (231.75,382.08) ;
\draw [color={rgb, 255:red, 0; green, 0; blue, 0 }  ,draw opacity=1 ][line width=1.2]    (249.5,364.33) -- (245.38,360.21) -- (231.75,346.58) ;
\draw [color={rgb, 255:red, 0; green, 0; blue, 0 }  ,draw opacity=1 ][line width=1.2]    (341,339.5) -- (341,326.54) -- (341,319.25) ;
\draw [color={rgb, 255:red, 0; green, 0; blue, 0 }  ,draw opacity=1 ][line width=1.2]    (341,339.5) -- (353.96,339.5) -- (361.25,339.5) ;
\draw [color={rgb, 255:red, 0; green, 0; blue, 0 }  ,draw opacity=1 ][line width=1.2]    (341,389.5) -- (353.96,389.5) -- (361.25,389.5) ;
\draw [color={rgb, 255:red, 0; green, 0; blue, 0 }  ,draw opacity=1 ][line width=1.2]    (341,409.75) -- (341,396.79) -- (341,389.5) ;
\draw [color={rgb, 255:red, 0; green, 0; blue, 0 }  ,draw opacity=1 ][line width=1.2]    (291,319.25) -- (291,332.21) -- (291,339.5) ;
\draw [color={rgb, 255:red, 0; green, 0; blue, 0 }  ,draw opacity=1 ][line width=1.2]    (291,389.5) -- (291,402.46) -- (291,409.75) ;
\draw  [line width=1.2]  (249.5,364.33) -- (291,339.5) -- (291,389.17) -- cycle ;
\draw    (263,372.5) -- (276.09,380.3) ;
\draw [shift={(278.67,381.83)}, rotate = 210.78] [fill={rgb, 255:red, 0; green, 0; blue, 0 }  ][line width=1.0]  [draw opacity=0] (8.93,-4.29) -- (0,0) -- (8.93,4.29) -- cycle    ;
\draw    (265.22,354.93) -- (283.33,343.83) ;
\draw [shift={(262.67,356.5)}, rotate = 328.5] [fill={rgb, 255:red, 0; green, 0; blue, 0 }  ][line width=1.0]  [draw opacity=0] (8.93,-4.29) -- (0,0) -- (8.93,4.29) -- cycle    ;
\draw    (290.67,375.17) -- (291.22,360.16) ;
\draw [shift={(291.33,357.17)}, rotate = 92.12] [fill={rgb, 255:red, 0; green, 0; blue, 0 }  ][line width=1.0]  [draw opacity=0] (8.93,-4.29) -- (0,0) -- (8.93,4.29) -- cycle    ;
\draw    (319,389.21) -- (298.67,389.5) ;
\draw [shift={(322,389.17)}, rotate = 179.18] [fill={rgb, 255:red, 0; green, 0; blue, 0 }  ][line width=1.0]  [draw opacity=0] (8.93,-4.29) -- (0,0) -- (8.93,4.29) -- cycle    ;
\draw    (334,339.17) -- (313.67,339.46) ;
\draw [shift={(310.67,339.5)}, rotate = 359.18] [fill={rgb, 255:red, 0; green, 0; blue, 0 }  ][line width=1.0]  [draw opacity=0] (8.93,-4.29) -- (0,0) -- (8.93,4.29) -- cycle    ;
\draw    (340.67,379.17) -- (340.95,362.83) ;
\draw [shift={(341,359.83)}, rotate = 90.99] [fill={rgb, 255:red, 0; green, 0; blue, 0 }  ][line width=1.0]  [draw opacity=0] (8.93,-4.29) -- (0,0) -- (8.93,4.29) -- cycle    ;

\draw (255.5,384.5) node [anchor=north west][inner sep=0.75pt]   [align=left] {$\displaystyle 1$};
\draw (254,330) node [anchor=north west][inner sep=0.75pt]   [align=left] {$\displaystyle 3$};
\draw (219.5,355.5) node [anchor=north west][inner sep=0.75pt]   [align=left] {$\displaystyle 2$};
\draw (311,315) node [anchor=north west][inner sep=0.75pt]   [align=left] {$\displaystyle 4$};
\draw (348,353) node [anchor=north west][inner sep=0.75pt]   [align=left] {$\displaystyle 6$};
\draw (311,400) node [anchor=north west][inner sep=0.75pt]   [align=left] {$\displaystyle 8$};
\draw (270,358) node [anchor=north west][inner sep=0.75pt]   [align=left] {$\displaystyle a$};
\draw (315,355) node [anchor=north west][inner sep=0.75pt]   [align=left] {$\displaystyle b$};

\end{tikzpicture}

}}}$.
\end{center}
\end{figure}

\noindent
The arrow in the loop indicates the future direction, where the energy is positive. So the fact that we don't get a closed loop is a reflection that the loop momentum cannot have positive energy everywhere in the loop. Thus kinematics $(B)$ should not be considered when checking cut constraints.


\newpage
\bibliographystyle{JHEP}
\bibliography{mybib.bib}{}

\end{document}